\documentclass[prx]{revtex4-2}

\usepackage{float}
\makeatletter
\let\newfloat\newfloat@ltx
\makeatother

\newcommand{\rhohatt}{\hat \rho_t}
\newcommand{\rhot}{\rho_t}
\newcommand{\rhohat}{\hat \rho}
\newcommand{\rhostar}{\rho_*}
\newcommand{\rhoB}{\rho_{\rm B}}

\newcommand{\tmax}{t_m}

\usepackage{amsfonts,amsmath,amssymb,graphicx,amsthm,bbm}
\usepackage{ulem}
\usepackage[margin=1in]{geometry}

\usepackage[x11names, rgb, html]{xcolor}

\definecolor{red}{HTML}{8C1515}
\definecolor{pink}{HTML}{F4795B}
\definecolor{orange}{HTML}{E98300}
\definecolor{yellow}{HTML}{FEDD5C}
\definecolor{green}{HTML}{175E54}
\definecolor{lightblue}{HTML}{009AB4}
\definecolor{darkblue}{HTML}{006B81}
\definecolor{tan}{HTML}{766253}
\definecolor{darktan}{HTML}{2F2424}
\definecolor{grey}{HTML}{c3ced0}
\definecolor{darkgrey}{HTML}{9dadc4}
\definecolor{black}{HTML}{110d1b}
\definecolor{white}{HTML}{f1f8f1}

\usepackage{hyperref}
\hypersetup{
  colorlinks = true,
  linkcolor = red,
  citecolor = darkblue,
  urlcolor = lightblue
}

\usepackage{algorithm}
\usepackage{algorithmicx}
\usepackage[noend]{algpseudocode}
\algrenewcommand{\algorithmiccomment}[1]{$\vartriangleright$ #1}
\algrenewcommand{\algorithmicreturn}{\textbf{Return: }}
\algnewcommand\algorithmicinput{\textbf{Input: }}
\algnewcommand\Input{\State \algorithmicinput}

\newcommand{\diff}[3]{\frac{\partial^{#3}#1}{\partial #2^{#3}}}

\renewcommand{\eqref}[1]{[\ref{#1}]}

\def\grad{\nabla}

\def\RR{\mathbb{R}} \def\NN{\mathbb{N}} 
\def\EE{\mathbb{E}}

\def\cN{\mathcal{N}}

\newcommand{\detjac}[2]{\det \left| \nabla_{#2} #1 \right|}

\def\<{\langle} \def\>{\rangle}

\DeclareRobustCommand{\argmin}{\operatorname*{argmin}}

\graphicspath{{figures/}}

\begin{document}
\title{Adaptive Monte Carlo augmented with normalizing flows}

\author{Marylou Gabri\'e}
\affiliation{Flatiron Institute, New York, NY 10010}
\affiliation{Center for Data Science, New York University, New York, NY 10011}
\email{mgabrie@nyu.edu}
\author{Grant M. Rotskoff}
\affiliation{Dept. of Chemistry, Stanford University, Stanford, CA 94305}
\email{rotskoff@stanford.edu}
\author{Eric Vanden-Eijnden}
\affiliation{Courant Institute, New York University, New York, NY 10012}
\email{eve2@cims.nyu.edu}

\begin{abstract}
Many problems in the physical sciences, machine learning, and statistical inference necessitate sampling from a high-dimensional, multi-modal probability distribution.
Markov Chain Monte Carlo (MCMC) algorithms, the ubiquitous tool for this task, typically rely on random local updates to propagate configurations of a given system in a way that ensures that generated configurations will be distributed according to a target probability distribution asymptotically.
In high-dimensional settings with multiple relevant metastable basins, local approaches require either immense computational effort or intricately designed importance sampling strategies to capture information about, for example, the relative populations of such basins. 
Here we analyze an adaptive MCMC which augments MCMC sampling with nonlocal transition kernels parameterized with generative models known as normalizing flows.
We focus on a setting where there is no preexisting data, as is commonly the case for problems in which MCMC is used. Our method uses: (i) a MCMC strategy that blends local moves obtained from any standard transition kernel with those from a generative model to accelerate the sampling and (ii) the data generated this way to adapt the generative model and improve its efficacy in the MCMC algorithm.
We provide a theoretical analysis of the convergence properties of this algorithm, and investigate numerically its efficiency, in particular in terms of its propensity to equilibrate fast between metastable modes whose rough location is known \textit{a~priori} but respective probability weight is not. We show that our algorithm can sample effectively across large free energy barriers, providing dramatic accelerations relative to traditional MCMC algorithms.
\end{abstract}

\maketitle

\section{Introduction}
\label{sec:intro}

Monte Carlo approximations are the method of choice to extract information from high-dimensional probability distributions encountered in the description of natural systems and statistical models. One generic feature of these distributions that is particularly challenging for sampling is multi-modality (or metastability); that is, when low-probability regions separate high-probability regions (or basins) of the state space. Markov Chain Monte Carlo algorithms, which are driven primarily by local dynamics such as Hamiltonian Monte Carlo or Langevin dynamics, typically struggle to transition between metastable basins leading to either extremely long correlation times along the chains and few effective independent samples, or even failure to equilibrate at all. As a result, slow relaxation and metastability plague sampling problems that arise in chemistry and biophysics~\cite{bolhuis_transition_2002}.

On the other hand, generative models that which have garnered much attention in the machine learning literature seem to efficiently sample complicated high-dimensional distributions, such as collections of images. Most of these generative models, including generative adversarial networks \cite{goodfellow_generative_2014} and variational autoencoders \cite{kingma_auto-encoding_2013}, rely on the transformation of samples from a simple and tractable base distribution through a map parametrized with neural networks. After learning, the map transforms samples from the base to mimic samples of a given empirical distribution.
This formulation allows drawing independent samples from the model at a negligible cost.
However, the conventional strategy for training a generative model requires an extensive data set of samples.
Arguably, these models have succeeded most dramatically in domains where the cost of generating and curating data is comparatively low (e.g., image recognition)~\cite{goodfellow_generative_2014, song_nice_2017, song_scorebased_2021}. 

In scientific computing applications, obtaining data from the distribution is the primary goal. 
Furthermore, the quality metrics used in traditional machine learning applications are \textit{a~priori} quite different from the efficacy and precision in sampling a target distribution. Hence, it is natural to ask whether traditional MCMC methods and generative models can be successfully combined to accelerate sampling of complicated high-dimensional distributions? 

The prospect of enhancing sampling with suitable generative models is an active area of inquiry~\cite{huang_accelerated_2017,sejdinovic_kernel_2014a,levy_generalizing_2018b, song_nice_2017, titsias_learning_2017, anh_autoencoding_2018,McNaughton2020}.
In particular, sampling via Metropolis-Hastings MCMC requires the computation of each transition generation probability and its inverse. 
As a result, the model architectures on which most generative neural networks rely are not conducive to Metropolis-Hasting MCMC. 
However, specific classes of neural networks have been designed with this in mind, allowing for efficient estimates of the probability of a generated sample, including auto-regressive models \cite{Germain2015} and normalizing flows (NF), which are expressive invertible function representations \cite{rezende_variational_2015,papamakarios_normalizing_2021}. 
At this point, NF have been investigated as transition operators in MCMC algorithms and variational ansatz in a variety of contexts in the physical sciences and Bayesian applications~\cite{ albergo_flow-based_2019,sidky_molecular_2020,sbailo_neural_2021, wu_solving_2019, nicoli_estimation_2021, DelDebbio2021,rezende_variational_2015}. These methods offer a promising speed-up for sampling unimodal distributions without requiring preexisting data samples by relying on a \emph{self-training} objective for the map (described in detail in Appendix \ref{app:transampl}). However, the multi-modal case requires prior knowledge about either the symmetries of the systems generating the degeneracy of the modes \cite{DelDebbio2021}, or the location of the metastable basins \cite{sbailo_neural_2021}. This necessity was noted in the influential work of No\'e \textit{et~al.} \cite{noe_boltzmann_2019} that proposes a training strategy for normalizing flows to generate low-energy configurations which are subsequently reweighted.

The aim of this paper is to propose an alternative class of adaptive MCMC algorithms augmented with a NF trained on the fly with the generated samples and also to carefully assess the prospects of these algorithms for accelerating sampling in cases where no extensive preexisting data set is available. Our main contributions are:
\begin{itemize}
    \item We introduce an adaptive Metropolis-Hastings MCMC algorithm that augments a chain performing local steps with nonlocal resampling steps proposed by a NF. The corresponding proposal distribution is adapted along sampling by training the NF via optimization of a forward Kullback-Leibler divergence estimated on the generated data. 
    \item As the adaptation of the map depends on the history of the chains, the convergence of the proposed algorithm, where training and sampling happen simultaneously, is not trivial. We show that the adaptive algorithm is akin to a nonlinear MCMC scheme \cite{andrieu_nonlinear_2011}, that we analyze in the continuous time limit. In this limit, we show that the algorithm converges asymptotically with an exponential rate that can be explicitly estimated.
    \item We test this adaptive MCMC approach on complex examples in high dimension (random fields, transition paths, and interacting particle systems at phase coexistence) and show that it dramatically accelerates the sampling. In particular we estimate the relative statistical weights of metastable states efficiently without constructing a specific pathway between the basins of interest.
\end{itemize}
Our results also emphasize some key determinants for the success of sampling augmented with learning:
\begin{itemize}
    \item One representative configuration per mode of interest in the target distribution must be known beforehand to initialize the chains. We critically assess the ability of using generative model proposals to discover unknown metastable states and show that this prospect is statistically unlikely without prior information about these states. 
    \item Blending generative sampling with a standard MCMC strategy is typically required to guarantee good sampling  
    of the target distribution: in particular we show that relying on generative sampling alone may not be sufficient because it requires that we learn the generative model to a degree of accuracy that is hard to achieve in practice, especially in high-dimensional examples.
    \item Finally, our analysis and numerical experiments show how scaling to high dimensions is also facilitated by parametrizations of normalizing flows that incorporate known structures of the target distributions, such as short-scale correlations. The possibility to inform the map, or the base distribution, with physical intuition alleviates the curse of dimensionality that would prevent general-purpose parametrizations from reaching the required level of precision with reasonably sized models as the dimension grows.
\end{itemize}

\section{Design Challenges in MCMC Methods}
The goal of sampling is to generate configurations $x \in \Omega\subset \RR^d$ in proportion to some probability measure $\nu_*(dx) = \rho_*(x)dx$ which we assume has probability density function $\rho_*$. 
In physical systems, we typically write this in Boltzmann form
\begin{equation}
    \rhostar(x) = Z_*^{-1} e^{- U_*(x)}
\end{equation}
where $U_* \propto -\log \rho_*$ is the potential energy function for the system.
We assume that we have an explicit model for $U_*$ and can efficiently evaluate this energy function, though we may have little \emph{a priori} information about the distribution of configurations associated with this energy and in general do not know the normalization constant $Z_*$.

MCMC algorithms avoid computing~$Z_*$ by generating a sequence $\{x(k)\}_{k\in \NN}$ of configurations with a transition kernel $\pi(x, y)$ with $\int_\Omega \pi(x,y)dy =1$ for all $x\in \Omega$, which quantifies the conditional probability density of a transition from state $x$ into state $y$.
Assume that the kernel $\pi(x,y)$ is irreducible and aperiodic \cite{meyn2012markov}, and satisfies the detailed balance relation
\begin{equation}
\label{eq:db}
    \rho_*(x) \pi(x, y) = \rho_*(y) \pi(y,x).
\end{equation}
Then the sequence $\{x(k)\}_{k\in\NN}$ will sample the target density $\rho_*$ in the sense that the empirical average of any suitable observable $\phi$ converges to its expectation over $\rho_*$, i.e., 
\begin{equation}
\label{eq:ergo}
    \lim_{N\to \infty} \frac1N \sum_{k=1}^N \phi(x(k)) = \int_\Omega \phi(x) \rho_*(x) dx. 
\end{equation}
Designing a transition kernel $\pi$ leading to fast convergence of the series in~\eqref{eq:ergo} is a generically challenging task for MCMC algorithms. 
In Metropolis-Hastings MCMC one constructs a proposal distribution that creates new samples that are then accepted or rejected according to a criterion that maintains~\eqref{eq:db}. 
For example, in the Metropolis adjusted Langevin algorithm (MALA)~\cite{roberts_exponential_1996} new configurations are proposed by approximating the solution of the Langevin equation propagated on a fixed time interval. 

Metropolis-Hastings MCMC algorithms, however, involve a trade-off between two requirements that are hard to fulfill simultaneously. Proposal distributions using local dynamics like MALA suffer from long decorrelation times when there is metastability in the target density $\rho_*.$
At the same time, seeking faster mixing times with non-local proposal distributions requires careful design to avoid high rejection rates. 
Recent work in the machine learning literature has suggested a \emph{data-driven} approach to constructing the transition kernel~\cite{levy_generalizing_2018b,titsias_learning_2017,song_nice_2017} that aids in this design challenge; these approaches originally were pioneered in the context of adaptive and nonlinear MCMC algorithms~\cite{andrieu_ergodicity_2006,haario_adaptive_2001,andrieu_nonlinear_2011,jasra_populationbased_2007}.
Here, we explore the use of normalizing flows to adaptively parameterize the transition kernel.

\section{MCMC Sampling with Normalizing Flows}
\label{sec:setup}
A normalizing flow (NF) is an invertible map $T:\Omega \to\Omega$ that is optimized to transport samples from a base measure $\nu_{\rm B}(dx) = \rho_{\rm B}(x)dx$ (for example a Gaussian with unit variance) to a given target distribution~\cite{papamakarios_normalizing_2021}.
The goal is to produce a map $T_*$ with inverse $\bar T_*$ such that an expectation of an observable with respect to $\rho_*$ can be estimated by transforming samples from the base density to the target, i.e. if $x_\text{B}$ is drawn from $\rho_{\rm B}(x)$ then $T_*(x_\text{B})$ is a sample from $\rho_*(x)$ so that for any suitable observable $\mathcal{O}$ we have
\begin{equation}
    \int_\Omega \mathcal{O}(T_*(x)) \rho_\text{B}(x) dx = 
    \int_\Omega \mathcal{O} (x) \rhostar(x) dx .
\end{equation}
The existence of such a map $T_*$ is guaranteed under general conditions on $\rhostar$ and $\rho_\text{B}$ investigated e.g. in the context of optimal transport theory \cite{villani_topics_2003,Santambrogio2015}.
Of course, in practice we do not have direct access to this ideal map $T_*$. Next we discuss how any approximation~$T$ of $T_*$ can in principle be used to perform exact sampling of the target via Metropolis-Hastings MCMC, and how the map $T$ can be improved via training.

\subsection{Metropolis-Hastings MCMC with NF} Throughout, we denote the push-forward of $\rho_\text{B}$ under the map $T$ simply by~$\rhohat$: it has the explicit form
\begin{equation}
    \label{eq:rhohatexplicit}
    \rhohat(x) = \rho_{\rm B}(\bar T(x)) \detjac{\bar T}{x},
\end{equation}
where $\bar T$ denotes the inverse map, i.e. $\bar T(T(x))=T(\bar T(x))=x$. In practice, the parametrization of the map $T$ must be designed carefully to evaluate this density efficiently, requiring easily estimable Jacobian determinants and inverses. This issue has been one of the main foci in the normalizing flow literature \cite{papamakarios_normalizing_2021} and is for instance solved using coupling layers \cite{Dinh2015, dinh_density_2017}.
Even if the map $T$ is not the optimal $T_*$, i.e. $\rhohat(x) \not=\rhostar(x)$, as long as $\rhohat$ and $\rho_*$ are either both positive or both zero at any point $x\in\Omega$, we can still generate configurations using $T$ with the correct statistical weight in the target distribution by using a Metropolis-Hastings MCMC algorithm with an accept-reject step: a proposed configuration $y=T(x_{\rm B})$ from a given configuration $x$ is accepted with probability
\begin{equation}
\label{eq:accept}
   \textrm{acc}(x,y) = \min \left[1, \frac{\rhohat(x) \rho_*(y)}{\rho_*(x)\rhohat(y)} \right].
\end{equation}
This procedure is equivalent to using the transition kernel
\begin{equation}
    \label{eq:pit}
    \pi_T(x,y) = \textrm{acc}(x,y) \rhohat(y) + \bigl(1 - r(x)\bigr) \delta(x-y)
\end{equation}
where $r(x) = \int_\Omega \textrm{acc}(x,y) \rhohat (y) dy$.
The formula in Eq.~\eqref{eq:accept} for the acceptance probability emphasizes that if the generated configurations do not have appreciable statistical weight in the target distribution (i.e. $\rho_*(y)$ is very small) few configurations will be accepted.
This problem can become fundamental in high-dimensional spaces because, unless care is taken to ensure otherwise, the push-forward measure and the target will not overlap-----for a discussion of this issue and a precise measure-theoretic formulation of MCMC with NF see  Appendix~\ref{app:measures}. 
In contrast, when the map yields an appreciable acceptance rate, the flow based proposals may mix much faster than proposals based on local moves as independent configurations $y$ can be directly sampled from $\rhohat$. We illustrate these features in numerical experiments presented below.

\subsection{Map Training} Improving the map $T$ requires that we optimize some objective function measuring the discrepancy between the $\rhohat(x)$ and $\rhostar(x)$: for example the Kullback-Leibler divergence of $\rhostar$ with respect to $\rhohat$, which is given by an expectation over~$\rhostar$
\begin{equation}
\label{eq:DKLstar}
    D_{\text{KL}}(\rhostar\|\rhohat) = C_* - \int_\Omega \log \rhohat(x) \rhostar(x) dx,
\end{equation}
where $C_* = \int_\Omega\log \rhostar(x) \rhostar(x) dx$ is a constant irrelevant for the optimization of this objective over $T$. 
Typically, this procedure is used in situations where a data set from $\rhostar$ is available beforehand~\cite{song_nice_2017, song_scorebased_2021} and can be used to construct an empirical approximation of Eq.~\eqref{eq:DKLstar}; in contrast, we are focused on situations where only limited data exists initially.
In this context, it has been suggested~\cite{rezende_variational_2015,noe_boltzmann_2019,albergo_flow-based_2019} to use the reverse KL divergence of $\rhohat$ with respect to $\rhostar$, since it can be expressed as an expectation over $\rhohat$:
\begin{equation}
\label{eq:revDKLstar}
    D_{\text{KL}}(\rhohat\|\rhostar) = -\log Z_* + \int_\Omega [U_*(x)+\log \rhohat(x)] \rhohat(x) dx.
\end{equation}
The (unknown) constant $\log Z_*$ is irrelevant for the optimization of this objective over $T$. This approach seems to alleviate altogether the need of preexisting samples from $\rho_*$: however, it rests on the possibility to discover relevant regions on $\rho_*$ via sampling $\rhohat$.  In practice, this may be very hard to achieve unless we have a good estimate of the ideal $T_*$ to begin with, which is typically not the case: for this reason, here we will resort to optimizing an approximation of the direct KL in Eq.~\eqref{eq:DKLstar}. This procedure, described in the next section, relies on a dynamical estimate of the forward KL divergence that uses data generated via an adaptive MCMC that synergistically takes advantage of the learning to produce samples of the target $\rhostar$ efficiently.

We stress that once the map $T$ becomes accurate enough, Eqs.~\eqref{eq:DKLstar} and~\eqref{eq:revDKLstar} can also be combined for further training, as was done e.g. in the related context of Boltzmann Generators \cite{noe_boltzmann_2019}---for a roadmap of the different possible strategies to train $T$ we refer the reader to Appendix~\ref{app:transampl}. We also stress that trainable generative models other than normalizing flows can be used as well, as long as they offer an easy way to sample some $\rhohat$ that can be adapted to the target $\rhostar$:  this feature is illustrated in the numerical examples presented below.

\section{Adaptive MCMC: Concurrent Sampling and Training}
\label{sec:nfs}

\begin{algorithm}[t!]
\caption{Adaptive MCMC: Concurrent MCMC sampling and map training \label{alg:concurrent}}   
\begin{algorithmic}[1]
    \State \textsc{SampleTrain}($U_*$, $T$, $\{x_i(0)\}_{i=1}^n$, $\tau$, $k_\text{max}$, $k_\text{Lang}$, $\epsilon$)
    \State {\bfseries Inputs:} $ U_*$ target energy, $T$ initial map, $\{x_i{(0)}\}_{i=1}^n$ initial data, $\tau>0$ time step, $k_\text{max}\in\NN$ total duration, $k_\text{Lang}\in\NN$ number of Langevin steps per resampling step, $\epsilon>0$ map training time step
    \State $k=0$
    \While{$k< k_\text{max}$}
    \For{$i=1,\dots, n$}
    \If{$k \mod k_\text{Lang}+1=0$}
    \State $x'_{\rm B, i} \sim \rho_{\rm B}$ \label{lst:resample1}
    \State $x_i' = T(x'_{{\rm B}, i})$ \Comment{push-forward via $T$} \label{lst:resample3}
    \State $x_i(k+1) = x_i'$ with probability ${\rm acc}(x_i(k),x_i')$, otherwise $x_i(k+1) = x_i(k)$ \Comment{resampling step}\label{lst:resample4}
    \Else
    \State $x_i' = x_i(k) - \tau  \nabla U_*(x_i(k)) + \sqrt{2 \tau}\, \eta_i$ with $\eta_i \sim \cN(0_d,I_d)$ \Comment{discretized Langevin step}
    \State $x_i(k+1) = x_i'$ with MALA acceptance probability or ULA, otherwise $x_i(k+1) = x_i(k)$
    \EndIf
    \EndFor
    \State $k\gets k+1$
    \State $\mathcal{L}[T] = - \frac1n \sum_{i=1}^n \log \rhohat(x_i(k))$ \Comment{evaluate $D_{\rm KL}(\rhot\|\rhohat)$ on sampled data}
    \State $T \gets T - \epsilon \nabla \mathcal{L}[T]$ \Comment{Update the map}
    \EndWhile
    \State {\bfseries return:} $\{x_i(k)\}_{k=0,i=1}^{k_{\rm max},n}$, $T$
\end{algorithmic}
\end{algorithm}

The adaptive MCMC we propose concurrently acquires new data by combining a local sampler with a nonlocal one based on a NF, and uses these data to further optimize the flow.
This procedure is summarized in Algorithm~\ref{alg:concurrent} with MALA as local MCMC algorithm, and it involves the following components:

\subsection{Sampling} Our algorithm combines MCMC steps using a local kernel $\pi$ with those obtained using the normalizing flow kernel $\pi_T$ in Eq.~\eqref{eq:pit}. As such it is consistent with the compounded transition kernel (assuming for simplicity of notation that we make consecutive steps which each kernel)
\begin{equation}
\label{eq:nfker}
    \hat\pi(x,y) = \int_\Omega \pi(x,z) \pi_T(z,y) dz
\end{equation}
which satisfies the detailed balance relation Eq.~\eqref{eq:db} because the transitions kernels $\pi$ and $\pi_T$ individually do. 
While the flow based kernel $\pi_T$ allows global mixing between modes once $T$ is sufficiently optimized, alternating with the local kernel~$\pi$ improves the robustness of the scheme, by ensuring sampling proceeds 
in places within the modes where the map is not-optimal. This is useful during the first iterations of the scheme when the map $T$ is almost untrained, as well as once training has converged, if the expressivness of the map parametrization is not sufficient to capture all the features of the target distribution. In Appendix~\ref{sec:a:wiggle}, we demonstrate numerically the benefit of retaining a local components to the sampling scheme (see  Fig.\ref{fig:wiggle}).
Let us also note that the convergence rate of a chain using~$\hat\pi(x,y)$ is necessarily faster than that of MCMC using $\pi(x,z)$ or $\pi_T(z,y)$ individually: if we assume existence of a spectral gap for both $\pi$ and $\pi_T$ and denote the leading eigenvalues of these kernels by $\hat\lambda<1$, $\lambda<1$, and $\lambda_T<1$, respectively, we have $\hat\lambda\le\lambda\lambda_T$. 
While we employ MALA here, any detailed balance MCMC method could be used in Eq.~\eqref{eq:nfker}. The transition kernel $\pi$ does not need to be local, it should, however, have satisfactory acceptance rates. 
Note that in the experiments that follow we used unadjusted Langevin dynamics (ULA) because the time steps were sufficiently small to ensure a high acceptance rate.

\subsection{Adaptation} The kernel $\pi_T$ of Algorithm~\ref{alg:concurrent} is adapted by using the newly sampled configurations as data to optimize the parameters of the normalizing flow $T$. Denoting by $\rho_k$ the probability density of the chain with kernel~$\hat\pi$ after $k\in \NN$ steps from initialization $\rho_0$, we minimize the KL-divergence of $\rho_k$ with respect to $\rhohat$, $D_{\text{KL}}(\rho_k\|\rhohat)$, instead of the KL-divergence of the unknown $\rhostar$ with respect to $\rhohat$ as in Eq.~\eqref{eq:DKLstar}. Denoting by $\{x_i(k)\}_{i=1}^n$ the sample of $n$ chains after $k \in \NN$ steps of MCMC, this amounts to using the following consistent estimator for $D_{\text{KL}}(\rho_k\|\rhohat)$, up to an irrelevant constant:
\begin{equation}
\label{eq:E1}
\begin{aligned}
    \mathcal{L}_n[T] & = -\frac1n \sum_{i=1}^n  \log \rhohat(x_i(k)) \\
    & = \frac1n \sum_{i=1}^n  \bigl( U_{\rm B}(\bar T(x_i(k)) - \log\det |\grad \bar T(x_i(k))| \bigr).
\end{aligned}
\end{equation}
In practice, we use stochastic gradient descent on this loss function to update the parameters of the normalizing flow (Algorithm \ref{alg:concurrent} line 11).
While the expression for the loss is written at iteration $k$, we can average gradients over multiple MCMC steps. Details of the maps parametrization's and training procedures for the experiments presented in the next sections are described in Appendix~\ref{sec:a:NF}.

\subsection{Initialization} To start the MCMC chains, we assume that we have configurations $\{x_i(0)\}_{i=1}^n$ in the different modes of the target but they are not necessarily drawn from $\rhostar$.
We emphasize that the method therefore applies in situations where the locations of the metastable states of interest are known \textit{a~priori} and one should not expect the procedure to find states in basins distinct from initialization.
We demonstrate that it is unlikely that the adaptive MCMC will discover new metastable basins without any initial information about their location in Appendix~\ref{sec:a:Gauss} on the example of a Gaussian mixture model (see Fig.~\ref{fig:gaussian_mixture}) and in  Appendix~\ref{app:exploration_bounds} for the random field example discussed  below.

We initialize the map $T$ as the identity transformation and propagate the initial data using $\hat\pi$. The initial sampling is essentially driven by the local MCMC, here Langevin dynamics, as the map is not adapted to the target. 
As the map improves, nonlocal moves start to be accepted, the autocorrelation time drops  and the Markov chains reallocate mass in proportion to the statistical weights of the different basins. These features are illustrated in Fig.~\ref{fig:gaussian_mixture} in the context of the Gaussian mixture model discussed in Appendix~\ref{sec:a:Gauss} and in Figs.~\ref{fig:phi4} and \ref{fig:phi4-training-mixing} in the context of the random field example discussed below.

\subsection{Convergence} Two important questions arise regarding Algorithm~\ref{alg:concurrent}: first, does this scheme produce samples that converge in distribution towards the target and, if so, does the adaptive training of the map $T$ improve the rate of convergence to the target distribution?
To analyze the properties of a transition operator that combines nonlocal moves with the normalizing flow and a local MCMC algorithm, we consider our approach in the continuous-time limit. In this limit, when using Langevin dynamics as local sampler, the density of the evolving $\rhot$ with respect to the target $\rho_*$, defined as $g_t = \rho_t/\rho_*$, satisfies
\begin{equation}
\begin{aligned}
  \label{eq:58}
  \partial_t g_t &= -\nabla U_*\cdot \nabla g_t + \Delta g_t\\
  &+  \alpha \int_\Omega \min(\hat g_t(x),\hat g_t(y))  \left( g_t(y)  - g_t(x)\right) \rho_*(y) dy
  \end{aligned}
\end{equation}
where $\hat g_t = \hat \rho_t/\rho_*$ and $\alpha\ge 0$ is an adjustable parameter that measure the balance between the Langevin and the resampling parts of the dynamics. Setting $\alpha=0$ amounts to using the Langevin dynamics alone:  in that case, for any initial condition $\rho_0$, we have that $\rho_t \to \rho_*$ (i.e. $g_t\to1$) as $t\to \infty$, but this convergence  will be exponentially slow in general \cite{stroock_logarithmic_1993}. 
The situation changes if we include the resampling step, i.e. consider Eq.~\eqref{eq:58} with $\alpha>0$. In Appendix~\ref{sec:MCMCcl}, under various assumptions about $\hat g_t$ we derive convergence rates under the dynamics in Eq.~\eqref{eq:58} for the Pearson $\chi^2$-divergence of $\rho_t$ with respect to $\rho_*$, which we denote as
\begin{equation}
    \label{eq:chi2}
    D_t = \int_\Omega \frac{\rho_t^2}{\rho_*} dx -1 = \int_\Omega g^2_t\rho_* dx -1 \ge 0.  
\end{equation}
In particular, we study the situation where $T$ learns the instantaneous distribution at all times, that is, $\rhohat = \rhot$ (and hence $\hat g_t = g_t$) for all $t\ge0$. 
While this is certainly a significant approximation, we observe in numerical experiments that there is a dramatic improvement in sampling once there is some mixing between metastable basins, which motivates this limiting scenario.
In this case, under the assumptions that there exists some $t_0\ge0$ such that  $D_{t_0}<\infty$ and
\begin{equation}
    G_{t_0} = \inf_{x\in \Omega} \frac{\rho_{t_0}(x)}{\rho_*(x) } = \inf_{x\in \Omega} g_{t_0}(x)> 0,
\end{equation}
we show that  
\begin{equation}
\label{eq:rateexp}
    \forall t \ge t_0 \ \ : \ \ D_t \le \frac{D_{t_0} }{\left(G_{t_0}(e^{\alpha (t-t_0)} -1) +1\right)^2}.
\end{equation}
This equation indicates that $D_t\le D_{t_0}$ remains approximately constant for $\alpha(t-t_0) \leq \log G_{t_0}^{-1}$, then decays exponentially with constant rate $2\alpha>0$ subsequently.
The derivation of Eq.~\eqref{eq:rateexp} also shows that the exponential rate is controlled by  the resampling step of the MCMC algorithm that relies on the normalizing flow, and this rate can only improve when we concurrently use Langevin dynamics steps.
In Appendix~\ref{app:bdfp}, we connect the sampling scheme we use to a birth-death Fokker-Planck equation~\cite{lu_accelerating_2019}, which could also be implemented in practice as a Markov jump process; again this analysis emphasizes the favorable convergence properties of the scheme.

\subsection{Scalability: Model-Informed Base Distributions and Maps, Mixtures, etc} 
As the dimension of the problem grows it becomes increasingly difficult to train a map to produce a push-forward distribution matching the target to a given level of accuracy. 
Before presenting numerical experiments, we emphasize a few additional ingredients easing the learning of generative models for the sampling of complex high-dimensional systems.

When training a normalizing flow to represent a target density for which a preexisting empirical dataset is available, a standardizing transformation or a ``whitening layer'' is typically added at the output \cite{noe_boltzmann_2019}. This layer centers and rescales the different input dimensions such that their covariance matches the identity covariance of the standard normal distribution usually used as base distribution. This operation, though it requires preexisting data, crucially improves the outcome of learning when the original covariance of the data is highly anisotropic. In the experiments below, we show that it is sometimes possible to rely on the knowledge of the target distribution to perform an operation akin to this whitening layer with no preexisting data samples. For instance, below we choose a base Gaussian distribution with covariance matching the short-scale correlations of equilibrium configurations of the system's known Hamiltonian. We can also design physics-informed base distributions that are more adapted to the problem at hand than a Gaussian distribution: for example, in the interacting particle system we used the uniform distribution of the particle in the domain, which is an ideal distribution in the gaseous phase. 

Using prior knowledge about the physics can also help designing the class of maps $T$ to optimize upon. For example, in the interacting particle system, we used maps that factorize in ways tailored to the system's features.  
Yet another way of easing learning, especially when modes have very different fine structures or statistical weights, is to rely on a mixture of maps, instead on a single map, training each component to represent a different mode. A similar idea was exploited in \cite{noe_boltzmann_2019} to compute free energy differences between basins after training. In practice, a map $T_m$ is pretrained for each mode indexed by $m$ using data generated with the local MCMC sampler initialized in the corresponding mode. Then, the adaptive MCMC procedure described in Algorithm~\ref{alg:concurrent} can be started. The nonlocal proposal is the mixture of the push-forwards $\rhohat_m$ with initial weights $p_m$. The adaptive part of the proposal then amounts to optimizing the mixture weights $p_m$ via Eq.~\eqref{eq:E1}, in a similar fashion as the parameters of the flow when using a single map\footnote{In principle, the parameters of each $T_m$ could also be further refined in this stage, but we have not tested this scenario in experiments yet.}. This mixture method requires that we train several maps but allows treatment of more complex systems, as demonstrated below.

\begin{figure*}[t!]
    \centering
    \includegraphics[width=\textwidth]{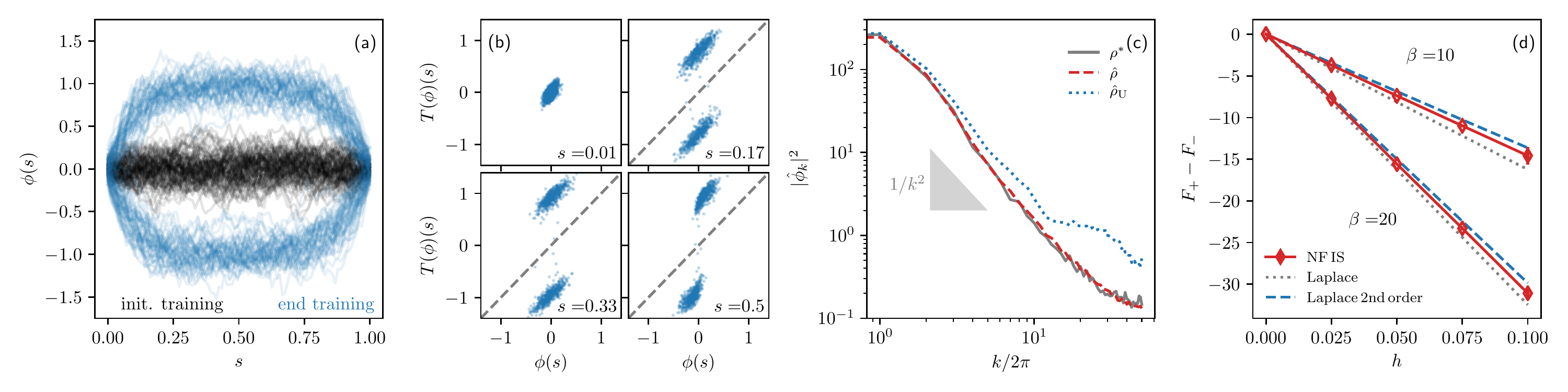}
    \caption{{\bf Sampling metastable states of the stochastic Allen-Cahn model with Langevin dynamics augmented with a normalizing flow.} (a) Configurations obtained by pushing independent samples from the informed base measure Eq.~\ref{eq:informed_base} through the flow $T$ at the beginning (black) and at the end of training (blue). Around $\sim60\%$ of generated configurations are accepted according to the Metropolis-Hasting criteria. (b) The learned map $T$ is local in space. (c) Fourier spectrum of the target samples, samples from a flow with informed base measure and uniformed base measure. An informed base measure is necessary to capture the higher frequency features of the target density. (d) Computation of the free energy differences between positive and negative modes with importance sampling from the normalizing flow as a function of a local biasing field added in the Hamiltonian Eq. \ref{eq:phi4-bias}. Results are reported for inverse temperature $\beta=20$, as in the rest of the plots, and for the same experiment repeated at temperature $\beta=10$. Errors bars computed from estimator variance are smaller than marker. }
    \label{fig:phi4}
\end{figure*}
\begin{figure*}
    \centering
    \includegraphics[width=\textwidth]{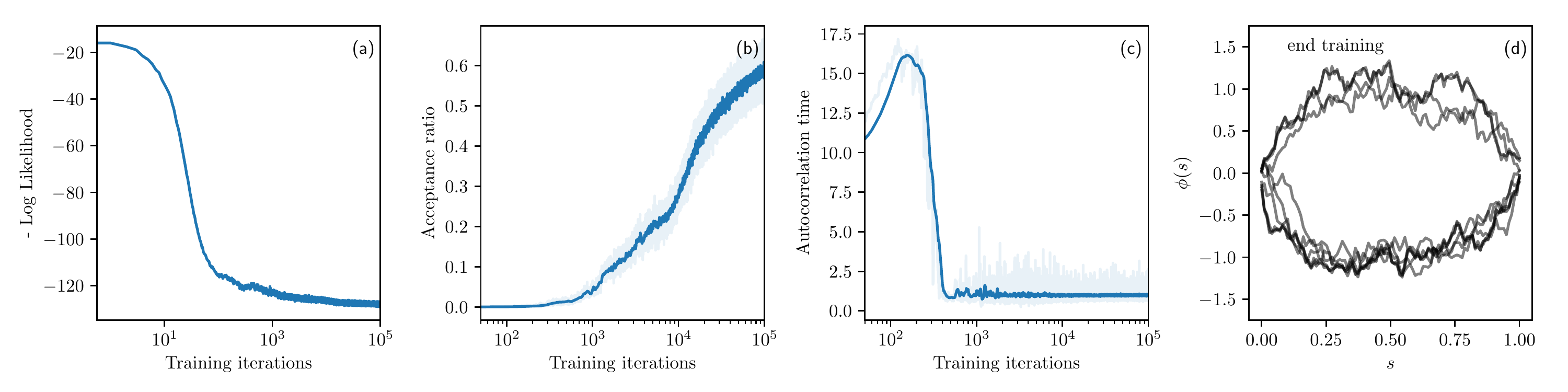}
    \caption{{\bf Concurrent training and sampling of the stochastic Allen-Cahn model with a Real-NVP normalizing flow.} (a) The stochastic gradient descent using samples generated by the procedure decreases the negative log-likelihood gradually. (b) As the training progresses the acceptance rate in the Metropolis-Hasting using proposals from the normalizing flow improves gradually, reaching levels well beyond $50\%$. Rolling average over the last 50 time steps is plotted in darker color. (c) As independent proposals from the flow starts getting accepted the Markov Chain autocorrelation times drops abruptly. (d) Fast mixing is illustrated by looking at the consecutive states of one walker updated with the transition kernel combining local Langevin updates and resampling with the push-forward. In 10 steps, the single walker has jumped between $\phi_+$ and $\phi_-$. }
    \label{fig:phi4-training-mixing}
\end{figure*}
\section{Numerical Experiments}
\subsection{Fast-mixing augmented MCMC for random fields}
As a first example to illustrate the efficacy of adaptive sampling, we consider a stochastic Allen-Cahn model, a canonical and ubiquitous model for the microscopic physics of phase transitions in condensed matter systems~\cite{berglund_eyring_2017}.
\paragraph{Field system.} The stochastic Allen-Cahn equation is defined in terms of a random field $\phi:[0,1]\to \RR$ that satisfies
\begin{equation}
    \label{eq:stochAC}
    \partial_t\phi = a \partial_s^2 \phi + a^{-1}(\phi-\phi^{3}) + \sqrt{2\beta^{-1}}\, \eta(t,s)
\end{equation}
where $a>0$ is a parameter, $\beta$ is the inverse temperature, $s\in[0,1]$ denotes the spatial variable, $\eta$ is a spatio-temporal white noise and we impose Dirichlet boundary conditions in which $\phi(s=0)=\phi(s=1)=0$ throughout. This stochastic partial differential equation (SPDE) is well-posed in one spatial dimension~\cite{faris_large_1982,marcus_parabolic_1974}, and its invariant measure is the Gibbs measure associated with the Hamiltonian
\begin{equation}
    \label{eq:phi4}
    U_*[\phi] = \beta \int_0^1   \left[ \frac{a}{2} (\partial_s \phi )^2 + \frac{1}{4a}\left(1-\phi^2(s)\right)^2  \right] ds .
\end{equation}
The first term in the Hamiltonian~\eqref{eq:phi4} is a spatial coupling that penalizes changes in $\phi$ and hence, at low temperature, has the effect of aligning the field in positive or negative direction. 
As a result the Hamiltonian~\eqref{eq:phi4} has two global minima, denoted by $\phi^+$ and $\phi^-$, in which the typical values of $\phi$ are $\pm 1$ (see Fig.~\ref{fig:phi4} (a)).  
Because there is a free energy barrier between $\phi^+$ and $\phi^-$, local updates via traditional MCMC based e.g. on using the stochastic Allen-Cahn equation~\eqref{eq:stochAC} will not mix, even on very long timescales.
Indeed, if we wanted to compute the free energy difference between these basins, we would need to construct a pathway through configuration space and use importance sampling techniques along the path~\cite{frenkel_understanding_2001}.
Our adaptive MCMC algorithm, augmented with a normalizing flow, offers an alternative approach.
Fig.~\ref{fig:phi4-training-mixing} demonstrates that a map $T$ can be trained to efficiently generate samples with high statistical weight in the target distribution enabling rapid mixing across the free energy barrier. 

\paragraph{Informed base measure.}
In order to learn the map robustly, a standard implementation of a normalizing flow model, with a standard Gaussian field with uncoupled spins as base measure, does not suffice in this instance.
Using a base measure that is ``informed'' alleviates this issue.
Explicitly, we sample the base measure corresponding to a Gaussian random field with a local coupling (a ``Ornstein-Uhlenbeck bridge"), which corresponds to a system with Hamiltonian
\begin{equation}
\label{eq:informed_base}
    U_{\rm B}[\phi] = \beta \int_0^1 \left[\frac{a}{2} (\partial_s \phi)^2 + \frac1{2a} \phi^2\right] ds.
\end{equation}
Importantly, this measure does not have any metastability and remains easy to sample.
As discussed in Appendix~\ref{app:measures}, at this continuous field level, we must choose this measure to ensure that the push-forward distribution has a non-vanishing statistical weight in the target distribution. 

\paragraph{Numerical implementation and results.} In practice, we must discretize the field on a grid, and throughout we take $N=100$ with a lattice spacing $\Delta s=1/N$ meaning that the map we must learn is high-dimensional $T:\RR^N\to \RR^N.$ 
We also use the associated Langevin equation as discretized version of the SPDE~\eqref{eq:stochAC} to generate samples as the local component of our compounded MCMC scheme.

We trained maps $T$ and $T_{\rm U}$ along our adpative MCMC with the informed base measure \eqref{eq:informed_base} and an uninformed Gaussian measure that lacked coupling term (Eq.~\ref{eq:uninformed_base} in the Appendix), respectively, using the same architecture, and compared their suitability for resampling after an equal number of iterations.
Typical configurations $\phi(x)$, in this case generated by the normalizing flow $T$,  are shown in Fig.~\ref{fig:phi4} (a). For comparison, we show in Fig.\ref{fig:phi4-uniformed-samples} samples generated with $T_U$. 

While $T$ generates samples which are accepted in the MCMC procedure with average acceptance rate approaching 60\% (Fig.~\ref{fig:phi4-training-mixing}), $T_{\rm U}$ fails to produce samples that have appreciable statistical weight in the target distribution. 
The evident difference is in the local structure of the random fields that are produced. Fig.~\ref{fig:phi4} (c) shows the Fourier spectrum of field $\phi$ computed with samples from the target measure (obtained using the proposed MCMC method after convergence) as well as from the push-forwards in the informed $\rhohat$ and uninformed $\rhohat_{\rm U}$ case.
While $\rhohat$ accurately captures the decay of the Fourier components at all scales, $T_{\rm U}$ fails to compensate for the uncoupled base measure and $\rhohat_{\rm U}$ does not accurately capture high-frequency oscillations of the field $\phi$, which subsequently leads to high rejection rates in the MCMC procedure.

While at the discrete level, the adequacy of the base measure is \textit{a~priori }less stringent than at the continuous level examined in Appendix~\ref{app:measures}, this experiment shows that at $N=100$ it is already highly beneficial to pre-adapt the covariance of the push-forward. In the absence of preexisting samples to compute an empirical withening transform, it is the role of the proposed ``informed'' base measure. 

\paragraph{Interpreting the map.} 
Examining the learned map $T$ reveals its simple underlying structure.
As shown in Fig.~\ref{fig:phi4} (b), the map is spatially local, transporting spins near the center of the domain to $\pm 1$ while spins near the boundary are mapped closer to 0.
It is again useful to examine the properties of a mapped configuration in Fourier space; the $k=0$ mode reveals that the mean value is transported substantially: $T(\hat\phi_0)$ is approximately $\pm 1,$ as shown in Fig.~\ref{fig:phi4-map-diagnostics} (c).
However, higher frequency modes are left invariant by the map, see Fig.~\ref{fig:phi4-map-diagnostics} (d,e).

\paragraph{Calculating free energy differences.} Perhaps most remarkably, the learned map $T$ can be used to evaluate free energy differences between the metastable basins $\phi^-$ and $\phi^+$, even in thermodynamic conditions distinct from those in which the map was trained.
Fig.~\ref{fig:phi4} (d) shows an estimate of the free energy difference between the positive and negative metastable basins as a function of an external field $h$, which enters the Hamiltonian as
\begin{equation}
    U_{*,h}[\phi] = \beta\int_0^1   \left[ \frac{a}{2} (\partial_s \phi )^2 + \frac1{4a}(1-\phi^2(s))^2 + h\phi(s) \right] ds.
    \label{eq:phi4-bias}
\end{equation}
These estimates are produced with importance sampling using $\rhohat$ as described in Appendix~\ref{app:deltaF}.
Analytical estimates at low temperature via a Laplace approximation reveal that the normalizing flow accurately recapitulates the free energy difference despite the fact that the map was optimized only with samples where $h=0.$
Similar generalization properties were observed in Refs.~\cite{noe_boltzmann_2019,Nicoli2020, Dibak2020}, where a map was used at temperatures distinct from the temperature at which training data was collected. This approach is valid in cases where the modified parameter, here the field $h$, distorts the relative populations of the metastable basins, but has a mild effect on the local structure of the field, which can be controlled by monitoring the variance of the estimator. \\

Additional tests for related applications are presented in Appendix. For this Stochastic Allen-Cahn system, we show that the method can be useful to sample configurations with domain walls by tilting the Hamiltonian (see Fig. \ref{fig:phi4-tilted} in Appendix \ref{sec:a:SAC}).
In Appendix~\ref{app:tps} we discuss a similar sampling problem that involves nonequilibrium transition path which we employ to illustrate the use of Brownian bridge base measures. This example is challenging as metastable basins have very different statistical weights: which is also the case for the particle systems discussed in the next section, where we demonstrate the usage of mixtures to tackle this circumstance.

\begin{figure*}[t!]
    \centering
    \includegraphics[width=0.9\textwidth]{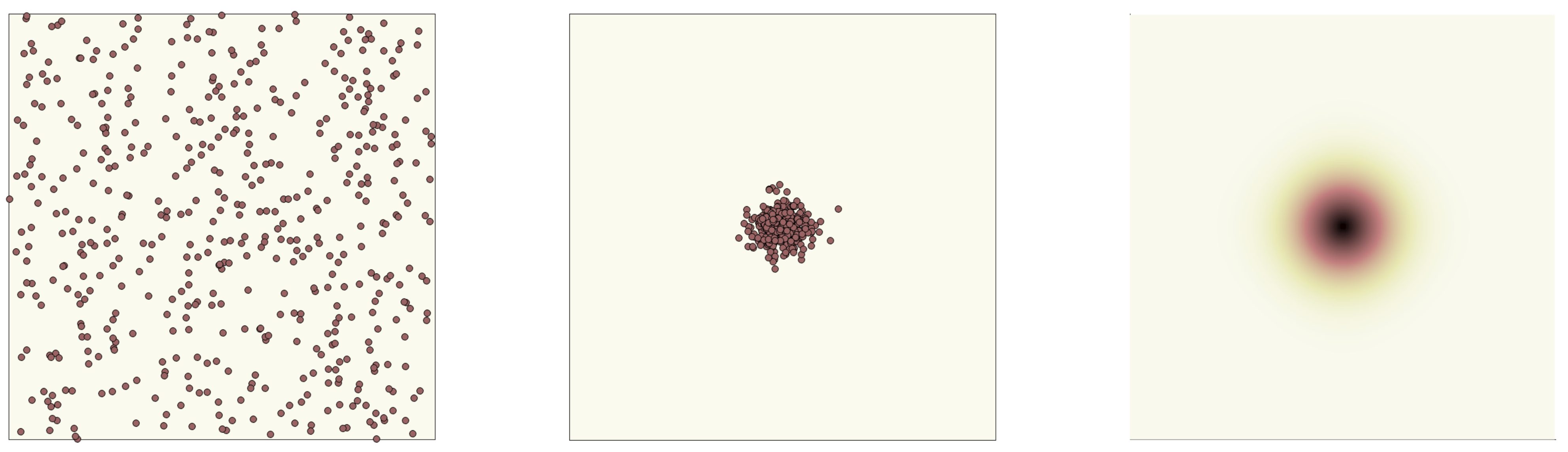}
    \caption{{\bf Detecting Phase Transitions in Interacting Particle Systems.} Left and middle panels: 200 particles seen in the gas and liquid phases, respectively, in  dimension $d=2$ at a temperature below the critical $\beta_c^{-1}$ at which both phases are metastable but the clustered one is thermodynamically preferred. Right panel: A contour plot of the local density $u_d(x)$ of the particles in the liquid phase, plotted in log-scale. }
    \label{fig:interpart1}
\end{figure*}

\subsection{Detecting Phase Transitions in Interacting Particle Systems}
Thermal systems undergoing a first order phase transition are archetypal examples of models displaying metastability. Near the transition point, ergodic mixing from the unstable to the stable phase is broken in the thermodynamic limit, leading to the well-known challenge of detecting these transitions with molecular dynamic simulations. In this section we show our method to be useful in this context.

\paragraph{Particle system and phase diagram.} As an example we consider a system of $N$ interacting particles evolving in a two-dimensional periodic box of lateral size $L$ according to the Langevin equation (here written in the overdamped limit):
\begin{equation}
    \label{eq:Lpart}
    dx_i = - \frac1N \sum_{j=1}^N \nabla W(x_i-x_j) dt + \sqrt{2\beta^{-1}}\, dW_i.
\end{equation}
The interaction $W(x)$ is a pair-wise attracting potential with range $a>0$
\begin{equation}
    \label{eq:shortr}
    W(x) = -\exp(L^2[1-\cos(2\pi|x|/L)^2]/(4\pi^2 a^2))
\end{equation}
which, when $a\ll L$, is well approximated by $W(x) = -\exp(- |x|^2/[2a^2])$.
These equations sample the Boltzmann-Gibbs distribution of the system:
\begin{equation}
    \label{eq:BGpart}
    \rho_*(X) = Z_*^{-1} \exp\left( - \frac{\beta}{2N} \sum_{i,j=1}^N W(x_i-x_j)\right)
\end{equation}
where we denote $X=(x_1,\ldots, x_N)\in [0,L]^{2N}$ the state of the $N$ particle system.

\begin{figure}[t!]
    \centering
    \includegraphics[width=.6\linewidth]{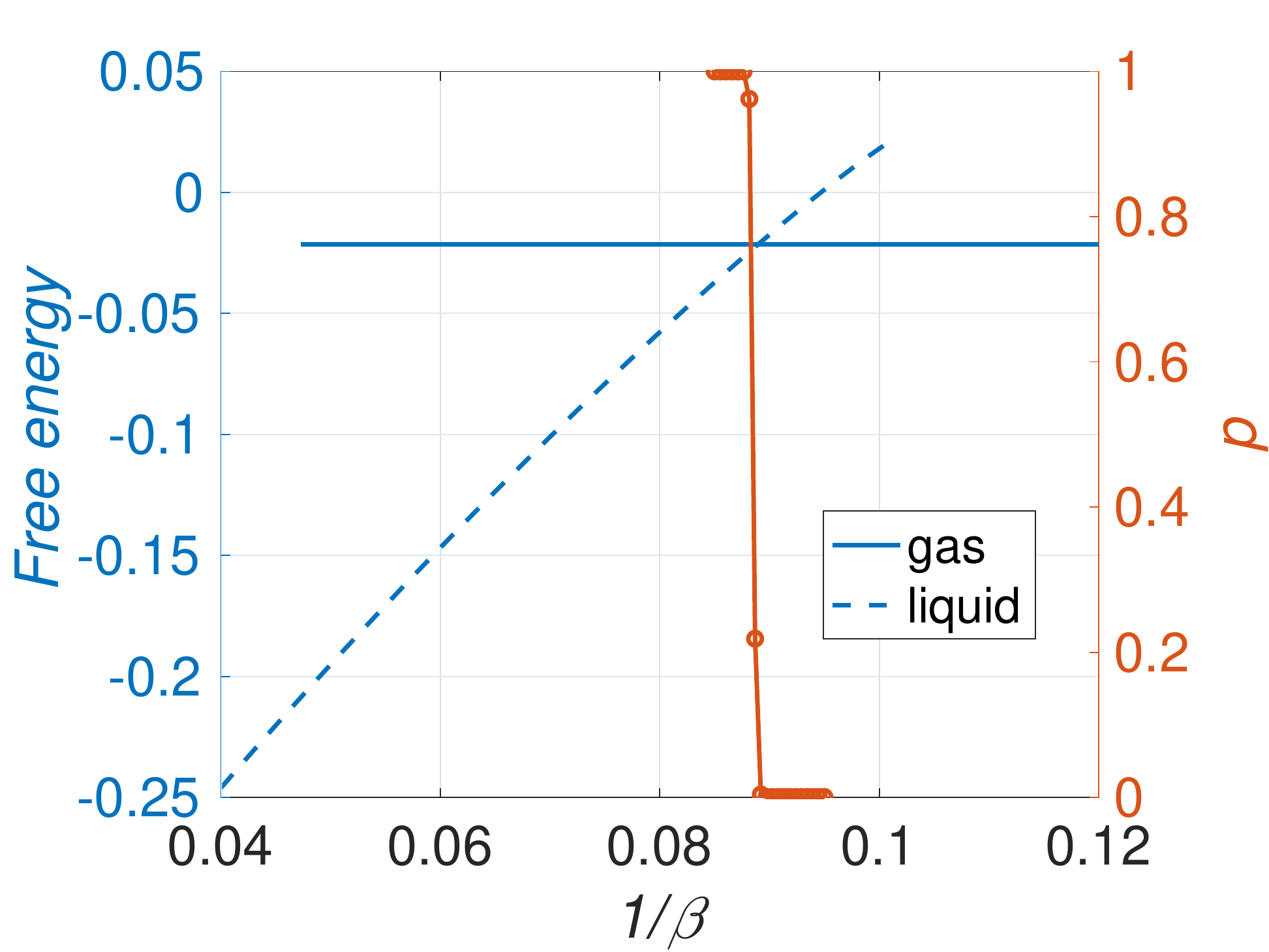}
    \caption{{\bf Detecting Phase Transitions in Interacting Particle Systems.} Blue curves and labels: the free energies of the gas and liquid phases, showing that a first order phase transition occurs at the critical $1/\beta_c \approx 0.089$. For temperatures around this value,  particles configurations in either the homogeneous or the clustered phase are highly metastable, and no transition between these states are observed in brute-force simulation of Eq.~\eqref{eq:Lpart}; Red curve and labels: the value of $p$ in the mixture~\eqref{eq:mixturepart} learned by our adaptive procedure augmented with a NF starting from $p(0)=1/2$;  the algorithm correctly learns the right value of $p$ in the mixture and thereby is able to detect the phase transition. }
    \label{fig:pd1}
\end{figure}

When $a$ is much smaller than $L$, in the thermodynamic limit ($N\gg1$), this system displays a first order phase transition between a gas-like phase, where the particles are uniformly distributed in the domain which is preferred at high temperatures, and a liquid-like phase, where they cluster in a droplet which is preferred at low temperatures. Typical particle configurations in these phases are shown in Fig.~\ref{fig:interpart1}. The phase diagram of the model can be estimated using a mean field approximation (see Appendix~\ref{sec:a:interacting} for details) and is shown in Fig.~\ref{fig:pd1}. 

Detecting this phase transition via brute-force simulation of Eq.~\eqref{eq:Lpart} is however challenging, because the particles stay trapped in whichever configuration they occupy (homogeneous or droplet) for very long periods of time: in fact, the transition times from one phase to the other in a parameter regime where they are both metastable can be estimated as
$t_{l\to g} \asymp \exp(N\beta F_{l\to g})$ and $t_{g\to l} \asymp \exp(N\beta F_{g\to l})$,
where $F_{g\to l}$ and $F_{l\to g}$ denote respectively to free energy barriers between the liquid  and the gas phase and vice-versa. Since these barriers are both independent of $N$, these transition times diverge exponentially with the number of particles $N$. 

\paragraph{Adaptive simulations augmented by nonlocal resampling.} 
Simulation of Eq.~\eqref{eq:Lpart} augmented by a nonlocal resampling map can detect the phase transition. As the two modes of interest have here very different structures and very different  statistical weights across the phase transition we resort to a parameterization of the nonlocal proposal density $\rhohat$ in terms of a mixture.
For the homogeneous phase mixture component, it is straightforward to draw particles configurations: we can simply pick each of their individual position uniformly in the box. For the droplet phase mixture component, it is natural to use as base distribution the uniform distribution $\rhoB(X) = 1/L^2$ that corresponds to the homogeneous phase, and train a map $T$ that then takes one such configuration and map it onto a droplet configuration whose local density is close to that of the particles in the liquid phase. Denoting this local density bu $u_d(x)$, we can also exploit the fact that the liquid droplet has no internal structure and factorize the map as $T(X) = (t(x_1),t(x_2),\ldots,t(x_N))$ with $t:[0,L]^2\to[0,L]^2$ such that $t(x)$ has density $u_d(x)$ if $x$ is drawn uniformly in $[0,L]^2$: that is, $u_d(x) = L^{-2}\det \nabla \bar t(x)$ where $\bar t$ is the inverse of the map $t$. 
All in all, this leads to a resampling mixture density $\rhohat$ that can be expressed as
\begin{equation}
    \label{eq:mixturepart}
    \rhohat(X) = p \prod_{i=1}^N u_d(x_i) + q L^{-2N}
\end{equation}
where $p\in[0,1]$ is a factor to be learned,  $q=1-p$, and the local density $u_d(x)$ needs to be estimated---in the simulations we simply used the mean-field approximation recalled in Appendix~\ref{sec:a:interacting} to calculate $u_d(x)$ numerically, but this density could also be estimated directly from the MD simulations. 

\paragraph{Implementation and results.} Consistent with Algorithm~\ref{alg:concurrent}, we run  Eq.~\eqref{eq:Lpart} used as local sampler for a fixed duration $t_L$, then attempt a resampling move by proposing a configuration from Eq.~\eqref{eq:mixturepart}.
This resampling step requires one to evaluate the Metropolis-Hastings probability in Eq.~\eqref{eq:accept} accurately, which is nontrivial when $N$ is large. Here we used as approximation $\log \rhohat(X) \approx S(X)$ with
\begin{equation}
    \label{eq:mixturepart:approx}
    S(X)= \min\left( \sum_{i=1}^N \log u_d(x_i) +\log p,  -2N \log L +\log q\right)
\end{equation}
leading to the following explicit approximation for the ratio $\rho_*(X)/\rhohat(X)$
\begin{equation}
    \label{eq:MHpartexplicit}
    \begin{aligned}
    &\frac{\rho_*(X)}{\rhohat(X)} \approx \exp\left( -\frac{\beta}{ 2N} \sum_{i,j=1}^N W(x_i-x_j) - S(X)  \right).
    \end{aligned}
\end{equation}
This expression shows that the presence $\rhohat(X)$ in the Metropolis-Hastings probability effectively accounts for the entropy of the particle configurations, as opposed to their energy accounted for by $\rhostar(X)$. Eq.~\eqref{eq:MHpartexplicit} also emphasizes the need for $\rhohat$ (i.e. the NF map in general) to be accurate enough: indeed, to get any significant probability of acceptance, even for a move aimed towards the thermodynamically preferred phase,  the factor in the exponentials in Eq.~\eqref{eq:MHpartexplicit} must be of order 1 in $N$. If the map fails to achieve this accuracy, and the factors remain of order $N$ (which is their typical scale for a map that is unadapted), the move would be systematically rejected (or accepted between configurations with little resemblance to permitted ones). This issue will be generic for problems where the system's energy is extensive.

In the context of the present example, once $u_d(x)$ has been estimated, the learning component of Algorithm~\ref{alg:concurrent} reduces to optimizing  the parameter~$p$. This is done by approximating the KL divergence $D_{\text{KL}}(\rhohat(X) ||\rho_*(X))$  using an estimator based on using the current state~$X(t)$ of the system. Specifically, we used $\log \rhohat(X(t))\approx S(X(t))$ as objective 
on which we performed gradient descent in $p$ concurrently with running the augmented MD strategy above. 

We applied the procedure above to a system of $N = 200$ particles drawn initially from the mixture density~\eqref{eq:mixturepart} using $p(0)=1/2$ as initial value. As can be seen on Fig.~\ref{fig:pd1}, this allowed us to train $p$ to values converging either to $p=0$ or $p=1$, in a way that detects the phase transition. That is, the augmented procedure correctly reweights the homogeneous and droplet configurations and determines which of the two is the most likely even in situations where brute-force MD simulations would observe no transitions between these configurations. 

We stress that a simplifying feature of this example is that the particles experience no short range repulsion, i.e. there is no order in the droplet phase. This is what allowed us to use the product of $u_d(x)$ in the mixture density in Eq.~\eqref{eq:mixturepart}. In systems with hardcore repulsions, this approximation is invalid, in which case more complicated mixtures (or equivalently more complicated maps in the normalizing flow) will have to be used. We leave investigation of such situations for future work.

\section{Conclusions}
\label{sec:conclu}

\subsection{Connections and Differences with Previous Works}  
Most of the methods that seek to train a normalizing flow to (approximately) sample from the Boltzmann distribution of a known target energy rely on the reverse-KL training using Eq.~\eqref{eq:revDKLstar} \cite{albergo_flow-based_2019,wu_solving_2019,Nicoli2020, Dibak2020}. However, this objective is known to be prone to ``mode-collaspse'' where the estimation concentrates on the bulk of one mode. This failure comes typically from the fact that the map may never explore modes far away from the core of the base distribution---such that these modes are missed altogether\footnote{Note that annealing of the target can help to catch multiple modes in some simple cases but offers no guarantees \cite{wu_solving_2019}.} \cite{hartnett_self-supervised_2020}. Additionally, the reverse KL objective is known to lead to underestimation of the tails \cite{Yao2018}. 

To alleviate the shortcomings of reverse-KL training, \cite{noe_boltzmann_2019} relied on initial short-trajectories to estimate and factor in the optimization objective the forward KL divergence. Closer to our proposition, the authors of \cite{Naesseth2020} propose Markov Score Climbing, an adaptive MCMC strategy using the same estimate of the forward-KL as Eq.~\eqref{eq:E1} to discover good variational approximators. Our method can be seen as an extension of Markov Score Climbing, introducing the additional alternation with a local sampler and the replacement of simple variational families by the more flexible normalizing flows. Along the same lines, \cite{Parno2018} investigated proposals parametrized by lower-triangular maps. Interestingly, \cite{noe_boltzmann_2019} also proposed an iteratively retrained variant of its Boltzmann generator that shares similarity with our proposition.  Note that an advantage of the adaptive MCMC over the consecutive training-then-sampling schemes with either reverse-KL \cite{albergo_flow-based_2019,wu_solving_2019,Nicoli2020} or forward-KL \cite{noe_boltzmann_2019,McNaughton2020} training objectives is to allow for real-time monitoring of the quality of training towards the final purpose of obtaining well-mixed samples. In the experiments that was done by monitoring, for instance, acceptance rates and auto-correlation of the chains across iterations.

The adaptive MCMC proposed here retains a local component in the sampling in the form of interleaved steps of a local sampler that brings robustness to the scheme. Albeit different, the adaptive MCMC proposed in \cite{noe_boltzmann_2019} also defines an intermediary between a purely global and a purely local procedure. The proposals consist in local steps in the latent space of the normalizing flow, while our query of the generative model yield completely independent resampling. To encourage transition across modes, the algorithm of \cite{noe_boltzmann_2019} was augmented with parallel-tempering in \cite{Dibak2020}, a step that does not appear to be necessary in our scheme.

\subsection{Outlook and Future Work}
As the use of data-driven methods from machine learning becomes increasingly routine in the physical sciences, we must carefully assess the cost of data acquisition and training to ensure that we can leverage ML methods in a productive fashion.
Sampling systems with complex local structure and multiple metastable basins is a generically challenging task in high-dimensional systems, and we have already seen that neural networks can contend with this challenge in nontrivial settings~\cite{albergo_flow-based_2019,noe_boltzmann_2019,huang_accelerated_2017, wu_solving_2019,song_scorebased_2021}.
Nonlocal transport in MCMC algorithms can significantly enhance mixing and normalizing flows provide a compelling framework for designing adaptive schemes, even in cases where no statistically representative data set is available at first. 
Nevertheless, we do not find that these methods enable discovery of unknown modes of a target distribution, emphasizing the importance of having some \emph{a priori} information about the metastable states of the system.

Many questions remain about how to ensure efficient learning in complex high-dimensional systems and encourage desirable properties of the map, such as locality and transferability. Incorporating known invariances and symmetries of target distributions into architectures is currently a key area of research, e.g. \cite{Kohler2019,Rezende2020}, that will help scaling further applications of sampling methods enhanced by learning.

\paragraph{GMR acknowledges generous support from the Terman Faculty Fellowship. EVE acknowledges partial support from the National Science Foundation (NSF) Materials Research Science and Engineering Center Program grant DMR-1420073, NSF DMS-1522767, and a Vannevar Bush Faculty Fellowship.}

\bibliography{refs}

\begin{thebibliography}{57}%
\makeatletter
\providecommand \@ifxundefined [1]{%
 \@ifx{#1\undefined}
}%
\providecommand \@ifnum [1]{%
 \ifnum #1\expandafter \@firstoftwo
 \else \expandafter \@secondoftwo
 \fi
}%
\providecommand \@ifx [1]{%
 \ifx #1\expandafter \@firstoftwo
 \else \expandafter \@secondoftwo
 \fi
}%
\providecommand \natexlab [1]{#1}%
\providecommand \enquote  [1]{``#1''}%
\providecommand \bibnamefont  [1]{#1}%
\providecommand \bibfnamefont [1]{#1}%
\providecommand \citenamefont [1]{#1}%
\providecommand \href@noop [0]{\@secondoftwo}%
\providecommand \href [0]{\begingroup \@sanitize@url \@href}%
\providecommand \@href[1]{\@@startlink{#1}\@@href}%
\providecommand \@@href[1]{\endgroup#1\@@endlink}%
\providecommand \@sanitize@url [0]{\catcode `\\12\catcode `\$12\catcode
  `\&12\catcode `\#12\catcode `\^12\catcode `\_12\catcode `\%12\relax}%
\providecommand \@@startlink[1]{}%
\providecommand \@@endlink[0]{}%
\providecommand \url  [0]{\begingroup\@sanitize@url \@url }%
\providecommand \@url [1]{\endgroup\@href {#1}{\urlprefix }}%
\providecommand \urlprefix  [0]{URL }%
\providecommand \Eprint [0]{\href }%
\providecommand \doibase [0]{https://doi.org/}%
\providecommand \selectlanguage [0]{\@gobble}%
\providecommand \bibinfo  [0]{\@secondoftwo}%
\providecommand \bibfield  [0]{\@secondoftwo}%
\providecommand \translation [1]{[#1]}%
\providecommand \BibitemOpen [0]{}%
\providecommand \bibitemStop [0]{}%
\providecommand \bibitemNoStop [0]{.\EOS\space}%
\providecommand \EOS [0]{\spacefactor3000\relax}%
\providecommand \BibitemShut  [1]{\csname bibitem#1\endcsname}%
\let\auto@bib@innerbib\@empty
\bibitem [{\citenamefont {Bolhuis}\ \emph {et~al.}(2002)\citenamefont
  {Bolhuis}, \citenamefont {Chandler}, \citenamefont {Dellago},\ and\
  \citenamefont {Geissler}}]{bolhuis_transition_2002}%
  \BibitemOpen
  \bibfield  {author} {\bibinfo {author} {\bibfnamefont {P.~G.}\ \bibnamefont
  {Bolhuis}}, \bibinfo {author} {\bibfnamefont {D.}~\bibnamefont {Chandler}},
  \bibinfo {author} {\bibfnamefont {C.}~\bibnamefont {Dellago}},\ and\ \bibinfo
  {author} {\bibfnamefont {P.~L.}\ \bibnamefont {Geissler}},\ }\bibfield
  {title} {\bibinfo {title} {Transition {{Path Sampling}}: {{Throwing Ropes
  Over Rough Mountain Passes}}, in the {{Dark}}},\ }\href
  {https://doi.org/10.1146/annurev.physchem.53.082301.113146} {\bibfield
  {journal} {\bibinfo  {journal} {Annual Review of Physical Chemistry}\
  }\textbf {\bibinfo {volume} {53}},\ \bibinfo {pages} {291} (\bibinfo {year}
  {2002})}\BibitemShut {NoStop}%
\bibitem [{\citenamefont {Goodfellow}\ \emph {et~al.}(2014)\citenamefont
  {Goodfellow}, \citenamefont {{Pouget-Abadie}}, \citenamefont {Mirza},
  \citenamefont {Xu}, \citenamefont {{Warde-Farley}}, \citenamefont {Ozair},
  \citenamefont {Courville},\ and\ \citenamefont
  {Bengio}}]{goodfellow_generative_2014}%
  \BibitemOpen
  \bibfield  {author} {\bibinfo {author} {\bibfnamefont {I.}~\bibnamefont
  {Goodfellow}}, \bibinfo {author} {\bibfnamefont {J.}~\bibnamefont
  {{Pouget-Abadie}}}, \bibinfo {author} {\bibfnamefont {M.}~\bibnamefont
  {Mirza}}, \bibinfo {author} {\bibfnamefont {B.}~\bibnamefont {Xu}}, \bibinfo
  {author} {\bibfnamefont {D.}~\bibnamefont {{Warde-Farley}}}, \bibinfo
  {author} {\bibfnamefont {S.}~\bibnamefont {Ozair}}, \bibinfo {author}
  {\bibfnamefont {A.}~\bibnamefont {Courville}},\ and\ \bibinfo {author}
  {\bibfnamefont {Y.}~\bibnamefont {Bengio}},\ }\bibfield  {title} {\bibinfo
  {title} {Generative {{Adversarial Nets}}},\ }in\ \href@noop {} {\emph
  {\bibinfo {booktitle} {Advances in {{Neural Information Processing Systems}}
  27}}},\ \bibinfo {editor} {edited by\ \bibinfo {editor} {\bibfnamefont
  {Z.}~\bibnamefont {Ghahramani}}, \bibinfo {editor} {\bibfnamefont
  {M.}~\bibnamefont {Welling}}, \bibinfo {editor} {\bibfnamefont
  {C.}~\bibnamefont {Cortes}}, \bibinfo {editor} {\bibfnamefont {N.~D.}\
  \bibnamefont {Lawrence}},\ and\ \bibinfo {editor} {\bibfnamefont {K.~Q.}\
  \bibnamefont {Weinberger}}}\ (\bibinfo  {publisher} {{Curran Associates,
  Inc.}},\ \bibinfo {year} {2014})\ pp.\ \bibinfo {pages}
  {2672--2680}\BibitemShut {NoStop}%
\bibitem [{\citenamefont {Kingma}\ and\ \citenamefont
  {Welling}(2013)}]{kingma_auto-encoding_2013}%
  \BibitemOpen
  \bibfield  {author} {\bibinfo {author} {\bibfnamefont {D.~P.}\ \bibnamefont
  {Kingma}}\ and\ \bibinfo {author} {\bibfnamefont {M.}~\bibnamefont
  {Welling}},\ }\bibfield  {title} {\bibinfo {title} {Auto-{{Encoding
  Variational Bayes}}},\ }\href {https://arxiv.org/1312.6114v10} {\bibfield
  {journal} {\bibinfo  {journal} {arXiv [Preprint]}\ }\textbf {\bibinfo
  {volume} {0}} (\bibinfo {year} {2013})},\ \Eprint
  {https://arxiv.org/abs/1312.6114v10} {arXiv:1312.6114v10} \BibitemShut
  {NoStop}%
\bibitem [{\citenamefont {Song}\ \emph {et~al.}(2017)\citenamefont {Song},
  \citenamefont {Zhao},\ and\ \citenamefont {Ermon}}]{song_nice_2017}%
  \BibitemOpen
  \bibfield  {author} {\bibinfo {author} {\bibfnamefont {J.}~\bibnamefont
  {Song}}, \bibinfo {author} {\bibfnamefont {S.}~\bibnamefont {Zhao}},\ and\
  \bibinfo {author} {\bibfnamefont {S.}~\bibnamefont {Ermon}},\ }\bibfield
  {title} {\bibinfo {title} {A-nice-mc: {{Adversarial}} training for
  {{MCMC}}},\ }in\ \href@noop {} {\emph {\bibinfo {booktitle} {Advances in
  Neural Information Processing Systems}}},\ Vol.~\bibinfo {volume} {30},\
  \bibinfo {editor} {edited by\ \bibinfo {editor} {\bibfnamefont
  {I.}~\bibnamefont {Guyon}}, \bibinfo {editor} {\bibfnamefont {U.~V.}\
  \bibnamefont {Luxburg}}, \bibinfo {editor} {\bibfnamefont {S.}~\bibnamefont
  {Bengio}}, \bibinfo {editor} {\bibfnamefont {H.}~\bibnamefont {Wallach}},
  \bibinfo {editor} {\bibfnamefont {R.}~\bibnamefont {Fergus}}, \bibinfo
  {editor} {\bibfnamefont {S.}~\bibnamefont {Vishwanathan}},\ and\ \bibinfo
  {editor} {\bibfnamefont {R.}~\bibnamefont {Garnett}}}\ (\bibinfo  {publisher}
  {{Curran Associates, Inc.}},\ \bibinfo {year} {2017})\BibitemShut {NoStop}%
\bibitem [{\citenamefont {Song}\ \emph {et~al.}(2021)\citenamefont {Song},
  \citenamefont {{Sohl-Dickstein}}, \citenamefont {Kingma}, \citenamefont
  {Kumar}, \citenamefont {Ermon},\ and\ \citenamefont
  {Poole}}]{song_scorebased_2021}%
  \BibitemOpen
  \bibfield  {author} {\bibinfo {author} {\bibfnamefont {Y.}~\bibnamefont
  {Song}}, \bibinfo {author} {\bibfnamefont {J.}~\bibnamefont
  {{Sohl-Dickstein}}}, \bibinfo {author} {\bibfnamefont {D.~P.}\ \bibnamefont
  {Kingma}}, \bibinfo {author} {\bibfnamefont {A.}~\bibnamefont {Kumar}},
  \bibinfo {author} {\bibfnamefont {S.}~\bibnamefont {Ermon}},\ and\ \bibinfo
  {author} {\bibfnamefont {B.}~\bibnamefont {Poole}},\ }\bibfield  {title}
  {\bibinfo {title} {Score-based generative modeling through stochastic
  differential equations},\ }in\ \href@noop {} {\emph {\bibinfo {booktitle}
  {International Conference on Learning Representations}}}\ (\bibinfo {year}
  {2021})\BibitemShut {NoStop}%
\bibitem [{\citenamefont {Huang}\ and\ \citenamefont
  {Wang}(2017)}]{huang_accelerated_2017}%
  \BibitemOpen
  \bibfield  {author} {\bibinfo {author} {\bibfnamefont {L.}~\bibnamefont
  {Huang}}\ and\ \bibinfo {author} {\bibfnamefont {L.}~\bibnamefont {Wang}},\
  }\bibfield  {title} {\bibinfo {title} {Accelerated {{Monte Carlo}}
  simulations with restricted {{Boltzmann}} machines},\ }\href
  {https://doi.org/10/gdqcqc} {\bibfield  {journal} {\bibinfo  {journal}
  {Physical Review B}\ }\textbf {\bibinfo {volume} {95}},\ \bibinfo {pages}
  {035105} (\bibinfo {year} {2017})}\BibitemShut {NoStop}%
\bibitem [{\citenamefont {Sejdinovic}\ \emph {et~al.}(2014)\citenamefont
  {Sejdinovic}, \citenamefont {Strathmann}, \citenamefont {Garcia},
  \citenamefont {Andrieu},\ and\ \citenamefont
  {Gretton}}]{sejdinovic_kernel_2014a}%
  \BibitemOpen
  \bibfield  {author} {\bibinfo {author} {\bibfnamefont {D.}~\bibnamefont
  {Sejdinovic}}, \bibinfo {author} {\bibfnamefont {H.}~\bibnamefont
  {Strathmann}}, \bibinfo {author} {\bibfnamefont {M.~L.}\ \bibnamefont
  {Garcia}}, \bibinfo {author} {\bibfnamefont {C.}~\bibnamefont {Andrieu}},\
  and\ \bibinfo {author} {\bibfnamefont {A.}~\bibnamefont {Gretton}},\
  }\bibfield  {title} {\bibinfo {title} {Kernel {{Adaptive
  Metropolis}}-{{Hastings}}},\ }in\ \href@noop {} {\emph {\bibinfo {booktitle}
  {International {{Conference}} on {{Machine Learning}}}}}\ (\bibinfo
  {publisher} {{PMLR}},\ \bibinfo {year} {2014})\ pp.\ \bibinfo {pages}
  {1665--1673}\BibitemShut {NoStop}%
\bibitem [{\citenamefont {Levy}\ \emph {et~al.}(2018)\citenamefont {Levy},
  \citenamefont {Hoffman},\ and\ \citenamefont
  {{Sohl-Dickstein}}}]{levy_generalizing_2018b}%
  \BibitemOpen
  \bibfield  {author} {\bibinfo {author} {\bibfnamefont {D.}~\bibnamefont
  {Levy}}, \bibinfo {author} {\bibfnamefont {M.~D.}\ \bibnamefont {Hoffman}},\
  and\ \bibinfo {author} {\bibfnamefont {J.}~\bibnamefont {{Sohl-Dickstein}}},\
  }\bibfield  {title} {\bibinfo {title} {Generalizing {{Hamiltonian Monte
  Carlo}} with {{Neural Networks}}},\ }in\ \href@noop {} {\emph {\bibinfo
  {booktitle} {International {{Conference}} on {{Learning Representations}}}}}\
  (\bibinfo {year} {2018})\BibitemShut {NoStop}%
\bibitem [{\citenamefont {Titsias}(2017)}]{titsias_learning_2017}%
  \BibitemOpen
  \bibfield  {author} {\bibinfo {author} {\bibfnamefont {M.~K.}\ \bibnamefont
  {Titsias}},\ }\bibfield  {title} {\bibinfo {title} {Learning {{Model
  Reparametrizations}}: {{Implicit Variational Inference}} by {{Fitting MCMC}}
  distributions},\ }\href@noop {} {\bibfield  {journal} {\bibinfo  {journal}
  {arXiv:1708.01529 [stat]}\ }\textbf {\bibinfo {volume} {0}} (\bibinfo {year}
  {2017})},\ \Eprint {https://arxiv.org/abs/1708.01529} {arXiv:1708.01529
  [stat]} \BibitemShut {NoStop}%
\bibitem [{\citenamefont {Le}\ \emph {et~al.}(2018)\citenamefont {Le},
  \citenamefont {Igl}, \citenamefont {Rainforth}, \citenamefont {Jin},\ and\
  \citenamefont {Wood}}]{anh_autoencoding_2018}%
  \BibitemOpen
  \bibfield  {author} {\bibinfo {author} {\bibfnamefont {T.~A.}\ \bibnamefont
  {Le}}, \bibinfo {author} {\bibfnamefont {M.}~\bibnamefont {Igl}}, \bibinfo
  {author} {\bibfnamefont {T.}~\bibnamefont {Rainforth}}, \bibinfo {author}
  {\bibfnamefont {T.}~\bibnamefont {Jin}},\ and\ \bibinfo {author}
  {\bibfnamefont {F.}~\bibnamefont {Wood}},\ }\bibfield  {title} {\bibinfo
  {title} {Auto-encoding sequential monte carlo},\ }in\ \href
  {https://openreview.net/forum?id=BJ8c3f-0b} {\emph {\bibinfo {booktitle}
  {International Conference on Learning Representations}}}\ (\bibinfo {year}
  {2018})\BibitemShut {NoStop}%
\bibitem [{\citenamefont {McNaughton}\ \emph {et~al.}(2020)\citenamefont
  {McNaughton}, \citenamefont {Milo{\v{s}}evi{\'{c}}}, \citenamefont {Perali},\
  and\ \citenamefont {Pilati}}]{McNaughton2020}%
  \BibitemOpen
  \bibfield  {author} {\bibinfo {author} {\bibfnamefont {B.}~\bibnamefont
  {McNaughton}}, \bibinfo {author} {\bibfnamefont {M.~V.}\ \bibnamefont
  {Milo{\v{s}}evi{\'{c}}}}, \bibinfo {author} {\bibfnamefont {A.}~\bibnamefont
  {Perali}},\ and\ \bibinfo {author} {\bibfnamefont {S.}~\bibnamefont
  {Pilati}},\ }\bibfield  {title} {\bibinfo {title} {{Boosting Monte Carlo
  simulations of spin glasses using autoregressive neural networks}},\ }\href
  {https://doi.org/10.1103/PhysRevE.101.053312} {\bibfield  {journal} {\bibinfo
   {journal} {Physical Review E}\ }\textbf {\bibinfo {volume} {101}},\ \bibinfo
  {pages} {1} (\bibinfo {year} {2020})},\ \Eprint
  {https://arxiv.org/abs/2002.04292} {arXiv:2002.04292} \BibitemShut {NoStop}%
\bibitem [{\citenamefont {Germain}\ \emph {et~al.}(2015)\citenamefont
  {Germain}, \citenamefont {Gregor}, \citenamefont {Murray},\ and\
  \citenamefont {Larochelle}}]{Germain2015}%
  \BibitemOpen
  \bibfield  {author} {\bibinfo {author} {\bibfnamefont {M.}~\bibnamefont
  {Germain}}, \bibinfo {author} {\bibfnamefont {K.}~\bibnamefont {Gregor}},
  \bibinfo {author} {\bibfnamefont {I.}~\bibnamefont {Murray}},\ and\ \bibinfo
  {author} {\bibfnamefont {H.}~\bibnamefont {Larochelle}},\ }\bibfield  {title}
  {\bibinfo {title} {{MADE: Masked autoencoder for distribution estimation}},\
  }in\ \href@noop {} {\emph {\bibinfo {booktitle} {32nd International
  Conference on Machine Learning, ICML 2015}}},\ Vol.~\bibinfo {volume} {2}\
  (\bibinfo {year} {2015})\ pp.\ \bibinfo {pages} {881--889}\BibitemShut
  {NoStop}%
\bibitem [{\citenamefont {Rezende}\ and\ \citenamefont
  {Mohamed}(2015)}]{rezende_variational_2015}%
  \BibitemOpen
  \bibfield  {author} {\bibinfo {author} {\bibfnamefont {D.}~\bibnamefont
  {Rezende}}\ and\ \bibinfo {author} {\bibfnamefont {S.}~\bibnamefont
  {Mohamed}},\ }\bibfield  {title} {\bibinfo {title} {Variational {{Inference}}
  with {{Normalizing Flows}}},\ }in\ \href@noop {} {\emph {\bibinfo {booktitle}
  {International {{Conference}} on {{Machine Learning}}}}}\ (\bibinfo
  {publisher} {{PMLR}},\ \bibinfo {year} {2015})\ pp.\ \bibinfo {pages}
  {1530--1538}\BibitemShut {NoStop}%
\bibitem [{\citenamefont {Papamakarios}\ \emph {et~al.}(2021)\citenamefont
  {Papamakarios}, \citenamefont {Nalisnick}, \citenamefont {Rezende},
  \citenamefont {Mohamed},\ and\ \citenamefont
  {Lakshminarayanan}}]{papamakarios_normalizing_2021}%
  \BibitemOpen
  \bibfield  {author} {\bibinfo {author} {\bibfnamefont {G.}~\bibnamefont
  {Papamakarios}}, \bibinfo {author} {\bibfnamefont {E.}~\bibnamefont
  {Nalisnick}}, \bibinfo {author} {\bibfnamefont {D.~J.}\ \bibnamefont
  {Rezende}}, \bibinfo {author} {\bibfnamefont {S.}~\bibnamefont {Mohamed}},\
  and\ \bibinfo {author} {\bibfnamefont {B.}~\bibnamefont {Lakshminarayanan}},\
  }\bibfield  {title} {\bibinfo {title} {Normalizing flows for probabilistic
  modeling and inference},\ }\href@noop {} {\bibfield  {journal} {\bibinfo
  {journal} {Journal of Machine Learning Research}\ }\textbf {\bibinfo {volume}
  {22}},\ \bibinfo {pages} {1} (\bibinfo {year} {2021})}\BibitemShut {NoStop}%
\bibitem [{\citenamefont {Albergo}\ \emph {et~al.}(2019)\citenamefont
  {Albergo}, \citenamefont {Kanwar},\ and\ \citenamefont
  {Shanahan}}]{albergo_flow-based_2019}%
  \BibitemOpen
  \bibfield  {author} {\bibinfo {author} {\bibfnamefont {M.~S.}\ \bibnamefont
  {Albergo}}, \bibinfo {author} {\bibfnamefont {G.}~\bibnamefont {Kanwar}},\
  and\ \bibinfo {author} {\bibfnamefont {P.~E.}\ \bibnamefont {Shanahan}},\
  }\bibfield  {title} {\bibinfo {title} {Flow-based generative models for
  {{Markov}} chain {{Monte Carlo}} in lattice field theory},\ }\href
  {https://doi.org/10.1103/PhysRevD.100.034515} {\bibfield  {journal} {\bibinfo
   {journal} {Physical Review D}\ }\textbf {\bibinfo {volume} {100}},\ \bibinfo
  {pages} {034515} (\bibinfo {year} {2019})}\BibitemShut {NoStop}%
\bibitem [{\citenamefont {Sidky}\ \emph {et~al.}(2020)\citenamefont {Sidky},
  \citenamefont {Chen},\ and\ \citenamefont {Ferguson}}]{sidky_molecular_2020}%
  \BibitemOpen
  \bibfield  {author} {\bibinfo {author} {\bibfnamefont {H.}~\bibnamefont
  {Sidky}}, \bibinfo {author} {\bibfnamefont {W.}~\bibnamefont {Chen}},\ and\
  \bibinfo {author} {\bibfnamefont {A.~L.}\ \bibnamefont {Ferguson}},\
  }\bibfield  {title} {\bibinfo {title} {Molecular latent space simulators},\
  }\href {https://doi.org/10.1039/D0SC03635H} {\bibfield  {journal} {\bibinfo
  {journal} {Chemical Science}\ }\textbf {\bibinfo {volume} {11}},\ \bibinfo
  {pages} {9459} (\bibinfo {year} {2020})}\BibitemShut {NoStop}%
\bibitem [{\citenamefont {Sbail{\`o}}\ \emph {et~al.}(2021)\citenamefont
  {Sbail{\`o}}, \citenamefont {Dibak},\ and\ \citenamefont
  {No{\'e}}}]{sbailo_neural_2021}%
  \BibitemOpen
  \bibfield  {author} {\bibinfo {author} {\bibfnamefont {L.}~\bibnamefont
  {Sbail{\`o}}}, \bibinfo {author} {\bibfnamefont {M.}~\bibnamefont {Dibak}},\
  and\ \bibinfo {author} {\bibfnamefont {F.}~\bibnamefont {No{\'e}}},\
  }\bibfield  {title} {\bibinfo {title} {Neural mode jump {{Monte Carlo}}},\
  }\href {https://doi.org/10.1063/5.0032346} {\bibfield  {journal} {\bibinfo
  {journal} {Journal of Chemical Physics}\ }\textbf {\bibinfo {volume} {154}},\
  \bibinfo {pages} {074101} (\bibinfo {year} {2021})}\BibitemShut {NoStop}%
\bibitem [{\citenamefont {Wu}\ \emph {et~al.}(2019)\citenamefont {Wu},
  \citenamefont {Wang},\ and\ \citenamefont {Zhang}}]{wu_solving_2019}%
  \BibitemOpen
  \bibfield  {author} {\bibinfo {author} {\bibfnamefont {D.}~\bibnamefont
  {Wu}}, \bibinfo {author} {\bibfnamefont {L.}~\bibnamefont {Wang}},\ and\
  \bibinfo {author} {\bibfnamefont {P.}~\bibnamefont {Zhang}},\ }\bibfield
  {title} {\bibinfo {title} {Solving {{Statistical Mechanics Using Variational
  Autoregressive Networks}}},\ }\href
  {https://doi.org/10.1103/PhysRevLett.122.080602} {\bibfield  {journal}
  {\bibinfo  {journal} {Physical Review Letters}\ }\textbf {\bibinfo {volume}
  {122}},\ \bibinfo {pages} {080602} (\bibinfo {year} {2019})}\BibitemShut
  {NoStop}%
\bibitem [{\citenamefont {Nicoli}\ \emph {et~al.}(2021)\citenamefont {Nicoli},
  \citenamefont {Anders}, \citenamefont {Funcke}, \citenamefont {Hartung},
  \citenamefont {Jansen}, \citenamefont {Kessel}, \citenamefont {Nakajima},\
  and\ \citenamefont {Stornati}}]{nicoli_estimation_2021}%
  \BibitemOpen
  \bibfield  {author} {\bibinfo {author} {\bibfnamefont {K.~A.}\ \bibnamefont
  {Nicoli}}, \bibinfo {author} {\bibfnamefont {C.~J.}\ \bibnamefont {Anders}},
  \bibinfo {author} {\bibfnamefont {L.}~\bibnamefont {Funcke}}, \bibinfo
  {author} {\bibfnamefont {T.}~\bibnamefont {Hartung}}, \bibinfo {author}
  {\bibfnamefont {K.}~\bibnamefont {Jansen}}, \bibinfo {author} {\bibfnamefont
  {P.}~\bibnamefont {Kessel}}, \bibinfo {author} {\bibfnamefont
  {S.}~\bibnamefont {Nakajima}},\ and\ \bibinfo {author} {\bibfnamefont
  {P.}~\bibnamefont {Stornati}},\ }\bibfield  {title} {\bibinfo {title}
  {Estimation of {{Thermodynamic Observables}} in {{Lattice Field Theories}}
  with {{Deep Generative Models}}},\ }\href
  {https://doi.org/10.1103/PhysRevLett.126.032001} {\bibfield  {journal}
  {\bibinfo  {journal} {Physical Review Letters}\ }\textbf {\bibinfo {volume}
  {126}},\ \bibinfo {pages} {032001} (\bibinfo {year} {2021})}\BibitemShut
  {NoStop}%
\bibitem [{\citenamefont {{Del Debbio}}\ \emph {et~al.}(2021)\citenamefont
  {{Del Debbio}}, \citenamefont {Rossney},\ and\ \citenamefont
  {Wilson}}]{DelDebbio2021}%
  \BibitemOpen
  \bibfield  {author} {\bibinfo {author} {\bibfnamefont {L.}~\bibnamefont {{Del
  Debbio}}}, \bibinfo {author} {\bibfnamefont {J.~M.}\ \bibnamefont
  {Rossney}},\ and\ \bibinfo {author} {\bibfnamefont {M.}~\bibnamefont
  {Wilson}},\ }\bibfield  {title} {\bibinfo {title} {{Efficient Modelling of
  Trivializing Maps for Lattice {$\phi^4$} Theory Using Normalizing Flows: A
  First Look at Scalability}},\ }\href {http://arxiv.org/abs/2105.12481}
  {\bibfield  {journal} {\bibinfo  {journal} {arXiv [Preprint]}\ }\textbf
  {\bibinfo {volume} {0}} (\bibinfo {year} {2021})},\ \Eprint
  {https://arxiv.org/abs/2105.12481} {arXiv:2105.12481} \BibitemShut {NoStop}%
\bibitem [{\citenamefont {No{\'e}}\ \emph {et~al.}(2019)\citenamefont
  {No{\'e}}, \citenamefont {Olsson}, \citenamefont {K{\"o}hler},\ and\
  \citenamefont {Wu}}]{noe_boltzmann_2019}%
  \BibitemOpen
  \bibfield  {author} {\bibinfo {author} {\bibfnamefont {F.}~\bibnamefont
  {No{\'e}}}, \bibinfo {author} {\bibfnamefont {S.}~\bibnamefont {Olsson}},
  \bibinfo {author} {\bibfnamefont {J.}~\bibnamefont {K{\"o}hler}},\ and\
  \bibinfo {author} {\bibfnamefont {H.}~\bibnamefont {Wu}},\ }\bibfield
  {title} {\bibinfo {title} {Boltzmann generators: {{Sampling}} equilibrium
  states of many-body systems with deep learning},\ }\href
  {https://doi.org/10.1126/science.aaw1147} {\bibfield  {journal} {\bibinfo
  {journal} {Science}\ }\textbf {\bibinfo {volume} {365}},\ \bibinfo {pages}
  {eaaw1147} (\bibinfo {year} {2019})}\BibitemShut {NoStop}%
\bibitem [{\citenamefont {Andrieu}\ \emph {et~al.}(2011)\citenamefont
  {Andrieu}, \citenamefont {Jasra}, \citenamefont {Doucet},\ and\ \citenamefont
  {Moral}}]{andrieu_nonlinear_2011}%
  \BibitemOpen
  \bibfield  {author} {\bibinfo {author} {\bibfnamefont {C.}~\bibnamefont
  {Andrieu}}, \bibinfo {author} {\bibfnamefont {A.}~\bibnamefont {Jasra}},
  \bibinfo {author} {\bibfnamefont {A.}~\bibnamefont {Doucet}},\ and\ \bibinfo
  {author} {\bibfnamefont {P.~D.}\ \bibnamefont {Moral}},\ }\bibfield  {title}
  {\bibinfo {title} {On nonlinear {{Markov}} chain {{Monte Carlo}}},\ }\href
  {https://doi.org/10.3150/10-BEJ307} {\bibfield  {journal} {\bibinfo
  {journal} {Bernoulli}\ }\textbf {\bibinfo {volume} {17}},\ \bibinfo {pages}
  {987} (\bibinfo {year} {2011})}\BibitemShut {NoStop}%
\bibitem [{\citenamefont {Meyn}\ and\ \citenamefont
  {Tweedie}(2012)}]{meyn2012markov}%
  \BibitemOpen
  \bibfield  {author} {\bibinfo {author} {\bibfnamefont {S.~P.}\ \bibnamefont
  {Meyn}}\ and\ \bibinfo {author} {\bibfnamefont {R.~L.}\ \bibnamefont
  {Tweedie}},\ }\href@noop {} {\emph {\bibinfo {title} {Markov chains and
  stochastic stability}}}\ (\bibinfo  {publisher} {Springer Science \& Business
  Media},\ \bibinfo {year} {2012})\BibitemShut {NoStop}%
\bibitem [{\citenamefont {Roberts}\ and\ \citenamefont
  {Tweedie}(1996)}]{roberts_exponential_1996}%
  \BibitemOpen
  \bibfield  {author} {\bibinfo {author} {\bibfnamefont {G.~O.}\ \bibnamefont
  {Roberts}}\ and\ \bibinfo {author} {\bibfnamefont {R.~L.}\ \bibnamefont
  {Tweedie}},\ }\bibfield  {title} {\bibinfo {title} {Exponential convergence
  of {{Langevin}} distributions and their discrete approximations},\
  }\href@noop {} {\bibfield  {journal} {\bibinfo  {journal} {Bernoulli}\
  }\textbf {\bibinfo {volume} {2}},\ \bibinfo {pages} {341} (\bibinfo {year}
  {1996})}\BibitemShut {NoStop}%
\bibitem [{\citenamefont {Andrieu}\ and\ \citenamefont
  {Moulines}(2006)}]{andrieu_ergodicity_2006}%
  \BibitemOpen
  \bibfield  {author} {\bibinfo {author} {\bibfnamefont {C.}~\bibnamefont
  {Andrieu}}\ and\ \bibinfo {author} {\bibfnamefont {{\'E}.}~\bibnamefont
  {Moulines}},\ }\bibfield  {title} {\bibinfo {title} {On the ergodicity
  properties of some adaptive {{MCMC}} algorithms},\ }\href
  {https://doi.org/10.1214/105051606000000286} {\bibfield  {journal} {\bibinfo
  {journal} {Annals of Applied Probability}\ }\textbf {\bibinfo {volume}
  {16}},\ \bibinfo {pages} {1462} (\bibinfo {year} {2006})}\BibitemShut
  {NoStop}%
\bibitem [{\citenamefont {Haario}\ \emph {et~al.}(2001)\citenamefont {Haario},
  \citenamefont {Saksman},\ and\ \citenamefont
  {Tamminen}}]{haario_adaptive_2001}%
  \BibitemOpen
  \bibfield  {author} {\bibinfo {author} {\bibfnamefont {H.}~\bibnamefont
  {Haario}}, \bibinfo {author} {\bibfnamefont {E.}~\bibnamefont {Saksman}},\
  and\ \bibinfo {author} {\bibfnamefont {J.}~\bibnamefont {Tamminen}},\
  }\bibfield  {title} {\bibinfo {title} {An adaptive {{Metropolis}}
  algorithm},\ }\href@noop {} {\bibfield  {journal} {\bibinfo  {journal}
  {Bernoulli}\ }\textbf {\bibinfo {volume} {7}},\ \bibinfo {pages} {223}
  (\bibinfo {year} {2001})}\BibitemShut {NoStop}%
\bibitem [{\citenamefont {Jasra}\ \emph {et~al.}(2007)\citenamefont {Jasra},
  \citenamefont {Stephens},\ and\ \citenamefont
  {Holmes}}]{jasra_populationbased_2007}%
  \BibitemOpen
  \bibfield  {author} {\bibinfo {author} {\bibfnamefont {A.}~\bibnamefont
  {Jasra}}, \bibinfo {author} {\bibfnamefont {D.~A.}\ \bibnamefont
  {Stephens}},\ and\ \bibinfo {author} {\bibfnamefont {C.~C.}\ \bibnamefont
  {Holmes}},\ }\bibfield  {title} {\bibinfo {title} {On population-based
  simulation for static inference},\ }\href
  {https://doi.org/10.1007/s11222-007-9028-9} {\bibfield  {journal} {\bibinfo
  {journal} {Statistics and Computing}\ }\textbf {\bibinfo {volume} {17}},\
  \bibinfo {pages} {263} (\bibinfo {year} {2007})}\BibitemShut {NoStop}%
\bibitem [{\citenamefont {Villani}(2003)}]{villani_topics_2003}%
  \BibitemOpen
  \bibfield  {author} {\bibinfo {author} {\bibfnamefont {C.}~\bibnamefont
  {Villani}},\ }\href@noop {} {\emph {\bibinfo {title} {Topics in {{Optimal
  Transportation}}}}},\ \bibinfo {series} {Graduate {{Studies}} in
  {{Mathematics}}}\ No.~\bibinfo {number} {58}\ (\bibinfo  {publisher}
  {{American Mathematical Society}},\ \bibinfo {address} {{Providence, Rhode
  Island}},\ \bibinfo {year} {2003})\BibitemShut {NoStop}%
\bibitem [{\citenamefont {Santambrogio}(2015)}]{Santambrogio2015}%
  \BibitemOpen
  \bibfield  {author} {\bibinfo {author} {\bibfnamefont {F.}~\bibnamefont
  {Santambrogio}},\ }\href {https://doi.org/10.1007/978-3-319-20828-2} {\emph
  {\bibinfo {title} {Birkh{\"{a}}user, Cham}}},\ \bibinfo {series} {Progress in
  Nonlinear Differential Equations and Their Applications}, Vol.~\bibinfo
  {volume} {87}\ (\bibinfo  {publisher} {Springer International Publishing},\
  \bibinfo {address} {Cham},\ \bibinfo {year} {2015})\ p.\ \bibinfo {pages}
  {353}\BibitemShut {NoStop}%
\bibitem [{\citenamefont {Dinh}\ \emph {et~al.}(2015)\citenamefont {Dinh},
  \citenamefont {Krueger},\ and\ \citenamefont {Bengio}}]{Dinh2015}%
  \BibitemOpen
  \bibfield  {author} {\bibinfo {author} {\bibfnamefont {L.}~\bibnamefont
  {Dinh}}, \bibinfo {author} {\bibfnamefont {D.}~\bibnamefont {Krueger}},\ and\
  \bibinfo {author} {\bibfnamefont {Y.}~\bibnamefont {Bengio}},\ }\bibfield
  {title} {\bibinfo {title} {{NICE: Non-linear independent components
  estimation}},\ }\href@noop {} {\bibfield  {journal} {\bibinfo  {journal} {3rd
  International Conference on Learning Representations, ICLR 2015 - Workshop
  Track Proceedings}\ }\textbf {\bibinfo {volume} {1}},\ \bibinfo {pages} {1}
  (\bibinfo {year} {2015})},\ \Eprint {https://arxiv.org/abs/1410.8516}
  {1410.8516} \BibitemShut {NoStop}%
\bibitem [{\citenamefont {Dinh}\ \emph {et~al.}(2017)\citenamefont {Dinh},
  \citenamefont {{Sohl-Dickstein}},\ and\ \citenamefont
  {Bengio}}]{dinh_density_2017}%
  \BibitemOpen
  \bibfield  {author} {\bibinfo {author} {\bibfnamefont {L.}~\bibnamefont
  {Dinh}}, \bibinfo {author} {\bibfnamefont {J.}~\bibnamefont
  {{Sohl-Dickstein}}},\ and\ \bibinfo {author} {\bibfnamefont {S.}~\bibnamefont
  {Bengio}},\ }\bibfield  {title} {\bibinfo {title} {Density {{Estimation Using
  Real NVP}}},\ }in\ \href@noop {} {\emph {\bibinfo {booktitle} {International
  Conference on Learning Representations}}}\ (\bibinfo {year} {2017})\
  p.~\bibinfo {pages} {32}\BibitemShut {NoStop}%
\bibitem [{\citenamefont {Stroock}(1993)}]{stroock_logarithmic_1993}%
  \BibitemOpen
  \bibfield  {author} {\bibinfo {author} {\bibfnamefont {D.~W.}\ \bibnamefont
  {Stroock}},\ }\bibfield  {title} {\bibinfo {title} {Logarithmic {{Sobolev}}
  inequalities for gibbs states},\ }in\ \href
  {https://doi.org/10.1007/BFb0074094} {\emph {\bibinfo {booktitle} {Dirichlet
  {{Forms}}: {{Lectures}} given at the 1st {{Session}} of the {{Centro
  Internazionale Matematico Estivo}} ({{C}}.{{I}}.{{M}}.{{E}}.) Held in
  {{Varenna}}, {{Italy}}, {{June}} 8\textendash 19, 1992}}},\ \bibinfo {series
  and number} {Lecture {{Notes}} in {{Mathematics}}},\ \bibinfo {editor}
  {edited by\ \bibinfo {editor} {\bibfnamefont {E.}~\bibnamefont {Fabes}},
  \bibinfo {editor} {\bibfnamefont {M.}~\bibnamefont {Fukushima}}, \bibinfo
  {editor} {\bibfnamefont {L.}~\bibnamefont {Gross}}, \bibinfo {editor}
  {\bibfnamefont {C.}~\bibnamefont {Kenig}}, \bibinfo {editor} {\bibfnamefont
  {M.}~\bibnamefont {R{\"o}ckner}}, \bibinfo {editor} {\bibfnamefont {D.~W.}\
  \bibnamefont {Stroock}}, \bibinfo {editor} {\bibfnamefont {G.}~\bibnamefont
  {Dell'Antonio}},\ and\ \bibinfo {editor} {\bibfnamefont {U.}~\bibnamefont
  {Mosco}}}\ (\bibinfo  {publisher} {{Springer}},\ \bibinfo {address} {{Berlin,
  Heidelberg}},\ \bibinfo {year} {1993})\ pp.\ \bibinfo {pages}
  {194--228}\BibitemShut {NoStop}%
\bibitem [{\citenamefont {Lu}\ \emph {et~al.}(2019)\citenamefont {Lu},
  \citenamefont {Lu},\ and\ \citenamefont {Nolen}}]{lu_accelerating_2019}%
  \BibitemOpen
  \bibfield  {author} {\bibinfo {author} {\bibfnamefont {Y.}~\bibnamefont
  {Lu}}, \bibinfo {author} {\bibfnamefont {J.}~\bibnamefont {Lu}},\ and\
  \bibinfo {author} {\bibfnamefont {J.}~\bibnamefont {Nolen}},\ }\bibfield
  {title} {\bibinfo {title} {Accelerating {{Langevin Sampling}} with
  {{Birth}}-death},\ }\href@noop {} {\bibfield  {journal} {\bibinfo  {journal}
  {arXiv:1905.09863 [cs, math, stat]}\ }\textbf {\bibinfo {volume} {0}}
  (\bibinfo {year} {2019})},\ \Eprint {https://arxiv.org/abs/1905.09863}
  {arXiv:1905.09863 [cs, math, stat]} \BibitemShut {NoStop}%
\bibitem [{Note1()}]{Note1}%
  \BibitemOpen
  \bibinfo {note} {In principle, the parameters of each $T_m$ could also be
  further refined in this stage, but we have not tested this scenario in
  experiments yet.}\BibitemShut {Stop}%
\bibitem [{\citenamefont {Berglund}\ \emph {et~al.}(2017)\citenamefont
  {Berglund}, \citenamefont {Ges{\`u}},\ and\ \citenamefont
  {Weber}}]{berglund_eyring_2017}%
  \BibitemOpen
  \bibfield  {author} {\bibinfo {author} {\bibfnamefont {N.}~\bibnamefont
  {Berglund}}, \bibinfo {author} {\bibfnamefont {G.~D.}\ \bibnamefont
  {Ges{\`u}}},\ and\ \bibinfo {author} {\bibfnamefont {H.}~\bibnamefont
  {Weber}},\ }\bibfield  {title} {\bibinfo {title} {An
  {{Eyring}}\textendash{{Kramers}} law for the stochastic
  {{Allen}}\textendash{{Cahn}} equation in dimension two},\ }\href
  {https://doi.org/10.1214/17-EJP60} {\bibfield  {journal} {\bibinfo  {journal}
  {Electronic Journal of Probability}\ }\textbf {\bibinfo {volume} {22}},\
  \bibinfo {pages} {1} (\bibinfo {year} {2017})}\BibitemShut {NoStop}%
\bibitem [{\citenamefont {Faris}\ and\ \citenamefont
  {{Jona-Lasinio}}(1982)}]{faris_large_1982}%
  \BibitemOpen
  \bibfield  {author} {\bibinfo {author} {\bibfnamefont {W.~G.}\ \bibnamefont
  {Faris}}\ and\ \bibinfo {author} {\bibfnamefont {G.}~\bibnamefont
  {{Jona-Lasinio}}},\ }\bibfield  {title} {\bibinfo {title} {Large fluctuations
  for a nonlinear heat equation with noise},\ }\href
  {https://doi.org/10.1088/0305-4470/15/10/011} {\bibfield  {journal} {\bibinfo
   {journal} {Journal of Physics A: Mathematical and General}\ }\textbf
  {\bibinfo {volume} {15}},\ \bibinfo {pages} {3025} (\bibinfo {year}
  {1982})}\BibitemShut {NoStop}%
\bibitem [{\citenamefont {Marcus}(1974)}]{marcus_parabolic_1974}%
  \BibitemOpen
  \bibfield  {author} {\bibinfo {author} {\bibfnamefont {R.}~\bibnamefont
  {Marcus}},\ }\bibfield  {title} {\bibinfo {title} {Parabolic {{Ito
  Equations}}},\ }\href {https://doi.org/10.2307/1996753} {\bibfield  {journal}
  {\bibinfo  {journal} {Transactions of the American Mathematical Society}\
  }\textbf {\bibinfo {volume} {198}},\ \bibinfo {pages} {177} (\bibinfo {year}
  {1974})}\BibitemShut {NoStop}%
\bibitem [{\citenamefont {Frenkel}\ and\ \citenamefont
  {Smit}(2001)}]{frenkel_understanding_2001}%
  \BibitemOpen
  \bibfield  {author} {\bibinfo {author} {\bibfnamefont {D.}~\bibnamefont
  {Frenkel}}\ and\ \bibinfo {author} {\bibfnamefont {B.}~\bibnamefont {Smit}},\
  }\href@noop {} {\emph {\bibinfo {title} {Understanding {{Molecular
  Simulation}}: {{From Algorithms}} to {{Applications}}}}}\ (\bibinfo
  {publisher} {{Elsevier}},\ \bibinfo {year} {2001})\BibitemShut {NoStop}%
\bibitem [{\citenamefont {Nicoli}\ \emph {et~al.}(2020)\citenamefont {Nicoli},
  \citenamefont {Nakajima}, \citenamefont {Strodthoff}, \citenamefont {Samek},
  \citenamefont {M{\"{u}}ller},\ and\ \citenamefont {Kessel}}]{Nicoli2020}%
  \BibitemOpen
  \bibfield  {author} {\bibinfo {author} {\bibfnamefont {K.~A.}\ \bibnamefont
  {Nicoli}}, \bibinfo {author} {\bibfnamefont {S.}~\bibnamefont {Nakajima}},
  \bibinfo {author} {\bibfnamefont {N.}~\bibnamefont {Strodthoff}}, \bibinfo
  {author} {\bibfnamefont {W.}~\bibnamefont {Samek}}, \bibinfo {author}
  {\bibfnamefont {K.~R.}\ \bibnamefont {M{\"{u}}ller}},\ and\ \bibinfo {author}
  {\bibfnamefont {P.}~\bibnamefont {Kessel}},\ }\bibfield  {title} {\bibinfo
  {title} {{Asymptotically unbiased estimation of physical observables with
  neural samplers}},\ }\bibfield  {journal} {\bibinfo  {journal} {Physical
  Review E}\ }\textbf {\bibinfo {volume} {101}},\ \href
  {https://doi.org/10.1103/PhysRevE.101.023304} {10.1103/PhysRevE.101.023304}
  (\bibinfo {year} {2020})\BibitemShut {NoStop}%
\bibitem [{\citenamefont {Dibak}\ \emph {et~al.}(2020)\citenamefont {Dibak},
  \citenamefont {Klein},\ and\ \citenamefont {No{\'{e}}}}]{Dibak2020}%
  \BibitemOpen
  \bibfield  {author} {\bibinfo {author} {\bibfnamefont {M.}~\bibnamefont
  {Dibak}}, \bibinfo {author} {\bibfnamefont {L.}~\bibnamefont {Klein}},\ and\
  \bibinfo {author} {\bibfnamefont {F.}~\bibnamefont {No{\'{e}}}},\ }\bibfield
  {title} {\bibinfo {title} {{Temperature-steerable flows}},\ }in\ \href
  {http://arxiv.org/abs/2012.00429} {\emph {\bibinfo {booktitle} {Third
  Workshop on Machine Learning and the Physical Sciences (NeurIPS 2020),
  Vancouver, Canada.}}}\ (\bibinfo {year} {2020})\ \Eprint
  {https://arxiv.org/abs/2012.00429} {arXiv:2012.00429} \BibitemShut {NoStop}%
\bibitem [{Note2()}]{Note2}%
  \BibitemOpen
  \bibinfo {note} {Note that annealing of the target can help to catch multiple
  modes in some simple cases but offers no guarantees \cite
  {wu_solving_2019}.}\BibitemShut {Stop}%
\bibitem [{\citenamefont {Hartnett}\ and\ \citenamefont
  {Mohseni}(2020)}]{hartnett_self-supervised_2020}%
  \BibitemOpen
  \bibfield  {author} {\bibinfo {author} {\bibfnamefont {G.~S.}\ \bibnamefont
  {Hartnett}}\ and\ \bibinfo {author} {\bibfnamefont {M.}~\bibnamefont
  {Mohseni}},\ }\bibfield  {title} {\bibinfo {title} {Self-{{Supervised
  Learning}} of {{Generative Spin}}-{{Glasses}} with {{Normalizing Flows}}},\
  }\href@noop {} {\bibfield  {journal} {\bibinfo  {journal} {arXiv:2001.00585
  [cond-mat, physics:quant-ph, stat]}\ }\textbf {\bibinfo {volume} {0}}
  (\bibinfo {year} {2020})},\ \Eprint {https://arxiv.org/abs/2001.00585}
  {arXiv:2001.00585 [cond-mat, physics:quant-ph, stat]} \BibitemShut {NoStop}%
\bibitem [{\citenamefont {Yao}\ \emph {et~al.}(2018)\citenamefont {Yao},
  \citenamefont {Vehtari}, \citenamefont {Simpson},\ and\ \citenamefont
  {Gelman}}]{Yao2018}%
  \BibitemOpen
  \bibfield  {author} {\bibinfo {author} {\bibfnamefont {Y.}~\bibnamefont
  {Yao}}, \bibinfo {author} {\bibfnamefont {A.}~\bibnamefont {Vehtari}},
  \bibinfo {author} {\bibfnamefont {D.}~\bibnamefont {Simpson}},\ and\ \bibinfo
  {author} {\bibfnamefont {A.}~\bibnamefont {Gelman}},\ }\href
  {http://arxiv.org/abs/1802.02538} {\bibinfo {title} {{Yes, but Did It Work?:
  Evaluating Variational Inference}}} (\bibinfo {year} {2018}),\ \Eprint
  {https://arxiv.org/abs/1802.02538} {arXiv:1802.02538} \BibitemShut {NoStop}%
\bibitem [{\citenamefont {Naesseth}\ \emph {et~al.}(2020)\citenamefont
  {Naesseth}, \citenamefont {Lindsten},\ and\ \citenamefont
  {Blei}}]{Naesseth2020}%
  \BibitemOpen
  \bibfield  {author} {\bibinfo {author} {\bibfnamefont {C.~A.}\ \bibnamefont
  {Naesseth}}, \bibinfo {author} {\bibfnamefont {F.}~\bibnamefont {Lindsten}},\
  and\ \bibinfo {author} {\bibfnamefont {D.}~\bibnamefont {Blei}},\ }\bibfield
  {title} {\bibinfo {title} {{Markovian score climbing: Variational inference
  with KL(p||q)}},\ }\href@noop {} {\bibfield  {journal} {\bibinfo  {journal}
  {Advances in Neural Information Processing Systems}\ }\textbf {\bibinfo
  {volume} {2020-Decem}} (\bibinfo {year} {2020})},\ \Eprint
  {https://arxiv.org/abs/2003.10374} {arXiv:2003.10374} \BibitemShut {NoStop}%
\bibitem [{\citenamefont {Parno}\ and\ \citenamefont
  {Marzouk}(2018)}]{Parno2018}%
  \BibitemOpen
  \bibfield  {author} {\bibinfo {author} {\bibfnamefont {M.~D.}\ \bibnamefont
  {Parno}}\ and\ \bibinfo {author} {\bibfnamefont {Y.~M.}\ \bibnamefont
  {Marzouk}},\ }\bibfield  {title} {\bibinfo {title} {{Transport map
  accelerated markov chain monte carlo}},\ }\href
  {https://doi.org/10.1137/17M1134640} {\bibfield  {journal} {\bibinfo
  {journal} {SIAM-ASA Journal on Uncertainty Quantification}\ }\textbf
  {\bibinfo {volume} {6}},\ \bibinfo {pages} {645} (\bibinfo {year} {2018})},\
  \Eprint {https://arxiv.org/abs/1412.5492} {arXiv:1412.5492} \BibitemShut
  {NoStop}%
\bibitem [{\citenamefont {K{\"{o}}hler}\ \emph {et~al.}(2020)\citenamefont
  {K{\"{o}}hler}, \citenamefont {Klein},\ and\ \citenamefont
  {No{\'{e}}}}]{Kohler2019}%
  \BibitemOpen
  \bibfield  {author} {\bibinfo {author} {\bibfnamefont {J.}~\bibnamefont
  {K{\"{o}}hler}}, \bibinfo {author} {\bibfnamefont {L.}~\bibnamefont
  {Klein}},\ and\ \bibinfo {author} {\bibfnamefont {F.}~\bibnamefont
  {No{\'{e}}}},\ }\bibfield  {title} {\bibinfo {title} {{Equivariant Flows:
  sampling configurations for multi-body systems with symmetric energies}},\
  }in\ \href {http://arxiv.org/abs/1910.00753} {\emph {\bibinfo {booktitle}
  {International Conference on Machine Learning (ICML)}}}\ (\bibinfo {year}
  {2020})\ \Eprint {https://arxiv.org/abs/1910.00753} {arXiv:1910.00753}
  \BibitemShut {NoStop}%
\bibitem [{\citenamefont {Rezende}\ \emph {et~al.}(2020)\citenamefont
  {Rezende}, \citenamefont {Papamakarios}, \citenamefont {Racani{\`{e}}re},
  \citenamefont {Albergo}, \citenamefont {Kanwar}, \citenamefont {Shanahan},\
  and\ \citenamefont {Cranmer}}]{Rezende2020}%
  \BibitemOpen
  \bibfield  {author} {\bibinfo {author} {\bibfnamefont {D.~J.}\ \bibnamefont
  {Rezende}}, \bibinfo {author} {\bibfnamefont {G.}~\bibnamefont
  {Papamakarios}}, \bibinfo {author} {\bibfnamefont {S.}~\bibnamefont
  {Racani{\`{e}}re}}, \bibinfo {author} {\bibfnamefont {M.~S.}\ \bibnamefont
  {Albergo}}, \bibinfo {author} {\bibfnamefont {G.}~\bibnamefont {Kanwar}},
  \bibinfo {author} {\bibfnamefont {P.~E.}\ \bibnamefont {Shanahan}},\ and\
  \bibinfo {author} {\bibfnamefont {K.}~\bibnamefont {Cranmer}},\ }\bibfield
  {title} {\bibinfo {title} {{Normalizing flows on tori and spheres}},\
  }\href@noop {} {\bibfield  {journal} {\bibinfo  {journal} {37th International
  Conference on Machine Learning, ICML 2020}\ }\textbf {\bibinfo {volume}
  {PartF16814}},\ \bibinfo {pages} {8039} (\bibinfo {year} {2020})},\ \Eprint
  {https://arxiv.org/abs/2002.02428} {arXiv:2002.02428} \BibitemShut {NoStop}%
\bibitem [{\citenamefont {Hairer}(2009)}]{hairer_introduction_2009}%
  \BibitemOpen
  \bibfield  {author} {\bibinfo {author} {\bibfnamefont {M.}~\bibnamefont
  {Hairer}},\ }\href@noop {} {\emph {\bibinfo {title} {An {{Introduction}} to
  {{Stochastic PDEs}}}}}\ (\bibinfo {year} {2009})\ p.~\bibinfo {pages}
  {78}\BibitemShut {NoStop}%
\bibitem [{\citenamefont {Wu}\ \emph {et~al.}(2020)\citenamefont {Wu},
  \citenamefont {K{\"o}hler},\ and\ \citenamefont
  {No\'e}}]{wu_stochastic_2020}%
  \BibitemOpen
  \bibfield  {author} {\bibinfo {author} {\bibfnamefont {H.}~\bibnamefont
  {Wu}}, \bibinfo {author} {\bibfnamefont {J.}~\bibnamefont {K{\"o}hler}},\
  and\ \bibinfo {author} {\bibfnamefont {F.}~\bibnamefont {No\'e}},\ }\bibfield
   {title} {\bibinfo {title} {Stochastic normalizing flows},\ }in\ \href@noop
  {} {\emph {\bibinfo {booktitle} {Advances in Neural Information Processing
  Systems}}},\ Vol.~\bibinfo {volume} {33},\ \bibinfo {editor} {edited by\
  \bibinfo {editor} {\bibfnamefont {H.}~\bibnamefont {Larochelle}}, \bibinfo
  {editor} {\bibfnamefont {M.}~\bibnamefont {Ranzato}}, \bibinfo {editor}
  {\bibfnamefont {R.}~\bibnamefont {Hadsell}}, \bibinfo {editor} {\bibfnamefont
  {M.~F.}\ \bibnamefont {Balcan}},\ and\ \bibinfo {editor} {\bibfnamefont
  {H.}~\bibnamefont {Lin}}}\ (\bibinfo  {publisher} {{Curran Associates,
  Inc.}},\ \bibinfo {year} {2020})\ pp.\ \bibinfo {pages}
  {5933--5944}\BibitemShut {NoStop}%
\bibitem [{\citenamefont {E}\ \emph {et~al.}(2005)\citenamefont {E},
  \citenamefont {Ren},\ and\ \citenamefont
  {{Vanden-Eijnden}}}]{e_transition_2005}%
  \BibitemOpen
  \bibfield  {author} {\bibinfo {author} {\bibfnamefont {W.}~\bibnamefont {E}},
  \bibinfo {author} {\bibfnamefont {W.}~\bibnamefont {Ren}},\ and\ \bibinfo
  {author} {\bibfnamefont {E.}~\bibnamefont {{Vanden-Eijnden}}},\ }\bibfield
  {title} {\bibinfo {title} {Transition pathways in complex systems:
  {{Reaction}} coordinates, isocommittor surfaces, and transition tubes},\
  }\href {https://doi.org/10.1016/j.cplett.2005.07.084} {\bibfield  {journal}
  {\bibinfo  {journal} {Chem. Phys. Lett.}\ }\textbf {\bibinfo {volume}
  {413}},\ \bibinfo {pages} {242} (\bibinfo {year} {2005})}\BibitemShut
  {NoStop}%
\bibitem [{\citenamefont {E}\ and\ \citenamefont
  {{Vanden-Eijnden}}(2010)}]{e_transition-path_2010}%
  \BibitemOpen
  \bibfield  {author} {\bibinfo {author} {\bibfnamefont {W.}~\bibnamefont {E}}\
  and\ \bibinfo {author} {\bibfnamefont {E.}~\bibnamefont {{Vanden-Eijnden}}},\
  }\bibfield  {title} {\bibinfo {title} {Transition-{{Path Theory}} and
  {{Path}}-{{Finding Algorithms}} for the {{Study}} of {{Rare Events}}},\
  }\href {https://doi.org/10.1146/annurev.physchem.040808.090412} {\bibfield
  {journal} {\bibinfo  {journal} {Annual Review of Physical Chemistry}\
  }\textbf {\bibinfo {volume} {61}},\ \bibinfo {pages} {391} (\bibinfo {year}
  {2010})}\BibitemShut {NoStop}%
\bibitem [{\citenamefont {Allen}\ \emph {et~al.}(2009)\citenamefont {Allen},
  \citenamefont {Valeriani},\ and\ \citenamefont {ten
  Wolde}}]{allen_forward_2009}%
  \BibitemOpen
  \bibfield  {author} {\bibinfo {author} {\bibfnamefont {R.~J.}\ \bibnamefont
  {Allen}}, \bibinfo {author} {\bibfnamefont {C.}~\bibnamefont {Valeriani}},\
  and\ \bibinfo {author} {\bibfnamefont {P.~R.}\ \bibnamefont {ten Wolde}},\
  }\bibfield  {title} {\bibinfo {title} {Forward flux sampling for rare event
  simulations},\ }\href {https://doi.org/10.1088/0953-8984/21/46/463102}
  {\bibfield  {journal} {\bibinfo  {journal} {Journal of Physics: Condensed
  Matter}\ }\textbf {\bibinfo {volume} {21}},\ \bibinfo {pages} {463102}
  (\bibinfo {year} {2009})}\BibitemShut {NoStop}%
\bibitem [{\citenamefont {Falasco}\ and\ \citenamefont
  {Esposito}(2020)}]{falasco_dissipationtime_2020}%
  \BibitemOpen
  \bibfield  {author} {\bibinfo {author} {\bibfnamefont {G.}~\bibnamefont
  {Falasco}}\ and\ \bibinfo {author} {\bibfnamefont {M.}~\bibnamefont
  {Esposito}},\ }\bibfield  {title} {\bibinfo {title} {Dissipation-{{Time
  Uncertainty Relation}}},\ }\href
  {https://doi.org/10.1103/PhysRevLett.125.120604} {\bibfield  {journal}
  {\bibinfo  {journal} {Physical Review Letters}\ }\textbf {\bibinfo {volume}
  {125}},\ \bibinfo {pages} {120604} (\bibinfo {year} {2020})}\BibitemShut
  {NoStop}%
\bibitem [{\citenamefont {{Kuznets-Speck}}\ and\ \citenamefont
  {Limmer}(2021)}]{kuznets-speck_dissipation_2021}%
  \BibitemOpen
  \bibfield  {author} {\bibinfo {author} {\bibfnamefont {B.}~\bibnamefont
  {{Kuznets-Speck}}}\ and\ \bibinfo {author} {\bibfnamefont {D.~T.}\
  \bibnamefont {Limmer}},\ }\bibfield  {title} {\bibinfo {title} {Dissipation
  bounds the amplification of transition rates far from equilibrium},\ }\href
  {https://doi.org/10.1073/pnas.2020863118} {\bibfield  {journal} {\bibinfo
  {journal} {Proceedings of the National Academy of Sciences}\ }\textbf
  {\bibinfo {volume} {118}},\ \bibinfo {pages} {e2020863118} (\bibinfo {year}
  {2021})}\BibitemShut {NoStop}%
\bibitem [{\citenamefont {Metzner}\ \emph {et~al.}(2006)\citenamefont
  {Metzner}, \citenamefont {Sch{\"u}tte},\ and\ \citenamefont
  {{Vanden-Eijnden}}}]{metzner_illustration_2006-1}%
  \BibitemOpen
  \bibfield  {author} {\bibinfo {author} {\bibfnamefont {P.}~\bibnamefont
  {Metzner}}, \bibinfo {author} {\bibfnamefont {C.}~\bibnamefont
  {Sch{\"u}tte}},\ and\ \bibinfo {author} {\bibfnamefont {E.}~\bibnamefont
  {{Vanden-Eijnden}}},\ }\bibfield  {title} {\bibinfo {title} {Illustration of
  transition path theory on a collection of simple examples},\ }\href
  {https://doi.org/10.1063/1.2335447} {\bibfield  {journal} {\bibinfo
  {journal} {The Journal of Chemical Physics}\ }\textbf {\bibinfo {volume}
  {125}},\ \bibinfo {pages} {084110} (\bibinfo {year} {2006})}\BibitemShut
  {NoStop}%
\bibitem [{\citenamefont {Park}\ \emph {et~al.}(2003)\citenamefont {Park},
  \citenamefont {Sener}, \citenamefont {Lu},\ and\ \citenamefont
  {Schulten}}]{park_reaction_2003}%
  \BibitemOpen
  \bibfield  {author} {\bibinfo {author} {\bibfnamefont {S.}~\bibnamefont
  {Park}}, \bibinfo {author} {\bibfnamefont {M.~K.}\ \bibnamefont {Sener}},
  \bibinfo {author} {\bibfnamefont {D.}~\bibnamefont {Lu}},\ and\ \bibinfo
  {author} {\bibfnamefont {K.}~\bibnamefont {Schulten}},\ }\bibfield  {title}
  {\bibinfo {title} {Reaction paths based on mean first-passage times},\ }\href
  {https://doi.org/10.1063/1.1570396} {\bibfield  {journal} {\bibinfo
  {journal} {The Journal of Chemical Physics}\ }\textbf {\bibinfo {volume}
  {119}},\ \bibinfo {pages} {1313} (\bibinfo {year} {2003})}\BibitemShut
  {NoStop}%
\bibitem [{\citenamefont {Carrillo}\ \emph {et~al.}(2019)\citenamefont
  {Carrillo}, \citenamefont {Gvalani}, \citenamefont {Pavliotis},\ and\
  \citenamefont {Schlichting}}]{carrillo2019long}%
  \BibitemOpen
  \bibfield  {author} {\bibinfo {author} {\bibfnamefont {J.~A.}\ \bibnamefont
  {Carrillo}}, \bibinfo {author} {\bibfnamefont {R.~S.}\ \bibnamefont
  {Gvalani}}, \bibinfo {author} {\bibfnamefont {G.~A.}\ \bibnamefont
  {Pavliotis}},\ and\ \bibinfo {author} {\bibfnamefont {A.}~\bibnamefont
  {Schlichting}},\ }\bibfield  {title} {\bibinfo {title} {Long-time behaviour
  and phase transitions for the mckean–vlasov equation on the torus},\ }\href
  {https://doi.org/10.1007/s00205-019-01430-4} {\bibfield  {journal} {\bibinfo
  {journal} {Archive for Rational Mechanics and Analysis}\ }\textbf {\bibinfo
  {volume} {235}},\ \bibinfo {pages} {635–690} (\bibinfo {year}
  {2019})}\BibitemShut {NoStop}%
\end{thebibliography}%

\clearpage
\appendix

\section{Comparing the Various Methods of Training and Sampling}
\label{app:transampl}

Let us briefly review the various methods of training a normalizing flow and using it for sampling, under different assumptions: specifically, we primarily distinguish situations in which data from the target density $\rhostar$ are available but no analytical information about this density is known, to those in which we have no or little preexisting data from $\rho_*$ but we know the shape of this  density up to an unknown normalization factor, i.e. we know $U_*(x)$ in $\rhostar(x) = Z_*^{-1} e^{-U_*(x)}$ but not $Z_*=\int_\Omega e^{-U_*(x)} dx$. Both situations naturally lead to different procedures of training and sampling. We discuss how to blend these two procedures within the adaptive MCMC strategy that we propose

\subsection{Data available, $\rho_*$ unknown}
\label{ass:sit1}

Assume that we have at our disposal a training set $\{x^*_i\}_{i=1}^n$ made of samples drawn from $\rho_*$ but have no other information available about this target density. In this case:

\paragraph{Training:} To train the map $T$ it is natural to use  the (direct) KL divergence of $\rhostar$ with respect to  $\hat\rho$, $D_\text{KL}(\rhostar||\hat \rho)$, since, up to a constant in the map, this divergence can be expressed as an expectation over $\rhostar$:
\begin{equation}
    \label{eq:ass:DKL}
    D_\text{KL}(\rhostar||\hat \rho) = C_* - \int_\Omega \log \rhohat (x) \rhostar(x) dx
\end{equation}
with $\rhohat (x)$ given by \eqref{eq:rhohatexplicit} (and hence a function of $T$) and $C_*= \int_\Omega \log \rhostar (x) \rhostar(x) dx$ (and hence independent of $T$). This leads to the following empirical loss, viewed as an objective for the map $T$,
\begin{equation}
    \label{eq:ass:DKL:emp}
    -\frac1n \sum_{i=1}^n  \log \rhohat (x^*_i)
\end{equation}
Minimizing this loss (possibly augmented by a regularizing term) over $T$ gives the optimal transport map given the data set. 

\paragraph{Sampling:} Once the optimal map $T$ minimizing~\eqref{eq:ass:DKL:emp} has been identified, new data can be generated by sampling points $x_B$ from the base density $\rho_B$ and pushing them through the map to obtain $T(x_B)$. Note that the quality of these new samples cannot be cross-validated without additional information, i.e. they are only as good as the map $T$ is (and in particular they cannot be added to the original data set to improve the quality of the map $T$ via training).

\subsection{No pre-existing data, $\rho_*$ known up to a normalizing factor}
\label{ass:sit2}

Assume that we know $U_*(x)$ in $\rhostar(x) = Z_*^{-1} e^{-U_*(x)}$ but not $Z_*=\int_\Omega e^{-U_*(x)} dx$, and have no samples from this target density available.
In this case: 

\paragraph{Training:} To train the map $T$ it is natural to use  the (reversed) KL divergence of $\rhohat$ with respect to  $\hat\rhostar$, $D_\text{KL}(\rhohat||\rhostar)$, since, up to a constant in the map, this divergence can be expressed as an expectation over~$\rhohat$:
\begin{equation}
    \label{eq:ass:DKLrev}
    D_\text{KL}(\rhohat||\rhostar) = \hat C + \int_\Omega \left(\log \rhohat (x) +U_*(x)\right) \rhohat(x) dx
\end{equation}
with $\rhohat (x)$ given by \eqref{eq:rhohatexplicit} (and hence a function of $T$) and $\hat C= - \int_\Omega \log Z_* \rhohat(x) dx = \log Z_*$ (and hence independent of $T$). This leads to the following empirical loss
\begin{equation}
    \label{eq:ass:DKL:emp:rev}
    \frac1n \sum_{i=1}^n  \left(\log \rhohat (T(x^B_i)) +U_*(T(x^B_i))\right)
\end{equation}
where $\{x^B_i\}_{i=1}^n$ is set of data points sampled from the base density $\rho_B$.
Minimizing this loss (possibly augmented by a regularizing term) over $T$ gives the optimal transport map given $U_*$. Note that in this situation, unlike in that considered in Sec.~\ref{ass:sit1}, we can easily enlarge the data set $\{x^B_i\}_{i=1}^n$ using as large a $n\in\NN$ as needed/possible, since generating samples from the base distribution is straightforward. This method has be sometimes called \emph{self-training}.

\paragraph{Sampling:} Once the optimal map $T$ minimizing~\eqref{eq:ass:DKL:emp} has been identified, new data can again be generated by sampling data points $x_B$ from the base density $\rho_B$ and pushing them through the map to obtain $T(x_B)$. Note that in this situation, unlike in that considered in Sec.~\ref{ass:sit1}, we can assess the quality of these new samples, for example by integrating them in the MCMC nonlocal resampling strategy described in main text, i.e. by accepting or rejecting the new samples according to the Metropolis-Hastings criterion with probability given in~\eqref{eq:accept}. These samples from $\rhostar$ can then potentially be used to further train the map by combining the estimator in~\eqref{eq:ass:DKL:emp} based on the direct KL divergence with the one in \eqref{eq:ass:DKL:emp:rev} based on the reversed KL divergence.

\subsection{Blending both procedures in an augmented MCMC strategy for training and sampling}
\label{ass:blend}

The procedure described in Sec.~\ref{ass:sit2} may at first sight look ideal when $U_*(x)$ is known since it seems to dispense us completely of the need for a pre-existing data set from the target density~$\rho_*$. This simplicity is however illusory in complex situations. Indeed, the procedure in Sec.~\ref{ass:sit2} relies entirely on sampling in the latent space of the base density $\rho_B$ and pushing these sample through~$T$. If the map $T$ that we use to initiate the training procedure is a poor approximation of the optimal $T_*$ such that $T_*(x_B) \sim \rhostar$, there is very small probability to generate samples that have any significant weight in the empirical loss~\eqref{eq:ass:DKL:emp:rev}: indeed these good samples have exponentially small weight in the base distribution initially. This implies that training will be hard or impossible unless we use prohibitively large data set from the bases density. In turn it also means that the new samples proposed in the MCMC resampling strategy will have a very small probability of being accepted.  

What we propose is a way out of this conundrum. If $U_*$ is available, and we have some prior information about the location of the regions with significant statistical weight on $\rho_*$, we can use any standard MCMC sampling strategy such as MALA with walkers started in these regions to start producing samples from $\rho*$. Because of the local nature of procedures like MALA, the quality of these samples will be locally good but globally poor: that is, the procedure will give us information conditional on being in regions of high likelihood on $\rho_*$, but not the relative statistical weight of these regions. This information, however, is sufficient to start the training of the map using the direct KL divergence in Eq.~\eqref{eq:ass:DKL}. In turn, this allows up to combine the local MCMC algorithm with a MCMC resampling step, thereby producing nonlocal moves. These nonlocal moves equilibrate the weights  between the high likelihood  regions  on $\rho_*$, which in turns allow us to refine the quality of the map, thereby producing better (i.e. more likely to be accepted) non local moves, etc. 

Note that if the map learned through this combined strategy becomes a good approximation of the optimal $T_*$ we could then forget altogether about the local MCMC sampling and revert to the procedure described in Sec.~\ref{ass:sit2}. In practice however, getting such a high quality of the map may be a tall order, and it is beneficial to keep the local MCMC sampling to capture the fine details of the target $\rho_*$  (see Fig. \ref{fig:wiggle}).

\section{Measure Theoretic Formulation of the Normalizing Flow augmented sampling algorithm}
\label{app:measures}

In the continuous limit, the models discussed in the main text do not have a well-defined probability density function.
For the sake of precision and completeness, we reformulate the discussion Sec.~\ref{sec:setup} in terms of arbitrary probability measures on a set~$\Omega$ with $\sigma$-algebra~$\mathcal{F}$. 
Let $(\Omega, \mathcal{F}, \nu_*)$ be the probability space associated with the target probability measure $\nu_*$.
Let us denote the transition kernel by $P (x,dy)$ (so that $P(x,dy) = \pi(x,y) dy$ if $P$ has a density) and write the detailed balance relation as (compare~\eqref{eq:db})
\begin{equation}
    \nu_*(dx) P(x,dy) = \nu_*(dy) P(y,dx),
\end{equation}
meaning that for all suitable test functions $\chi:\Omega \times\Omega\to\RR $ we have  (compare~\eqref{eq:ergo})
\begin{equation}
    \int_{\Omega\times\Omega} \chi(x,y)\left(\nu_*(dx) P(x,dy) - \nu_*(dy) P(y,dx)\right) = 0.
\end{equation}

In this general setup, the nonlocal map that we use to augment the local dynamics is an invertible transformation from $\Omega$ to itself,  i.e $T:\Omega\to \Omega$. The map $T$ is used to transport a base measure $\nu_\text{B}$ from which we can sample efficiently into the target measure $\nu_*$. In the ideal case, the map $T_*$ is constructed so that
\begin{equation}
    \nu_* = T_{*\sharp}\nu_\text{B}
\end{equation}
where the equality above holds in the weak sense, i.e. for a test function $\phi: \Omega \to \RR$,
\begin{equation}
     \int_\Omega \phi(x) \nu_*(dx) =  \int_\Omega \phi\bigl( T_*(x) \bigr) \nu_\text{B}(dx).
\end{equation}
In practice, instead of $T_*$ we only have at our disposal some approximation $T : \Omega \to \Omega$, also invertible. In order to use this imperfect map in a Metropolis-Hastings procedure, we must require that the target measure, $\nu_*$, and  the push-forward of the base measure measure, $T_\sharp\nu_\text{B}$, be mutually absolutely continuous with respect to one another.
 If that is the case, denoting by $(d\nu_*/dT_\sharp\nu_{\rm B})(x)$ and $(dT_\sharp\nu_{\rm B}/d\nu_*)(x) = 1/(d\nu_*/dT_\sharp\nu_{\rm B})(x)$ the Radon-Nikodym derivatives of $T_\sharp\nu_{\rm B}$ with respect to $\nu_*$ and of $\nu_*$ with respect to $T_\sharp\nu_{\rm B}$, the MCMC algorithm that resamples from $T_\sharp\nu_\text{B}$ proceeds as follows
\begin{enumerate}
    \item Given $x(k)$, sample $x_\text{B}\sim \nu_{\rm B}$ and let $y=T(x_\text{B})$
    \item Set $x(k+1) = y$ with probability
    \begin{equation}
        {\rm acc}(x(k), y) = \min\left[ 1, \frac{d\nu_*}{dT_\sharp\nu_{\rm B}}(y)\frac{dT_\sharp\nu_{\rm B}}{d\nu_*}(x(k)) \right]
    \end{equation}
    and otherwise set $x(k+1) = x(k).$
\end{enumerate}

In our approach, the map $T$ is evolved by minimizing some objective function, for example the KL divergence, measuring the discrepancy between the pushforward $T_\sharp\nu_\text{B}$ and the target $\nu_*$ or its approximation after $k$ step of MCMC sampling, $\nu_k$. In this case it is crucial to choose a base measure $\nu_B$ such that the KL divergence is well-defined initially: indeed if it is finite at the start of the procedure, it will remain so during optimization since the KL divergence can only decrease during gradient descent. Such a choice is not automatic for the SPDE examples we consider in text, because the infinite-dimensional measures associated with the solution of these equations have a strong propensity to be mutually orthogonal~\cite{hairer_introduction_2009}.

\subsection{Stochastic AC equation}
In the case of the stochastic Allen-Cahn equation, as base measure it is convenient to use the measure associated with the Hamiltonian~\eqref{eq:informed_base}. This is the Gaussian measure of an Ornstein-Uhlenbeck bridge, i.e. the process with mean zero and covariance (this is the inverse of the operator $-\beta a \partial_s^2 + \beta a^{-1}$ with zero Dirichlet boundary conditions):
\begin{equation}
    C_\text{B}(s,s') = \frac{a}{2\beta} \frac{e^{-|s-s'|/a+1}+e^{-|s-s'|/a-1}-e^{-(s+s')/a+1}-e^{(s+s')/a-1}}{(e^1-e^{-1})}.
\end{equation}
Denoting a realization of this bridge by $B\in C([0,1],\RR)$ we then have
\begin{equation}
\label{eq:RDdef}
   \frac{d\nu_*}{d\nu_\text{B}}(B) = Z_V^{-1} \exp\left(-\beta\textstyle\int_0^1
     V(B(s))ds\right), \qquad V(B) = \frac1{4a} (1-B^2)^2
\end{equation}
where
\begin{equation}
\label{eq:ZV}
     Z_V= \EE_{\nu_\text{B}} \exp\left(-\beta\textstyle\int_0^1
     V(B(s))ds\right) 
\end{equation}
The Radon-Nikodym derivative~\eqref{eq:RDdef} is well-defined if $Z_V<\infty$, which is the case for the potential energy $V(B) $\cite{faris_large_1982}. The KL divergence of $\nu_*$ with respect to $\nu_\text{B}$ can therefore be expressed as 
\begin{equation}
\label{eq:DKLstochAC}
   \begin{aligned}
     &D_{\text{KL}} (\nu_*\|\nu_\text{B})= -\log Z_V- \beta Z_V^{-1} \EE_{\nu_\text{B}} \left(
       \bar V(B) e^{-\beta\bar V(B)}\right)
   \end{aligned}
\end{equation}
where we denote $\bar V(B) =\int_0^1 V(B(s))ds$. We can also invert these relations and express them in terms of expectation over the target $\nu_*$. Denoting by $\phi\in C([0,1],\RR)$ a random sample of this measure, we have
\begin{equation}
\label{eq:ZVs}
     Z_V= \left(\EE_{\nu_*} \exp\left(\beta\textstyle\int_0^1
     V(\phi(s))ds\right) \right)^{-1}
\end{equation}
and
\begin{equation}
\label{eq:DKLstochACs}
   \begin{aligned}
     &D_{\text{KL}} (\nu_*\|\nu_\text{B})= -\log Z_V- \beta  \EE_{\nu_*} \left(
       \textstyle\int_0^1
     V(\phi(s))ds\right).
   \end{aligned}
\end{equation}
The quantities in these last two formulas can be estimated by replacing the expectation over $\nu_*$ by an empirical average over samples drawn from this measure or, more generally, drawn from the current approximation of this measure in the MCMC, $\nu_k$. This allows us to start the training procedure of the map $T$. In practice, of course, the fields in the formula above are discretized on a grid with $N$ points and the integrals are replaced by Riemann sums. Correspondingly, the map $T$ is approximated by a map from $\RR^N$ to $\RR^N$, using the RealNVP described in Appendix~\ref{sec:a:NF}.

\subsection{Transition Paths}

A similar procedure can be used to construct the base measure in the context of transition path sampling. In this case, as base measure we can take the Gaussian measure associated with the bridge process connecting $x_A$ to $x_B$: this process can be expressed as
\begin{equation}
    X_B(t) = x_A (1-t/\tmax) + x_B t/\tmax + \sigma B(t/\tmax)
\end{equation}
where $B\in C([0,1],\RR^d)$ is the standard Brownian bridge in $\RR^d$, i.e. the Gaussian process with mean zero and covariance
\begin{equation}
    C(u,u') = \begin{cases}
    u'(1-u)\,\text{Id}  \qquad &\text{if} \ \  u' \le  u\\
    u(1-u')\, \text{Id} \qquad &\text{if} \ \  u \le u'
    \end{cases} 
\end{equation}
Using this bridge process as base measure leads to  Radon-Nikodym derivative
\begin{equation}
\label{eq:martingale}
    \frac{d\nu_*}{d\nu_\text{B}}(X_\text{B}) = \exp \left( \frac1{\sigma^2}\int_0^{\tmax} \!\!\!\!b(X_\text{B}(t)) dX_\text{B}(t) - \frac1{2\sigma^2} \int_0^{\tmax} \!\!\!\!|b(X_\text{B}(t))|^2 dt \right)
\end{equation}
which is a Martingale with respect to the filtration $\mathcal{F}_{t_m}$ whenever the drift satisfies the Novikov condition, namely,
\begin{equation}
    \EE \exp \left( - \frac1{2\sigma^2} \int_0^{\tmax} |b(X_\text{B}(t))|^2 dt\right) < +\infty.
\end{equation}

\section{Convergence analysis of the MCMC scheme in the continuous limit }
\label{sec:MCMCcl}

\subsection{Chapman-Kolmogorov equation} Written in terms of the densities $\rho_*(x)$ and $\hat \rho(y)$ (assumed to be fixed for now) the transition kernel in~\eqref{eq:nfker} reads
\begin{equation}
   \label{eq:48}
   \pi_T(x,y) = a(x,y) \hat \rho(y) + (1-b(x)) \delta(x-y)
\end{equation}
where
\begin{equation}
   \label{eq:49}
      \begin{aligned}
   a(x,y) &= \min \left(\frac{\hat \rho(x) \rho_*(y)}{\hat \rho(y) \rho_*(x)},1\right),
   \\
   b(x) &= \int_\Omega a(x,y) \hat \rho(y)dy.
      \end{aligned}
\end{equation}
Denoting as $\{\rho_k(x)\}_{k\in\NN}$ the updated probability density of the walker in the Markov chain associated with the kernel $\pi_T(x,y)$ alone, this density satisfies the Chapman-Kolmogorov equation
\begin{equation}
   \label{eq:50}
    \rho_{k+1} (x) = \int_\Omega \rho_k(y) \pi_T(y,x) dy.
\end{equation}
Using the explicit form of $\pi_T(x,y)$ in~\eqref{eq:48}, after some simple reorganization this equation can be written as
\begin{equation}
   \label{eq:51}
   \rho_{k+1} (x) = \rho_k(x) +  \int_\Omega R(x,y) \left(\rho_*(x) \rho_k(y) -  \rho_k(x) \rho_*(y)\right) dy
 \end{equation}
where we defined
\begin{equation}
  \label{eq:52b}
   R(x,y) = R(y,x) = \min\left(\frac{\hat \rho(x)}{\rho_*(x)},
     \frac{\hat \rho(y)}{\rho_*(y)} \right).
\end{equation}
Note that if we had $\hat \rho = \rho_*$, then $R(x,y) =1$ and \eqref{eq:51} would reach equilibrium in one step, $\rho_{k+1} = \rho_*$ whatever $\rho_k$.

\begin{algorithm*}[htb]
\caption{MCMC with partial resampling steps, given an evolving map \label{alg:concurrentA}}   
\begin{algorithmic}[1]
    \State \textsc{Sample}($U_*$, $\{T_t\}_{t\ge0}$, $\{x_i(0)\}_{i=1}^n$, $\tau$, $k_\text{max}$, $\alpha$)
    \State {\bfseries Inputs:} $ U_*$ target energy, $\{T_t\}_{t\ge0}$  evolving maps, $\{x_i{(0)}\}_{i=1}^n$ initial data, $\tau>0$ time step, $T_\text{max}>0$ total duration, $\alpha>0$
    \State $k=0$
    \While{$k< T_\text{max}/\tau$}
    \For{$i=1,\dots, n$}
    \If{$k \mod 2=0$}
    \State $x'_{\rm B, i} \sim \rho_{\rm B}$ \label{lst:resample1A}
    \State $x_i' = T_{k\tau} (x'_{{\rm B}, i})$ \Comment{push-forward via $T$} \label{lst:resample3A}
    \State $x_i(k+1) = x_i'$ with probability $\min(\alpha\tau \, {\rm acc}(x_i(k),x_i'),1)$, otherwise $x_i(k+1) = x_i(k)$ \Comment{partial resampling step}\label{lst:resample4A}
    \Else
    \State $x_i(k+1) = x_i(k) - \tau  \nabla U_*(x_i(k)) + \sqrt{2 \tau}\, \eta_i$ with $\eta_i \sim \cN(0_d,I_d)$ \Comment{discretized Langevin step}
    \EndIf
    \EndFor
    \State $k\gets k+1$
    \EndWhile
    \State {\bfseries return:} $\{x_i(k)\}_{k=0,i=1}^{k_{\rm max},n}$
\end{algorithmic}
\end{algorithm*}

\subsection{Continuous limit}
To take the continuous limit of~\eqref{eq:51}, we modify this equation in a way that the update of the density is only partial. Specifically, denoting $\rho_t$ the value of the density at time $t\ge0$, we turn this equation into
\begin{equation}
  \label{eq:53}
\begin{aligned}
  \rho_{t+\tau} (x) = \rho_t(x) &+ \alpha \tau \int_\Omega R(x,y)  \left(\rho_*(x) \rho_t(y) -\rho_t(x)   \rho_*(y)\right) dy
\end{aligned}
\end{equation}
where $\alpha>0$ and $\tau>0$ are parameters. This will allow us to make the MCMC resampling updates on par with those of MALA, using $\tau>0$ as timestep in both (see Algorithm~\ref{alg:concurrentA}). Subtracting $\rho_t(x)$ from both sides of~\eqref{eq:53}, dividing by~$\tau$, and letting $\tau\to0$ gives
\begin{equation}
  \label{eq:54}
  \partial_t \rho_{t} (x) = \alpha  \int_\Omega R(x,y)  \left(\rho_*(x)\rho_t(y) -\rho_t(x) \rho_*(y)\right) dy.
\end{equation}
We can now add the Langevin terms that arise in the continuous limit of the compounded MCMC scheme that we use, to arrive at
\begin{equation}
\begin{aligned}
  \label{eq:58b}
  \partial_t \rho_t &= \nabla \cdot(\rho_t \nabla U_* + \nabla \rho_t)+ \alpha  \int_\Omega R(x,y)  \left(\rho_*(x)\rho_t(y) -\rho_t(x) \rho_*(y)\right) dy
  \end{aligned}
\end{equation}
where $\alpha>0$ measures the separation of time scale between the Langevin and the resampling terms. This equation arises in the limit as $\tau\to0$ for the evolution specified in Algorithm~\ref{alg:concurrentA}. Written in term of $g_t = \rho_t/\rho_*$ and $\hat g_t = \hat \rho_t/\rho_*$ (now also allowed to vary with time) Eq.~\eqref{eq:58b} reads
\begin{equation}
  \label{eq:59}
  \begin{aligned}
    \partial_t g_t &= -\nabla U_* \cdot \nabla g_t + \Delta g_t + \alpha \int_\Omega \min(\hat g_t(x),\hat g_t(y)) \left(g_t(y)-g_t(x)\right)\rho_*(y)  dy
  \end{aligned}
\end{equation}

\subsection{Convergence rate} 
Consider the evolution of the Pearson $\chi^2$-divergence of $\rho_t$ with respect to $\rho_*$ defined in~\eqref{eq:chi2} assuming that $D_0<\infty$. Using~\eqref{eq:59} we deduce
\begin{equation}
    \label{eq:chi2dt1}
    \begin{aligned}
    \frac{dD_t}{dt} & = 2\int_\Omega g_t(x) \partial_t g_t (x) \rho_*(x) dx\\
    & = 2\int_\Omega g_t(x) \nabla \cdot(\rho_*(x) \nabla g_t(x))  dx + 2\alpha \int_{\Omega^2} \min(\hat g_t(x),\hat g_t(y)) \left(g_t(y)-g_t(x)\right)g_t(x)\rho_*(x)\rho_*(y)  dxdy\\
    & = -2\int_\Omega  |\nabla g_t(x)|^2 \rho_*(x)  dx - \alpha \int_{\Omega^2} \min(\hat g_t(x),\hat g_t(y)) \left|g_t(y)-g_t(x)\right|^2\rho_*(x)\rho_*(y)  dxdy\\
    &\le - \alpha \int_{\Omega^2} \min(\hat g_t(x),\hat g_t(y)) \left|g_t(y)-g_t(x)\right|^2\rho_*(x)\rho_*(y)  dxdy
    \end{aligned}
\end{equation}
where we used $(-\nabla U_* \cdot \nabla g_t + \Delta g_t)\rho_* = \nabla \cdot(\rho_* \nabla g_t)$ to reexpress the first integral in the second equality. If we denote $\hat G_t = \inf_{x\in\Omega}\hat g_t(x) \in [0,1]$, \eqref{eq:chi2dt1} implies
\begin{equation}
    \label{eq:chi2dt}
    \begin{aligned}
    \frac{dD_t}{dt} & \le - \alpha \hat G_t \int_{\Omega^2} \left|g_t(y)-g_t(x)\right|^2\rho_*(x)\rho_*(y)  dxdy= -2\alpha \hat G_t D_t,
    \end{aligned}
\end{equation}
where we used the normalization conditions $\int_\Omega g_t(x) \rho_*(x)dx  =\int_\Omega \rhohat(x)dx =1$. As a result, using Gronwall inequality we deduce
\begin{equation}
    \label{eq:chi2dts1}
    D_t \le D_0 e^{-\alpha \int_0^t \hat G_s ds}.
\end{equation}
This equation indicates that $D_t\to0$ as $t\to\infty$ as long as $\int_0^t \hat G_s ds \to \infty$. That is, convergence can only fail if $\hat G_t= o(t^{-1})$ as $t\to\infty$, and it is guaranteed otherwise. Convergence is also exponential asymptotically, as long as $\hat G_t$ remains bounded away from 0 as $t\to\infty$.

To get a more explicit convergence rate, let us analyze~\eqref{eq:chi2dts1} in two subcases. First let us assume that the map is not trained, i.e. $\hat g_t(x)= \hat g(x)$ is fixed, and denote $\hat G = \inf_{x\in\Omega}\hat g(x) \in [0,1]$. In this case, \eqref{eq:chi2dts1} reduces to
\begin{equation}
    \label{eq:chi2dtbs}
    D_t \le D_0 e^{-2\alpha \hat G t} \qquad (\hat g_t = \hat g \ \ \text{fixed} ).
\end{equation}
Note that this bound is only nontrivial if $\hat G>0$. Even if that is the case, the rate in~\eqref{eq:chi2dtbs} can be pretty poor if $\hat G$ is very small (e.g exponentially small in the input dimension $d$), which is to be expected if the map is not trained. The best case scenario is of course the idealized situation when $\hat G=1$, which requires that $\hat g =1$ (i.e. $\hat \rho=\rho_*$) because of the normalization conditions $\int_\Omega \hat{g}(x) \rho_*(x)dx  =\int_\Omega \rhohat(x)dx =1$: this case is the continuous equivalent of the one step convergence of the discrete MCMC scheme with resampling from $\rho_*$.

Second let us assume that $\hat g_t = g_t$, that is the trained distribution instantaneously follows the walkers distribution at all times.
In this case,
\eqref{eq:chi2dts1} reduces to
\begin{equation}
    \label{eq:chi2dtcs}
    D_t \le D_0 e^{-2\alpha \int_0^t G_sds} \qquad (\hat g = g_t ),
\end{equation}
where we denote
\begin{equation}
  \label{eq:64}
  G_t=  \inf_{x\in \Omega} \left(\frac{\rho_t(x)}{\rho_*(x)}\right)= \inf_{x\in\Omega} g_t(x)\in [0,1].
\end{equation}
To make this bound explicit, let us consider the evolution of $G_t$. Denoting  $x_t= \argmin_{x\in \Omega}g_t(x)$ so that $G_t = g_t(x_t)$, and using
$\min(g_t(x_t),g_t(y)) = g_t(x_t) = G_t$, $\nabla g_t(x_t)=0$, and $\Delta g_t(x_t) \ge0$ by definition of $x_t$, from~\eqref{eq:59} we have
\begin{equation}
  \label{eq:56}
  \begin{aligned}
    \frac{d G_t}{dt} & = \partial_tg_t(x_t) + \dot x_t
    \cdot \nabla g_t(x_t) \\
    & = \Delta g_t(x_t) + \alpha G_t\int_\Omega \left( g_t(y) -G_t\right) \rho_*(y) dy\\
    & =\Delta g_t(x_t) +\alpha G_t-\alpha G_t^2\\
    & \ge \alpha G_t-\alpha G_t^2
  \end{aligned}
\end{equation}
where we used again the normalization conditions $\int_\Omega  g_t(y) \rho_*(y) dy=\int_\Omega \rho_*(y) dy=1$.  Eq.~\eqref{eq:56} implies that
\begin{equation}
  \label{eq:61}
  \frac{1}{G_t-G_t^2} \frac{d G_t}{dt} \ge \alpha
\end{equation}
which after integration gives
\begin{equation}
  \label{eq:62}
 \log \left(
    \frac{G_t (1-G_0)}{G_0(1-G_t)}\right) \ge \alpha t
\end{equation}
This means that we have
\begin{equation}
  \label{eq:63}
  G_t \ge \frac{G_0}{G_0+(1-G_0)e^{-\alpha t}}.
\end{equation}
Inserting this equation in~\eqref{eq:chi2dtcs} and performing the integral explicitly gives
\begin{equation}
\label{eq:rateexps}
     D_t \le \frac{D_{0} }{\left(G_{0}(e^{\alpha t} -1) +1\right)^2}.
\end{equation}
This bound is only nontrivial if  $G_0\in(0,1]$. By performing all the time integration on $[t_0,t]$ we can obtain \eqref{eq:rateexp}, which is now nontrivial if $G_{t_0}\in(0,1]$.

\section{Augmentation by Resampling via  Birth-Death}
\label{app:bdfp}
 
In the augmented MCMC procedure described in main text, we mix Langevin and Metropolized resampling with the normalizing flow.  Alternatively, we can perform this resampling  via a birth-death process where mass is continuously  transported non-locally by resampling configurations with the push-forward $\rhohat$. This yields the following evolution equation for the sample density $\rho_t$
\begin{equation}
\label{eq:bd0}
    \partial_t\rhot = \nabla\cdot\left( \rhot \nabla U_* + \nabla \rhot \right) - \alpha ( V(\rhot)- \bar V_t) \rhohatt 
\end{equation}
where we defined
\begin{equation}
V(\rho_t) = \frac{\rhot}{\rho_*}, \qquad \text{and} \qquad \bar V_t = \int_\Omega V(\rho_t)\rhohatt dx= \int_\Omega \frac{\rhot\rhohatt}{\rho_*}  dx.
\end{equation}
The interpretation of the added term proportional to $\alpha$ is somewhat subtle: this term amounts to adding or removing mass from $\rhot$ when $\rhot/\rhostar$ is respectively above or below its mean with respect to $\rhohatt$, a process that only stops when $V(\rho_t) = \bar V_t$, i.e. $\rhot=\rhostar$. In practice, this process can be implemented by: \begin{enumerate}
    \item removing samples from our empirical representation of $\rhot$ with rate $\rhohatt/\rhostar$, which accounts for the effect of the term involving $V(\rho_t)$;
    \item reinjecting into $\rhot$ the mass lost this way by generating samples from $\rhohatt$, which accounts for the effect of the term involving $\bar V_t$.
\end{enumerate} 
In terms of $g_t=\rhot/\rhostar$ and $\hat g_t = \rhohatt/\rhostar$, \eqref{eq:bd0} reads
\begin{equation}
\label{eq:bd}
\begin{aligned}
    \partial_t g_t &= -\nabla U_* \cdot\nabla g_t + \Delta  g_t - \alpha ( g_t- \bar V_t) \hat g_t, \qquad  \bar V_t =  \int_\Omega \hat g_t g_t \rho_*  dx,
\end{aligned}
\end{equation}
and we can derive a convergence rate for the solution to this equation by looking at the  evolution of the Pearson $\chi^2$-divergence of $\rho_t$ with respect to $\rho_*$ defined in \eqref{eq:chi2}. Treating the Langevin terms as we did in Eq.~\eqref{eq:chi2dt1} we obtain
\begin{equation}
    \label{eq:chi2bd}
    \frac{dD_t}{dt} \le -2\alpha \int_\Omega (g_t - \bar V_t)  g_t\hat g_t \rho_* dx = -2\alpha \int_\Omega   g^2_t\hat g_t \rho_* dx + 2\alpha \bar V_t^2 = -2\alpha \int_\Omega   \left|g_t-\bar V_t\right|^2 \hat g_t \rho_*dx.
\end{equation}
Since $\int_\Omega \bar V_t ( g_t -1)\rho_* dx =0$, the Pearson $\chi^2$-divergence can also be expressed as
\begin{equation}
    D_t = \int_\Omega (g_t-\bar V_t) ( g_t -1)\rho_* dx
\end{equation}
using Cauchy-Schwartz inequality we obtain
\begin{equation}
    D^2_t \le E_t \int_\Omega |g_t-\bar V_t|^2 \hat g_t\rho_* dx  \qquad \text{where} \qquad E_t = \int_\Omega  \frac{| g_t -1|^2}{\hat g_t}\rho_* dx.
\end{equation}
Using this inequality in~\eqref{eq:chi2bd} we deduce
\begin{equation}
    \label{eq:chi2bd2}
    \frac{dD_t}{dt} \le -2\alpha E_t^{-1}  D_t^2 
\end{equation}
which implies
\begin{equation}
    \label{eq:chi2bd3}
    D_t \le \frac{D_0}{1+2\alpha D_0 \int_0^t E_s^{-1} ds}.
\end{equation}
This equation implies convergence as long as $\int_0^t E_s^{-1} ds\to\infty $ as $t\to\infty$. In particular, if $\sup_{t\ge0} E_t \le M < \infty$, assuming $M>0$, we have
\begin{equation}
    \label{eq:chi2bd4}
    D_t \le \frac{D_0}{1+2\alpha D_0 M^{-1} t}.
\end{equation}
Since
\begin{equation}
    E_t \le \int_\Omega  \frac{ g^2_t }{\hat g_t}\rho_* dx + \int_\Omega  \frac{ 1 }{\hat g_t}\rho_* dx = 
    \int_\Omega  \frac{ \rho^2_t }{\hat \rho_t} dx + \int_\Omega  \frac{ \rho_*^2 }{\hat \rho_t} dx
\end{equation}
we see that Eq.~\eqref{eq:chi2bd4} holds as long as the Pearson $\chi^2$-divergences of $\rho_t$ with respect to $\hat \rho_t$ and $\rhostar$ with respect to $\hat \rho_t$ remain bounded at all times.

A more concrete convergence rate can be obtained if we assume again that $\rhohatt=\rhot$ (i.e. $\hat g_t = g_t$). In this case
\begin{equation}
    E_t = \int_\Omega \frac{\rho_*}{g_t} dx - 1 = \int_\Omega \frac{\rho^2_*}{\rho_t} dx - 1
\end{equation}
i.e. it is the Pearson $\chi^2$-divergences of $\rhostar$ with respect to $\rhot$.  Since this divergence is known to be non-increasing if $\rho_t$ solves a master equation with $\rho_*$ as fixed point (see below for a proof), we have $E_t\le E_0$ for $t\ge0$, which implies that 
\begin{equation}
\label{eq:rate}
    D_t \le \frac{D_0}{1+2\alpha D_0 E^{-1}_0 t}.
\end{equation} 
This bound is only nontrivial if $D_0<\infty $ and $E_0<\infty$. More generally if there exists a $t_0\ge0$ such that $D_{t_0}<\infty $ and $E_{t_0}<\infty$, a similar argument gives
\begin{equation}
\label{eq:ratet0}
   \forall t\ge t_0 \quad : \quad  D_t \le \frac{D_{t_0}}{1+2\alpha D_{t_0} E^{-1}_{t_0} (t-t_0)}.
\end{equation} 
This uniform bound asymptotically reduces to
\begin{equation}
\label{eq:rateasympt}
    D_t \leq \tfrac12 E_{t_0} (\alpha t)^{-1} \quad \text{for} \ \ 2\alpha D_0 E^{-1}_{t_0} t \gg1.
\end{equation}
The constant $E_{t_0}$ emphasizes the importance of having initial samples close to metastable basins of the target distribution to guarantee that $E_{t_0} <\infty$.  This further highlights that the methods we explore here are not generically suitable for exploring distributions for which the metastable basins are not known (cf. Appendix~\ref{app:exploration_bounds}).

\paragraph{Verification that $E_t=\int_\Omega \rho_*^2/\rho_tdx$ is non-increasing.} We have
\begin{equation}
  \label{eq:43aa}
  \begin{aligned}
    \frac{dE_t}{dt}  & = -2
    \int_\Omega V^{-2}(\rho_t) \partial_t \rho_t dx\\
    & = -2
    \int_\Omega V^{-2}(\rho_t) \nabla \cdot(\rho_* \nabla V(\rho_t))
    dx + 2 \alpha \int_\Omega V^{-2}(\rho_t) (V(\rho_t) - \bar V_t)
    \hat \rho_t dx\\
    & = -4
    \int_\Omega V^{-3}(\rho_t) |\nabla V(\rho_t)|^2 \rho_*
    dx - 2 \alpha \int_\Omega V^{-1}(\rho_t) \hat \rho_t dx  + 2\alpha
    \bar V_t \int_\Omega V^{-2}(\rho_t)\hat \rho_t dx\\
    & \le  2 \alpha \int_\Omega V^{-1}(\rho_t) \hat \rho_t dx  - 2\alpha
    \bar V_t \int_\Omega V^{-2}(\rho_t)\hat \rho_t dx.
  \end{aligned}
\end{equation}
Using Jensen's inequality, 
\begin{equation}
  \label{eq:41aa}
  \int_\Omega V^{-2}(\rho_t)\hat \rho_t dx\ge \left(\int_\Omega
    V^{-1}(\rho_t)
    \hat \rho_t dx\right)^2
\end{equation}
and so we can deduce from the last inequality in~\eqref{eq:43aa} that
\begin{equation}
  \label{eq:44aa}
  \begin{aligned}
    \frac{d E_t}{dt} & \le 2 \alpha
    \int_\Omega V^{-1}(\rho_t) \hat \rho_t dx - 2\alpha \bar V_t
    \left(\int_\Omega V^{-1}(\rho_t)\hat \rho_t dx\right)^2\\
    & =  2 \alpha\left(
    \int_\Omega V^{-1}(\rho_t) \hat \rho_t dx\right)\left(1 - \bar V_t
    \int_\Omega V^{-1}(\rho_t)\hat \rho_t dx\right).
  \end{aligned}
\end{equation}
Using Jensen's again, 
\begin{equation}
  \label{eq:40aa}
  \int_\Omega V^{-1}(\rho_t) \hat \rho_t dx \ge \left(\int_\Omega
    V(\rho_t) \hat \rho_t dx\right)^{-1} =\bar V_t^{-1}
\end{equation}
which implies
\begin{equation}
  \label{eq:40baa}
  1-\bar V_t \int_\Omega V^{-1}(\rho_t) \hat \rho_t dx \le 0.
\end{equation}
Using this equation in~\eqref{eq:44aa}, we deduce
\begin{equation}
  \label{eq:45aa}
  \frac{d E_t}{dt}   \le 0.
\end{equation}

\begin{figure}
    \centering
    \includegraphics[width=0.95\linewidth]{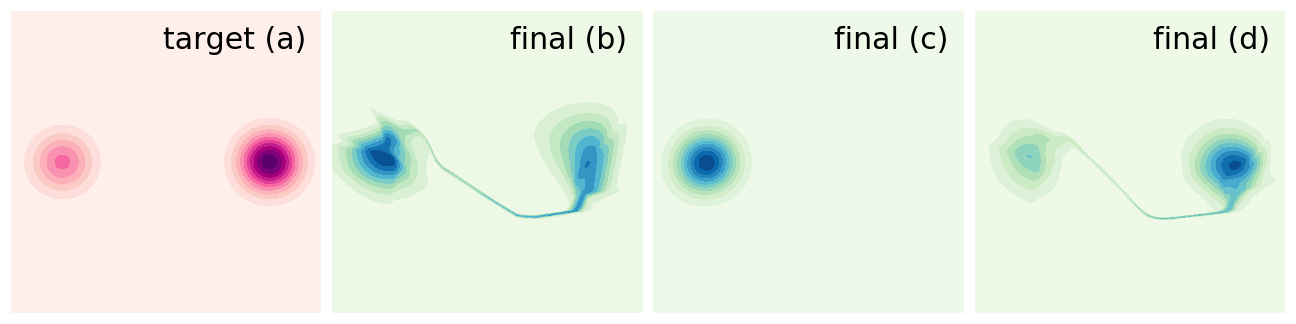}
    \includegraphics[width=0.3245\linewidth]{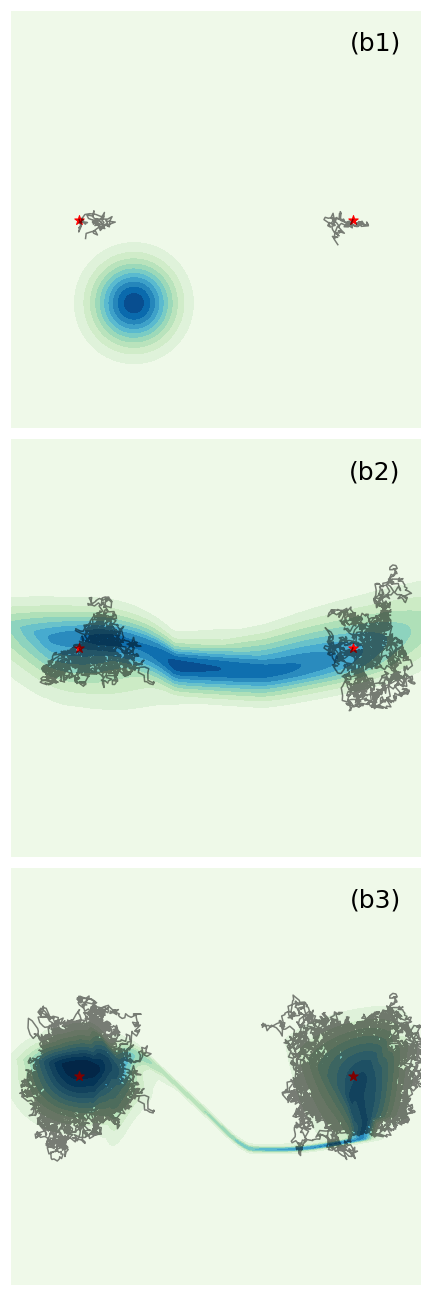}
    \includegraphics[width=0.3245\linewidth]{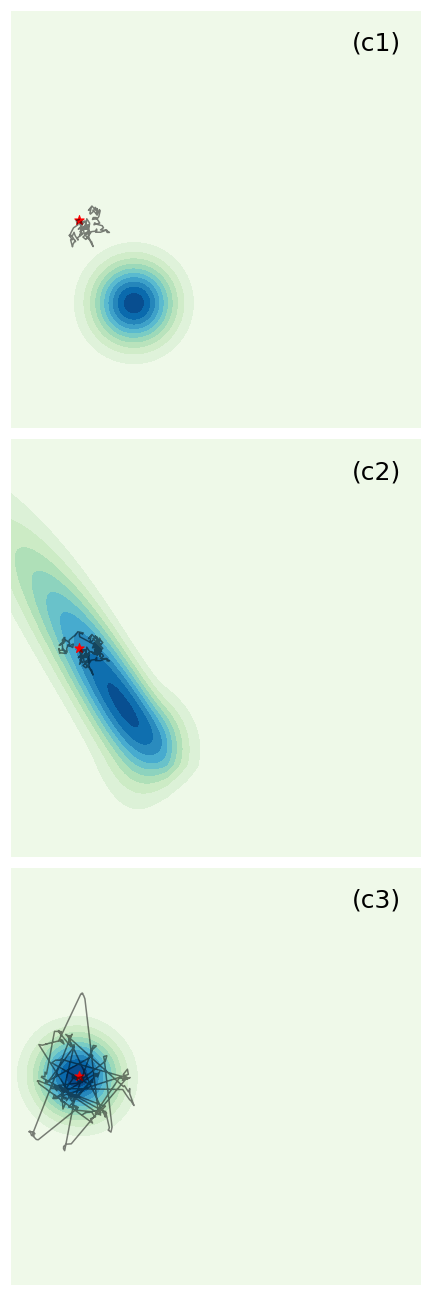}
    \includegraphics[width=0.3245\linewidth]{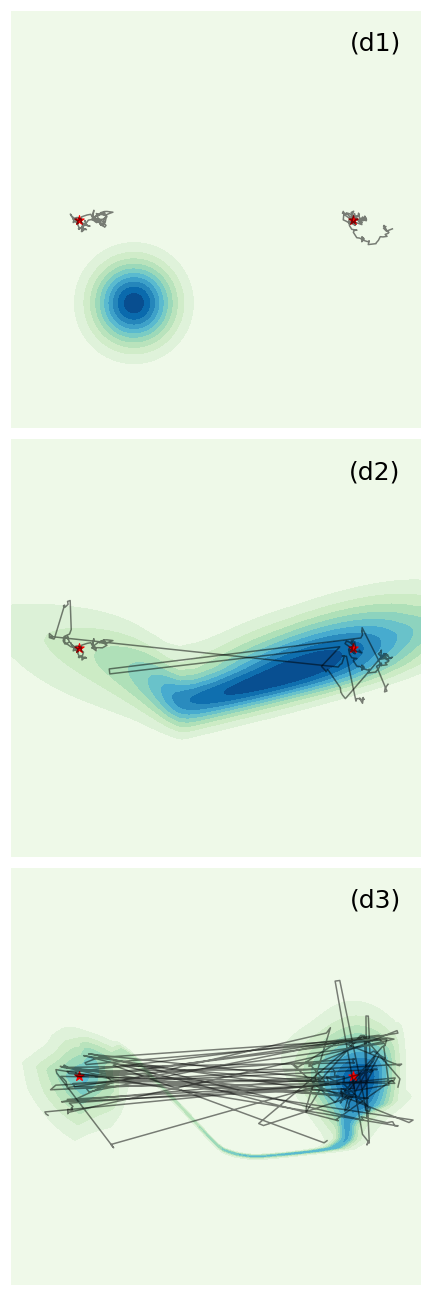}
    \caption{{\bf Advantage of resampling and necessity of good initialization.} (a) Target 2d mixture of two Gaussians. In all other panels, the push-forward density modelled by the normalizing flow is shown in blue. The trajectory of representative walkers of the concurrent sampling and training is plotted in black from the initialization marked by a red star. See Appendix \ref{sec:a:Gauss} for details on the scenarios. (b) Relative weights of components are not recovered in final density learned using only local sampling with equal number of walkers starting in both components. (b1) through (b3) Early, middle, and later stage of training in this setting. The trajectory of walkers starting at the red star does not mix between modes. (c) Even if the combined sampling method is used, with nonlocal resampling, if walkers are initialized only in one mode, the second one is not learned by the flow. (c1) through (c3) As the flow learns to model one mode some non local moves are accepted by the walker but no proposition allows mixing with and modelling of the second mode. (d) When using the nonlocal resampling and proper initialization in both modes, the final learned density properly models the statistical weights of both modes. (d1) through (d3) The progressive learning of both modes by the flow allows more and more efficient mixing along the training.}
    \label{fig:gaussian_mixture}
\end{figure}

\section{Sampling unvisited metastable states}
\label{app:exploration_bounds}

Once the normalizing flow $T$ is trained to sample a given metastable basin, it is generally not suitable to discover unvisited metastable basins (i.e., those not in the data set).
We illustrate this phenomenon by considering the probability of sampling a configuration near $\phi_{-}$, the negative spin solution of the stochastic Allen-Cahn model with a map $T_+$ that has been training only on samples from the $\phi_+$ metastable solution.
First, we note that our analysis of the maps (Fig.~\ref{fig:phi4}) illustrates that the map is local; in this case, it means that when $N$ is large (and hence we neglect the boundary terms, which will not change the overall scaling of the calculation that follows), the map will essentially be additive
\begin{equation}
    T_+(\phi_i) = \phi_i + 1.
\end{equation}
With this trained map $T_+$, it is straightforward to see that the probability of sampling a configuration in the $\phi_-$ basin is actually worse than it would be with the untrained identity map.
Denoting a configuration $\phi = \{ \phi_i \}_{i=1}^N$, we have
\begin{equation}
    P_{\rm B}(\bar T_+(\phi)) = \exp \left(-\frac{a \beta}{2\Delta s} \sum_{i=1}^{N+1} (\phi_i - \phi_{i-1})^2 - \frac{\beta \Delta s}{4a} \sum_{i=1}^N (\phi_i-1)^2 \right),
\end{equation}
where the inverse map in this case simply subtracts 1 and has unit Jacobian.
When $N$ is large, the probability of sampling in the $\phi_-$ metastable basin can be estimated by a Laplace approximation and it is easy to see that the integral over the possible states in the basin will be dominated by the configuration in which all the spins are aligned. 
Hence, we can estimate this probability by computing the probability of the configuration in which $u_i = -1$ for all $i=1, \dots, N,$ which we call $u_-.$
This probability decays exponentially with the dimension, 
\begin{equation}
    P_{\rm B}(\bar T_+(\phi_-)) = \exp \left(- \beta a^{-1} \Delta s N \right).
\end{equation}
In the stochastic Allen-Cahn model, we take $\Delta s = 1/N$ so that the free energy barrier between the two metastable configurations is intensive (it should be noted that free energy barriers are typically extensive with system size for phase transitions).
When $\beta b$ is large ($\beta=20$, $a=1/10$ in our examples) it requires a rare event to spontaneously sample a configuration away from the metastable basin present in the original data. 
While this calculation is specific to the Allen-Cahn model, we have observed the same trends in other test systems, suggesting that, generically, learning about a given metastable basin reinforces bias for those samples and could hinder exploration of other basins of a target probability distribution. 
Taken together, these observations emphasize the importance of \emph{a priori} identification of metastable states of interest.

\begin{figure}
\begin{minipage}{\linewidth}
    \centering
    \includegraphics[width=0.99\textwidth]{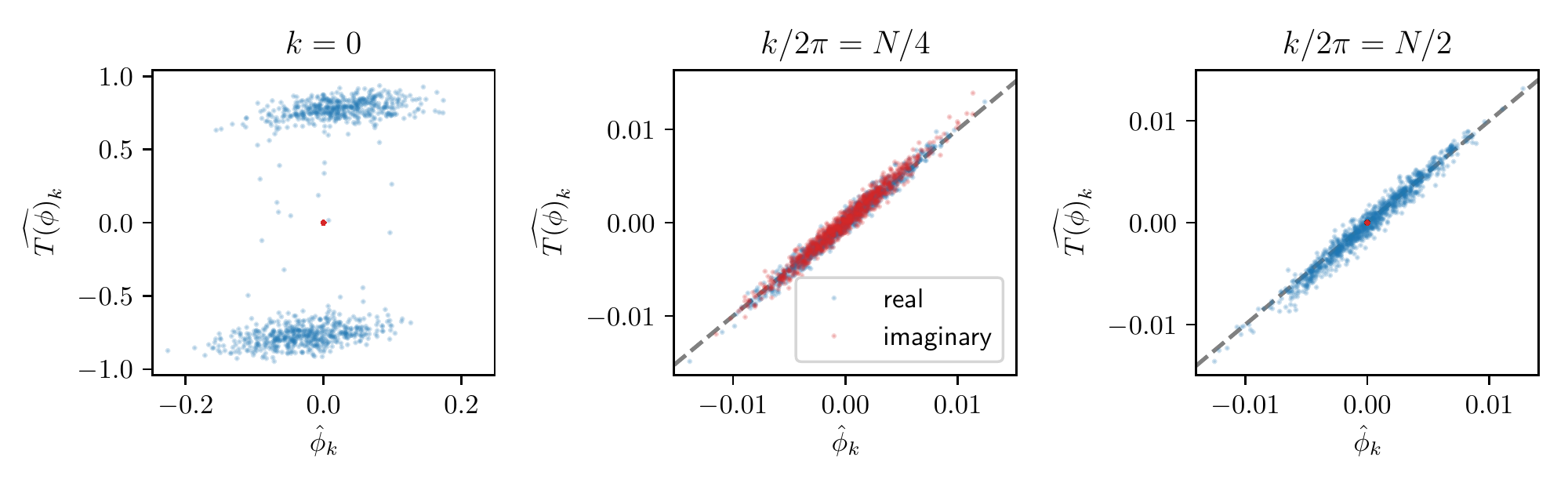}
    \caption{{\bf Diagnostic of the learned map $T$ to sample the stochastic Allen-Cahn model on Fourier modes.} Correspondence between Fourier coefficients of base samples $\hat \phi_k$ with Fourier coefficients of the corresponding mapped fields $\hat T(\phi)_k$, at low $k = 0$, intermediate $k= N\pi/2$ and high $k= N\pi$ frequency. Plots are done with 1000 independent samples from the base measure.}
    \label{fig:phi4-map-diagnostics}
\end{minipage}
\begin{minipage}{\linewidth}
    \centering
    \includegraphics[width=0.99\linewidth]{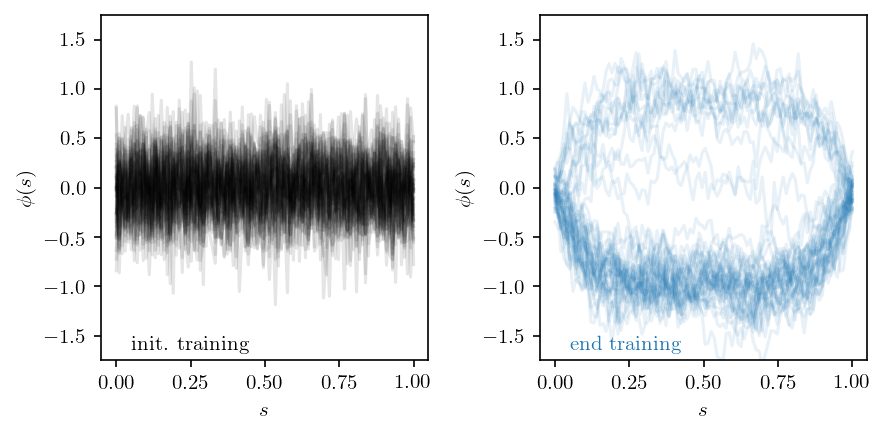}
    \caption{{\bf Failure of the uniformed prior.} Push-forward samples from the ``uniformed'' base distribution through the map $T$ at the beginning and at the end of training. The initial walkers where initialized at $10\%$ in $\phi_+$ and $90\%$ in $\phi_-$. Final samples are roughly locating $\phi_+$ and $\phi_-$. Yet there is no rebalancing of the modes ($50\%$ in each mode because of the Hamiltonian invariance to $\phi \to -\phi$) since almost no generated samples is accepted. }
    \label{fig:phi4-uniformed-samples}
\end{minipage}
\end{figure}

\section{Normalizing flow architecture}
\label{sec:a:NF}
We parametrize the map $T$ as a RealNVP \cite{dinh_density_2017}, for which inverse and Jacobian determinant can be computed efficiently. Its building block is an invertible affine coupling layer updating half of the variables, 
\begin{align}
    x^{(k+1)}_{1:d/2} = e^{s\left(x^{(k)}_{d/2:d}\right)} * x^{k}_{1:d/2} + t\left(x^{(k)}_{d/2:d}\right)
\end{align}
where $s(\cdot)$ and $t(\cdot)$ are learnable neural networks from $\mathbb{R}^{d/2}\to \mathbb{R}^{d/2}$, $*$ is a component-wise multiplication, and $k$ indexes the depth of the network. 

In all the scenarios of the Gaussian mixture model (Sec.~\ref{sec:a:Gauss}), we used RealNVPs with 6 pairs of coupling layers. All the translation $t(.)$ and scaling neural networks $s(.)$ are multi-layer perceptrons with depth 3 and hidden layers of size 100. For the optimization, we used batches of 400 samples from 40 independent walkers updated 10 times to compute gradients and update paramters with Adam for 1500 iterations with a learning rate of 0.005.

For the unimodal wiggle experiment (Sec.~\ref{sec:a:wiggle}), we wish to verify what happens when the normalizing flow parmaetrization lacks expressiveness. We used RealNVPs with 1 pair of coupling layers, each with 2 multilayer-perceptrons with 2 hidden layers of 10 units as translation and scaling networks. The base distribution is an Gaussian with small covariance compared to the scale of the wiggle bulk density. At initialization the wiggle is shifted from zero such that there is ``no overlap'' between push-forward and target. 

Experiments reported on the Allen-Cahn model (`Numerical experiments' section in main text and Sec.~\ref{sec:a:SAC}) use 
RealNVPs with 10 pairs of coupling layers and again ReLU multi-layer perceptrons with depth 3 and width 100 for scalings and translations.
During concurrent training/sampling, we gather batches of 1000 samples from 100 independent walkers updated 10 times to compute each gradient step on the parameters of the normalizing flow. The optimization is done using Adam for $10^5$ iterations with a learning rate of 0.001. The parameters of the RealNVPS used to generate Fig. \ref{fig:phi4}, Fig. \ref{fig:phi4-training-mixing}, Fig. \ref{fig:phi4-map-diagnostics}, Fig. \ref{fig:phi4-uniformed-samples} and Fig. \ref{fig:phi4-tilted} are given in the git repository listed below.

For the nonequilibrium transition path sampling experiments (`Numerical experiments' section in main text and Sec.~\ref{sec:a:TPS}), we used a RealNVP architecture with 10 pairs of coupling layers and MLPs of depth 3 with hidden layers of 100 neurons.
The parameters of the models used to generate configurations in Fig.~\ref{fig:noneq}, Fig.~\ref{fig:resampling}, and Fig.~\ref{fig:resampling} are included in the git repository.
We optimized the network using a batch size of 200 and 1 Langevin step per normalizing flow step.
We optimized the parameters using the Adam optimizer with a learning rate of $5\times 10^{-4}$ with parameters $\beta_1 = 0.9$, $\beta_2 = 0.999$ and $\epsilon=1\times10^{-8}.$
We carried out the optimization for a total of $10^5$ steps.

\section{Additional tests}
\label{sec:a:add}

\subsection{Two-dimensional Gaussian mixture example}
\label{sec:a:Gauss}
We illustrate success and failures of concurrent training/sampling on the simple example of a Gaussian mixture in two dimensions with two modes (see Fig. \ref{fig:gaussian_mixture}(a)). Both components have covariance $\sigma I$ with $\sigma=1$ and their means are separated by 10$\sigma$. The rightmost component is twice as likely as the leftmost. We choose the standard normal distribution as base measure $\rhoB$. At initialization the push-forward $\rhohat$ is approximately equal to $\rhoB$ and does not overlap with either of the target modes.  
We compare three scenarios: (b), (c) and (d) in Fig. \ref{fig:gaussian_mixture}. In scenario (b), we run Alg. \ref{alg:concurrent} omitting lines \ref{lst:resample1}-\ref{lst:resample3}, that is using exclusively the local sampling kernel, starting from an equal number of chains initialized in each mixture component. As the local chains fail to mix between modes, neither the learned density nor the samples reflect the relative statistical weights of the target. In scenario (c), Alg. \ref{alg:concurrent} is run properly but with initial chains only in the left-most mode. The right-most mode is never sampled by the procedure. Finally in scenario (d) Alg. \ref{alg:concurrent} is run with chains initialized in both modes and accurate and efficient sampling is achieved.
For all cases above, batches of 400 samples from 40 independent walkers updated 10 times are used to compute each gradient step on the parameters of the normalizing flow. The optimization is done using Adam for 1500 iterations with a learning rate of 0.005. In scenario (d), acceptance rate of the push-forward proposals reaches $80-85\%$. The residual connection between modes in (d3) originates from the transformation of the uni-modal base measure into the bi-modal target. Additional iterations would make it thinner. These artefacts can be eliminated using stochastic normalizing flows \cite{wu_stochastic_2020}, for which however the push-forward probability is not analytically tractable anymore.

\subsection{Two-dimensional wiggle}
\label{sec:a:wiggle}
\begin{figure}
    \centering
    \includegraphics[width=0.32\linewidth]{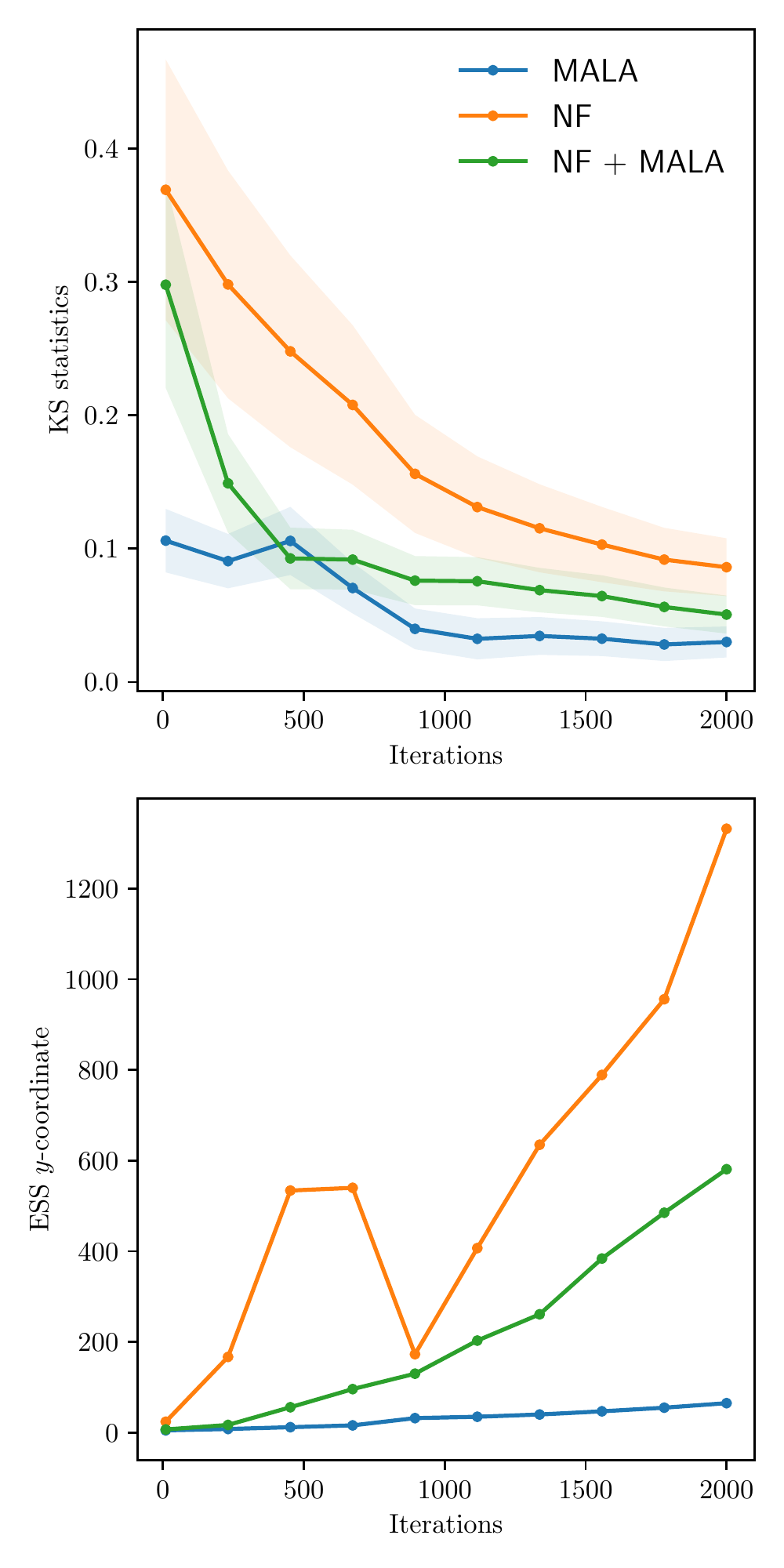}
    \includegraphics[width=0.67\linewidth]{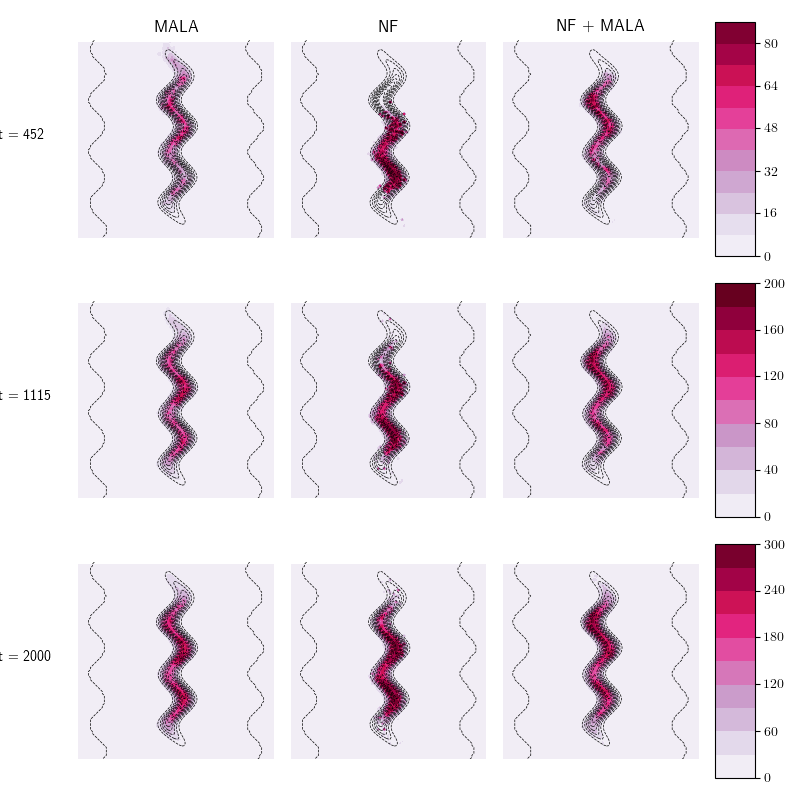}
    \caption{{\bf Advantage of keeping a local sampler along with the global normalizing flow based kernel.} The performance of three sampling kernels for a 2d structured unimodal distribution: Metropolis Adjusted Langevin (MALA),  adaptive MCMC with independent Metropolis normalizing-flow based kernel (NF), and finally (NF + MALA) which alternated between 1 MALA update and 1 NF update. At the top left, Kolmogorov-Smirnov statistics (KS) track absolute maximum difference between empirical 1d marginal densities and ground truth over $100$ random projections. (NF + MALA) has lower KS than (NF) for same iteration number emphasizing the advantage of retaining a local part to the sampling when the learned push-forward is not precise enough. At the bottom left, the estimate of effective sample size ESS-$y$ (keeping all samples expect first $100$ burn-in samples) suggests that the nonlocal kernel decorrelates the samples faster than (MALA) even for unimodal distribution. However, note that estimating ESS before chains convergence can be misleading, as the correction experienced by the (NF) ESS when the upper half is discovered reminds us. We also report 2d histograms of samples for the three methods at three different iteration numbers. Color levels corresponds to number of samples per bin (see colorbars at the right for each iteration number), and dotted lines correspond to contour lines for target density. (NF) is taking many more iterations to visit the tails of the distribution compared to (NF + MALA). }
    \label{fig:wiggle}
\end{figure}

To further investigate the benefit of combining the retaining update steps with a local sampler in between updates with the global samplers based on a normalizing flow we examine another visual two-dimensional example. The target distribution is a unimodal two-dimensional ``wiggle'' (see densities panels in Fig.~\ref{fig:wiggle}). Using for each of them $10$ walkers initialized at the middle of the ``wiggle'', we compare again three scenarios : (MALA) where the walkers are evolving only under the local MALA kernel, (NF) where the MALA steps from the adaptive MCMC are switched off and any jump comes from proposals of a learning normalizing flow, and finally (MALA + NF) which corresponds our adaptive MCMC where the walkers are alternatively updated with MALA (1 iteration) and the normalizing flow proposal (1 iteration). The architecture and initialization of the normalizing flows used in (NF) and (MALA +NF) are identical and relatively modest with only one pair of coupling layers, and starting with a low variance initial push-forward that does not overlap the ``wiggle'' core probability-mass region. This configuration emulates the common situation in higher-dimensions that the learned push-forward density imperfectly matches the target density. 

We monitor the quality of samples using Kolmogorov-Smirnov (KS) statistics, an estimated effective sample size (ESS) of the $y$-coordinate of the chains and 2d histograms of the empirical densities. We evaluate these observables at iteration $t$ by considering all the visited states up to $t$, discarding the first $100$ iterations of the chains. The KS statistics corresponds to the absolute maximum difference between empirical 1d marginal densities and ground truth marginalized 1d densities (computed by quadrature) over $100$ random projections. We report means and standard deviation over these random directions. A perfect sampling will correspond to KS statistics concentrating around 0. 

Comparing the three scenarios, we can see that (MALA) more accurately samples the 2d distribution (KS statistics and empirical densities). Seemingly, (NF) has the largest effective sample size. However the estimation of the ESS is not stable until the chains become stationary and the ESS can be a biased measure of success. Looking at (NF) empirical densities, middle column, the walkers have only visited at first a lower portion of the wiggle and the ESS is overestimated. It drops when the walkers start visiting the upper region and the estimation gets corrected (orange line in ESS chart).  On the other hand, (NF+MALA) allows a better coverage of the tails than (NF) at the same iteration number. As it gets close to stationarity , the ESS estimate gets more credible suggesting that (NF+MALA) produces effectively less correlated samples than (MALA).

\section{Stochastic Allen-Cahn experiments}
\label{sec:a:SAC}
In our experiments the energy~\eqref{eq:phi4} is discretized on a grid with $N$ points. The configuration is then a set of coupled continuous spins $\{ \phi_i \}_{i=1}^N$ according to the discretized Hamiltonian
\begin{equation}
    U_*(\phi) = \frac{a\beta}{2\Delta s}  \sum_{i=1}^{N+1} (\phi_i-\phi_{i-1})^2 + \frac{\beta b \Delta s}4 \sum_{i=1}^N (1-\phi^2)^2
\end{equation}
with $\Delta s = 1/N$,
and Dirichlet boundary conditions $\phi_0 = \phi_{N+1} = 0$. The discretized Hamiltonian of the base distribution similarly reads
\begin{equation}
    U_{\rm B}(\phi) = \frac{\beta a}{2 \Delta s} \sum_{i=1}^{N+1} (\phi_i-\phi_{i-1})^2 + \frac{\beta b \Delta s}{2} \sum_{i=1}^N \phi_i^2,
\end{equation}
with the same boundary conditions. Unless otherwise stated, we take $a=0.1$, $b=1/a=10$, inverse temperature $\beta=20$ and discretization $N=100$.

\begin{figure*} 
    \centering
    \includegraphics[width=\textwidth]{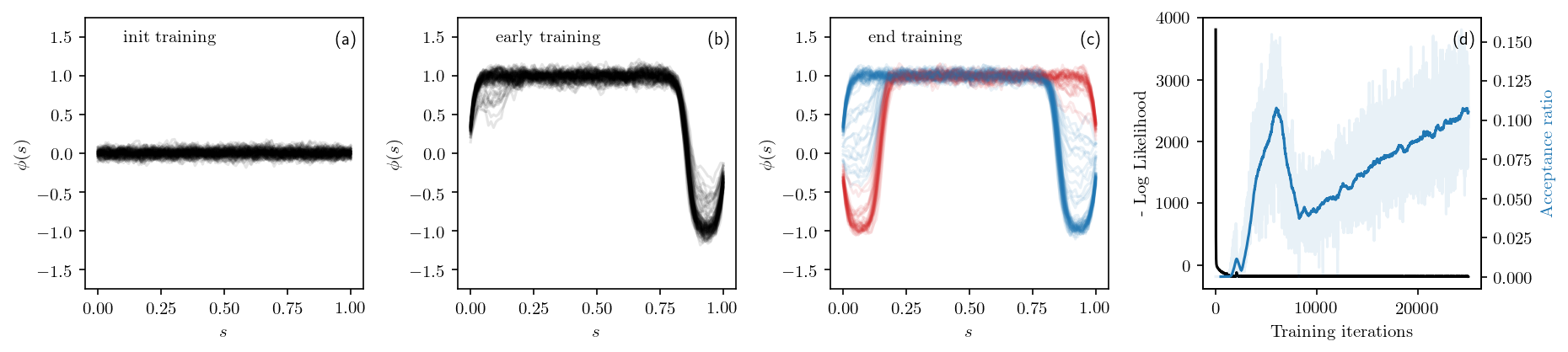}
    \caption{{\bf Concurrent training and sampling of the stochastic Allen-Cahn model with tilted Hamiltonian (\ref{eq:phi4-tilt}).} (a,b,c) 100 independent samples from the push-forward $\rhohat$ at different stage of training. The map gradually adjusts from generating samples close to the base measure samples to generating samples featuring a domain wall corresponding to the two modes with $\bar \phi = 0.7$. Colors are used to help distinguish between them. (d) In black, left axis, the negative log likelihood decreases rapidly along the gradient descent steps. In blue, right axis, the acceptance ratio features a drop in the middle of training when the map shifts from concentrating on a single mode to properly capture both states. The acceptance rate is of roughly $10\%$ at the end of this run and could likely be further improved by pursuing the gradient descent.}
    \label{fig:phi4-tilted}
\end{figure*}

During concurrent training/sampling, we gather batches of 1000 samples from 100 independent walkers updated 10 times to compute each gradient step on the parameters of the normalizing flow. The optimization is done using Adam for $10^5$ iterations with a learning rate of 0.001. In Fig.~\ref{fig:phi4-training-mixing} we report the evolution of the objective function, the acceptance ratio in the Metropolis steps and the autocorrelation time of the sampling chains during training, as well as the configurations visited by a single walkers in 10 steps at the end of training, demonstrating fast mixing. Fig.~\ref{fig:phi4-map-diagnostics} examines Fourier modes of samples from the push-forward $\hat \rho$, at low and high frequencies, with respect to the Fourier decomposition of the base measure sample from which they originate. At high frequencies, the map implements the identity.

The most common implementations of normalizing flow use instead a standard Gaussian distribution for the base. In the continuous limit, the Hamiltonian of this ``uniformed'' base distribution is
\begin{align}
    \label{eq:uninformed_base}
    U_{\rm BU} = \frac {\beta } {2a}\int_0^1  \phi^2(s) ds,
\end{align}
with the coupling term of the target~\eqref{eq:phi4}. This defines a white-in-space process which gives back the standard multivariate Gaussian when discretized as $U_{\rm BU}(\phi) = \frac12\beta a^{-1}\Delta s \sum_{i=1}^N \phi_i^2$. Poor performance are obtained when using $U_{\rm BU}$ as shown on Fig.~\ref{fig:phi4-uniformed-samples} for $N=100$. As the discretization is refined, it gets harder and harder for the map to account for the fine scale structure when it is not encoded in the base distribution. Even after a long run of Alg.~\ref{alg:concurrent} generated samples are not probable. As a result almost none is accepted in the resampling step and there is no mixing between modes.

Finally, we also demonstrate the possibility to sample unlikely configurations with the presented algorithm using a biased measure of the stochastic Allen-Cahn model. As an example we consider the Allen-Cahn Hamiltonian (\ref{eq:phi4}) tilted towards configurations with a given spatial average $\bar \phi$, 
\begin{equation}
    U_{*, \lambda, \bar \phi}[\phi] = U_{*}[\phi] + \lambda \beta \left(\int_0^1 \phi (x) dx - \bar \phi\right)^2,
    \label{eq:phi4-tilt}
\end{equation}
where $\lambda$ is a positive Lagrange multiplier. By setting $\bar \phi$ away from $\pm 1$, the tilt favors configurations with domain walls. In the experiment displayed on Fig. \ref{fig:phi4-tilted}, for which $a=0.2$, $b=1/a$, $\beta=200$, $\bar \phi = 0.7$ and $\lambda = 10^4$, we can see that the normalizing flow learns to generate the two symmetric configurations with one domain wall thermodynamically favored by the tilted measure.

\section{Nonequilibrium transition path sampling experiments}
\label{app:tps}

The success of the algorithm on the Allen-Cahn SPDE model suggests that it could be a useful strategy for sampling transition paths on complex potential energy surfaces even in the presence of nonconservative driving forces.
Transition path theory~\cite{e_transition_2005,e_transition-path_2010} and path sampling techniques~\cite{bolhuis_transition_2002,allen_forward_2009} allow for space-time local sampling, but it is generically difficult to compute relative weights between transition paths, especially when multiple transition channels exist.  Nevertheless, these problems are important for determining reaction mechanisms and reaction rates for chemical reactions, a topic which increasingly is being studied in nonequilibrium settings~\cite{falasco_dissipationtime_2020,kuznets-speck_dissipation_2021}. 
Here, we investigate the ability of our augmented Monte Carlo dynamics to sample the transition paths of reactions involving multiple channels.

\paragraph{Path measure.} We consider a system whose instantaneous position $x\in\RR^d$ is the solution of the stochastic differential equation
\begin{equation}
\label{eq:nsde}
   dx(t) = b(x(t)) dt + \sigma dW(t),
\end{equation}
where $b$ is the drift, $W(t)\in \RR^d$ is a Wiener process, and $\sigma>0$ is a parameters measuring the intensity of the noise.
When $b$ is non-conservative, the dynamics is not in detailed balance and the stationary distribution of the dynamics is non-Boltzmann. Few robust methods are currently available to compare relative path weights when there is metastability in space-time~\cite{metzner_illustration_2006-1}.

\label{sec:noneq}
\begin{figure}
    \centering
    \includegraphics[width=\linewidth]{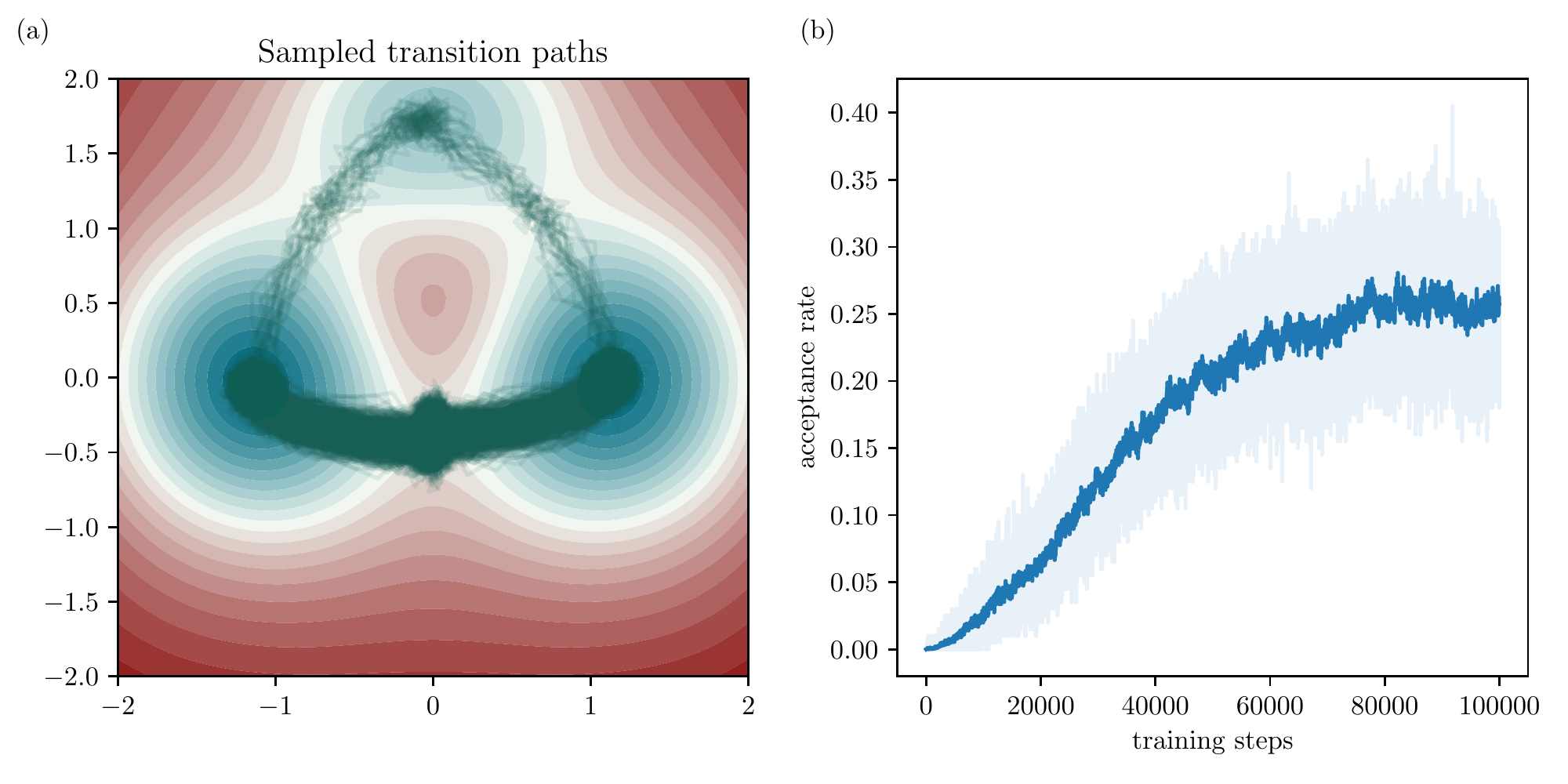}
    \caption{{\bf Sampling nonequilibrium transition paths over an entropic path-space barrier with augmented path-space MCMC.} (a) 1000 paths (solutions to \eqref{eq:nsde} with end-point conditioning) sampled with $\hat\pi.$ The potential energy surface is shown as contours. (b) The acceptance rate of samples generated with the flow map as a function of optimization time is shown with the rolling average over the last 50 time steps plotted above.}
    \label{fig:noneq}
\end{figure}

Denoting by  $x[0,\tmax]$ a trajectory on $t\in [0,\tmax]$, we seek to sample such trajectories conditioning on the end points $x(0)=x_A$ and $x(\tmax) = x_B$.
Using a path integral type expression for the probability density in path space, we can formally write 
\begin{equation}
    \mathbb{P}_{*}(x[0,\tmax]) \propto \exp\left[-\frac1{2\sigma^2}
    \int_0^{\tmax} |\dot{x}(t) - b(x(t))|^2 dt \right].
    \label{eq:pm}
\end{equation}
We emphasize here that while the above expression is only formal, the relative path measures that we work with can be made rigorous using the Girsanov theorem, cf. Appendix~\ref{app:measures}. In fact, it is easy to see that the path measure in Eq.~\eqref{eq:pm} is similar to the one for the random field considered in the previous section: the particle position $x(t)$ plays  the role of the field $\phi(s)$ and the time $t\in[0,\tmax]$ that of the spatial-coordinate $s\in[0,1]$. As in the case of the stochastic Allen-Cahn model, we must therefore take care to ensure that the base measure is well-adapted to the target.
In the infinite-dimensional Hilbert space on which the path measure is defined, the naive base measure typically used for a normalizing flow will be orthogonal to Eq.~\eqref{eq:pm}. As a consequence, the relative entropy used to train the map $T$ (Eq.~\eqref{eq:E1}) can be infinite~\cite{hairer_introduction_2009}.

\paragraph{Brownian bridge as base measure.}
The probability measure associated with the Brownian bridge process offers a simple solution to this problem.
We write the path integral expression for the statistical weight of a Brownian bridge path $x_{[0,T]}:$
\begin{equation}
    \mathbb{P}_{\rm B}(x_{[0,\tmax]}) \propto \exp\left[-\frac1{2\sigma^2} \int_0^{\tmax} |\dot{x}_t|^2 dt \right],
\end{equation}
with the boundary conditions $x_0=x_A$ and $x_{\tmax} = x_B$.
The choice fulfills the key requirement  that the nonequilibrium path measure is absolutely continuous with respect to the base measure (cf. Appendix~\ref{app:measures}).

\paragraph{Numerical implementation and results.} To carry out the simulations, we must discretize the path into $N$ points with a time step $\Delta t$. 
We then obtain the target path action
\begin{equation}
    S_* = \frac{\Delta t}{2\sigma^2} \sum_{i=1}^{N-1} \left| \frac{x_{i+1}-x_i}{\Delta t} - b(x_i)\right|^2
\end{equation}
where the factor of $\Delta t$ in the effective temperature ensures proper scaling in the limit $\Delta t \to 0$; the  action $S_*$ plays the role of $U_*$ in the example above.
We impose the boundary conditions $x_0=x_A$ and $x_N=x_B$.
An analogous expression for the path action associated with the base measure simply lacks the contribution from the drift, $b(x_i)$. 
We update the paths locally in space-time by carrying out Langevin dynamics on the path action, which can be written pointwise for $i=1,\ldots, N-1$ as
\begin{equation}
    x_{i}(k+1) = x_i(k) - \diff{S_*(x(k))}{x_i}{} \tau + \sqrt{2\tau} \eta_i(k),
\end{equation}
where $\tau>0$ is the time step of the path-space Langevin dynamics.
Just as in the examples considered above, if there is metastability in the space of paths, the space-time local Langevin dynamics is metastable and inter-conversion between transition paths will be prohibitively slow for a purely local sampling algorithm. 

Fig.~\ref{fig:noneq} shows that the adaptive MCMC algorithm can be used to efficiently and directly sample transition paths from separated metastable basins.
This potential energy surface has been used as a model of entropic switching, originally in Ref.~\cite{park_reaction_2003} (cf. Appendix~\ref{sec:a:TPS}).
The drift term $b$ is the sum of a conservative force $-\nabla V$ and a nonconverservative drift given by the vector field $c(-x_2,x_1)^T,$ leading to a nonequilibrium dynamics.
We condition on the endpoints $x_A = (-1,0)$ and $x_B = (1,0)$, which are the locations of the global minima of the potential and thereby also the center of the two most metastable states on this potential at low temperature.
While techniques based on transition path theory typically require an equilibrium dynamics, the present method applies equally well to problems in which there is a nonequilibrium drift.
The dimensionality of the underlying potential is low in this example, nevertheless, the normalizing flow accepts a $2N$ dimensional input, where $N$ is the number of discretization points, leading to an effective dimension of 200 in this example. 
The results emphasize that metastability in path space can be probed directly even when there is an appreciable gap between the two basins in path space and a substantially different local structure to the typical transition paths. 

\begin{figure}
    \centering
    \includegraphics[width=\linewidth]{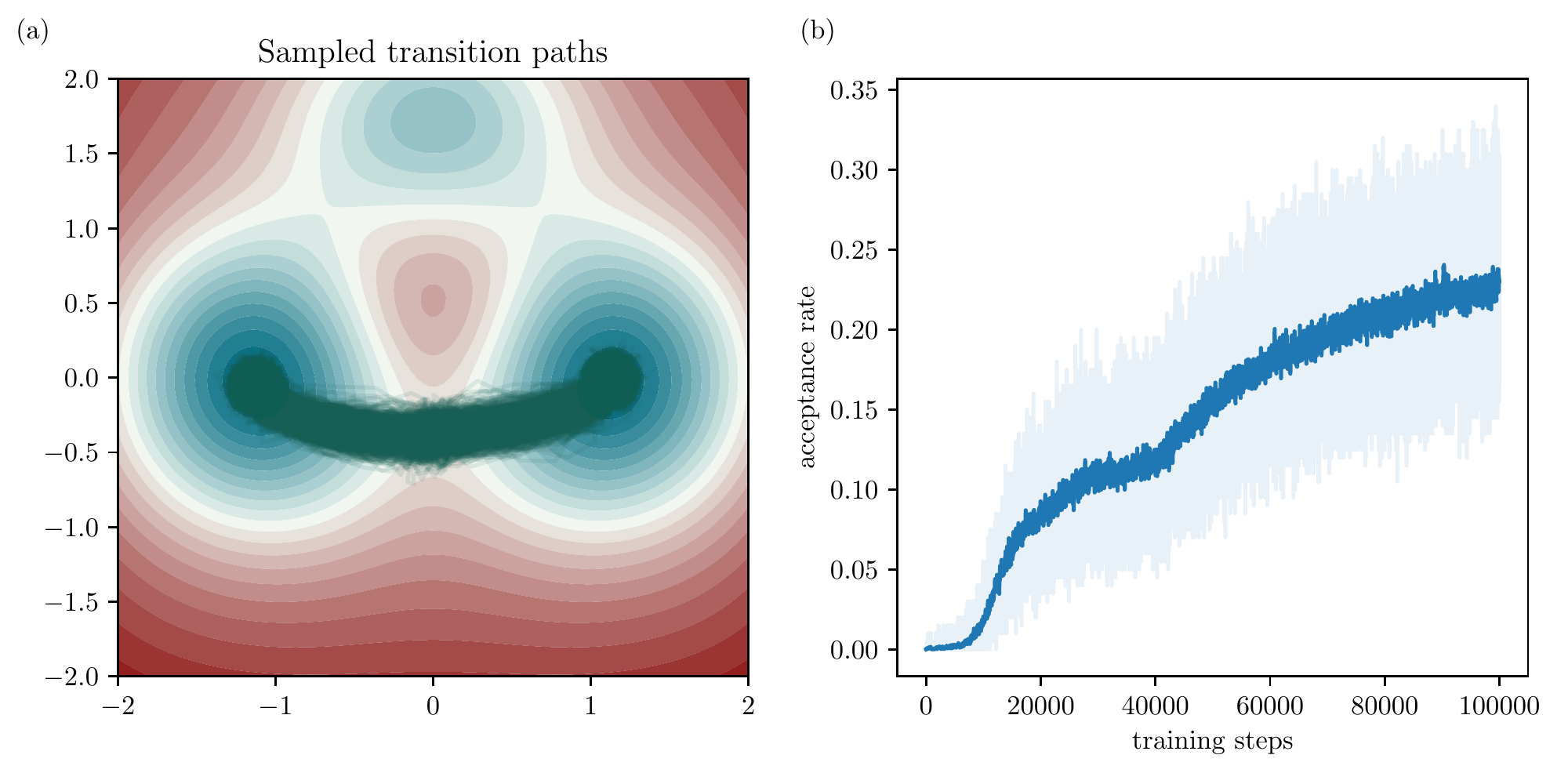}
    \caption{{\bf The map produces transition paths corresponding only to the lower channel when trained with resampling at each training step and a finite batch size }(as discussed in the main text). The resulting generator produces samples shown in $(a)$. The acceptance rate is shown as a function of training time in $(b)$. The rate of increase in acceptance plateaus when the initial samples from the upper channel have been eliminated. }
    \label{fig:resampling}
\end{figure}

The paths shown in Fig.~\ref{fig:noneq} were obtained by training the normalizing flow with Langevin alone, the limit of slow resampling during the training. 
This was done to mitigate a potential failure mechanism of the algorithm in which a higher probability mode $A$ is learned by the normalizing flow before a lower probability mode $B$ (for a detailed discussion cf. Appendix~\ref{sec:a:TPS} and Fig.~\ref{fig:resampling}).
With a finite batch size, if a $A$ is learned initially, the Langevin walkers from $B$ can be transported to $A$, potentially eliminating all data from $B$. 
Although it comes with a higher training cost, a systematic way to avoid this failure is to rely on a mixture of learned generators (one for each metastable basin), as we discuss in our interacting particle system example.

\begin{figure}
    \centering
    \includegraphics[width=0.8\linewidth]{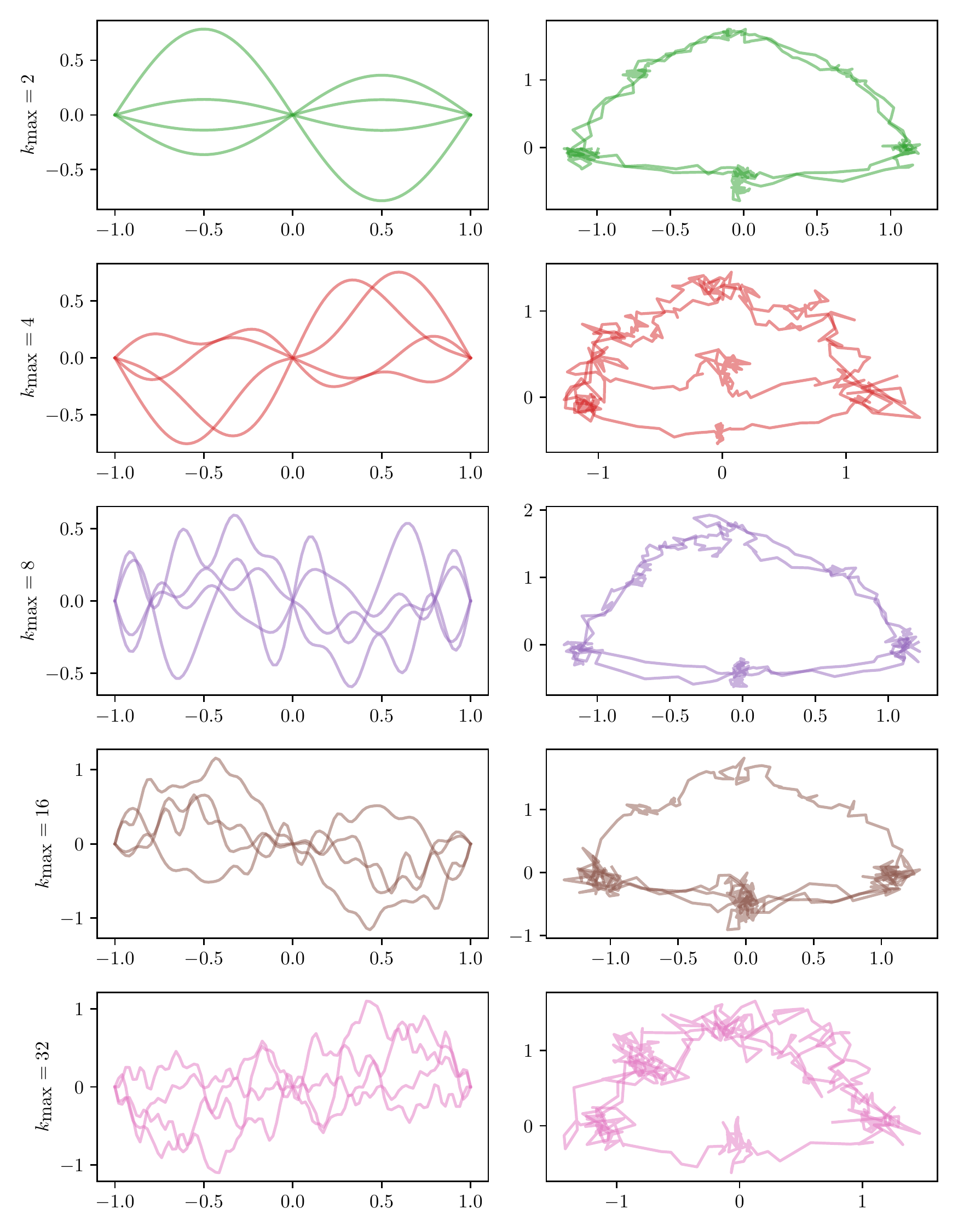}
    \caption{The first four terms of the stochastic Karhunen-Lo\`eve expansion of the Brownian bridge are plotted over 5 realizations in $(a)$. The map $T$, with parameters optimized from the training procedure with no resampling (cf. Fig~\ref{fig:noneq}), applied to the terms of this expansion are shown in $(b)$.}
    \label{fig:karhunen}
\end{figure}

As in the case of the stochastic Allen-Cahn model, the map learns low frequency information and preserves high-frequency.
We illustrate this fact by examining the behavior of the map on a truncated Karhunen-Lo\`eve expansion of the Brownian bridge, shown in Fig.~\ref{fig:karhunen}.
The low order terms of this expansion capture the large scale structure of the bridge process whereas higher order terms include higher frequency oscillations. 
The map recapitulates the overall structure of the paths for an expansion of the bridge process truncated at $k_{\rm max}=2$, but the paths themselves are comparatively smooth compared to solutions of the conditioned SDE~\eqref{eq:nsde}.
Including higher order terms does not change the overall shape of the paths, furthermore, doing so restores the higher frequency oscillations in the transition paths. 

\label{sec:a:TPS}

To sample~\eqref{eq:nsde}, as described in the main text, we discretize the path into $N$ points with a time step $\Delta t$. 
We then carry out Langevin dynamics on the target path action
\begin{equation}
    S_* = \beta \Delta t \sum_{i=1}^{N-1} \left| \frac{x_{i+1}-x_i}{\Delta t} - b(x_i)\right|^2
\end{equation}
with an effective inverse temperature of $\beta\Delta t$, where the factor of $\Delta t$ in the effective temperature ensures the proper scaling in the limit $\Delta t \to 0$ and the boundary conditions $x_0=x_A$ and $x_N=x_B$ are imposed.
The base measure has action
\begin{equation}
    S_{\rm B} = \beta \Delta t \sum_{i=1}^{N-1} \left| \frac{x_{i+1}-x_i}{\Delta t} \right|^2.
\end{equation}
Langevin dynamics on the path action, written pointwise for $i=1,\ldots, N-1$ evolves as  
\begin{equation}
    x_{i}(k+1) = x_i(k) - \diff{S_*(x(k))}{x_i}{} \tau + \sqrt{2\tau} \eta_i(k),
\end{equation}
where $\tau>0$ is the time step of the path-space Langevin dynamics.
In all experiments $\beta=4$, $N=100$, $\Delta t = 6\times 10^{-3}$, $\tau = 5\times 10^{-5}$ and
\begin{equation}
    b(x) = -\grad V(x) + f(x)
\end{equation}
with 
\begin{equation}
    V(x) = \sum_{i=1}^4 A_i e^{(x-\mu_i)^2}
\end{equation}
with $A_1 = 30$, $A_2=-30$, $A_3=-50$, $A_4=-50$ and $\mu_1=(0,1/3)^T$ $\mu_1=(0,5/3)^T$, $\mu_1=(-1,0)^T$, $\mu_1=(1,0)^T$.
The nonconservative part is given by 
\begin{equation}
    f(x) = c (-x_2, x_1)^T
\end{equation}
with $c=2.5.$
We examine transition paths conditioned with $x_A = (-1,0)$ and $x_B = (0,1)$ and take the Brownian bridge process connecting these two points as the corresponding base measure. 

To assess the smoothness and locality of the map, we plotted $T(B_i^{(k_{\rm max})})$ for a set of paths $B_i$ $i=1,\dots,4$ which come from a truncated Karhunen-Lo\`eve expansion of the Brownian bridge process. 
Shown in Fig.~\ref{fig:karhunen}, the left hand side plots
\begin{equation}
    B_i^{(k_{\rm max})} = \sum_{k=1}^{k_{\rm max}} Z_k \frac{\sqrt{2}\sin(\pi k x)}{\pi k}
\end{equation}
with $Z_k\sim \mathcal{N}(0,1)$ for increasing values of $k_{\rm max}.$
These plots illustrate that the high frequency fluctuations visible at larger values of $k_{\rm max}$ lead to path fluctuations of higher frequency.

\section{Interacting particle system experiments}
\label{sec:a:interacting} Given the instantaneous particles positions $\{x_i(t)\}_{i=1}^N$ with $x_i \in \Omega = [0,L]^d$, the microscopic particle density is defined as
\begin{equation}
    \label{eq:meandensmicro}
    u^{(N)}(t,x) = \frac1N \sum_{i=1}^N \delta (x-x_i(t))
\end{equation}
If the particle positions evolve according to  Eq.~\eqref{eq:Lpart}, in the mean field limit as $N\to\infty$, $u^{(N)}(t,x)\to u(t,x)$ that satisfies the McKean-Vlasov equation
\begin{equation}
    \label{eq:McKV}
    \partial_t u = \nabla \cdot \left ( u \int_\Omega \nabla W(x-y) u(t,y) dy \right) + \beta^{-1} \Delta u
\end{equation}
This equation in the Wasserstein gradient flow over the free energy~\eqref{eq:MFenergy}, i.e. it can be written as 
\begin{equation}
    \label{eq:McKV2}
    \partial_t u = \nabla \cdot \left ( u \frac{\delta F(u)}{\delta u} \right)
\end{equation}
where $F(u)$ is the free energy for the local particle density $u(x)$:
\begin{equation}
    \label{eq:MFenergy}
    F(u) = \frac12\int_{\Omega\times \Omega} \! \! \! W(x-y) u(x) u(y) dx dy + \beta^{-1} \int_\Omega u(x)\log u(x)  dx
\end{equation}
The second term proportional to $\beta^{-1}$ in $F(u)$ accounts for the entropy of the particle configurations. 
Writing Eq.~\eqref{eq:McKV} as in Eq.~\eqref{eq:McKV2} shows that the local minimizers of the free energy $F(u)$ are stable fixed points of these equations; these minimizers  satisfy Euler-Lagrange equation 
\begin{equation}
    \label{eq:ELMcV}
    \int_\Omega W(x-y) u(y) dy + \beta^{-1} \log u = K
\end{equation}
where the constant $K$ comes from the Lagrange multiplier used to enforce $\int_\Omega u(x) dx = 1$. This equation can also be reorganized into
\begin{equation}
    \label{eq:drop}
    u(x) = C \exp\left(-\beta \int_{\Omega} W(x-y) u(y)dy\right)
\end{equation}
where  $C$ is a constant ensuring that $\int_\Omega u(x) = 1$.
Classifying all the minimizers of Eq.~\eqref{eq:MFenergy} analytically via solution of Eq.~\eqref{eq:drop} is a complicated problem~\cite{carrillo2019long} but it can be done numerically (see below for details). These calculations show that at low enough $\beta^{-1}$, the energy term in Eq.~\eqref{eq:MFenergy} dominates, and the global minimizer of the free energy $F(u)$ is a nonconstant solution $u_d(x)$ to Eq.~\eqref{eq:drop};  when $a$ is small enough compared to $L$, this profile can be well approximated by replacing $u$ at the right hand of side of Eq.~\eqref{eq:drop} by a Dirac, which gives the droplet local density
\begin{equation}
    \label{eq:drop2}
    u_d(x) \simeq C \exp\left(-\beta W(x)\right) \;, \qquad a\ll L \;.
\end{equation}
At high enough $\beta^{-1}$, however, the global minimizer of Eq.~\eqref{eq:MFenergy} is simply the homogeneous $u_h(x) = L^{-d}$. The transition between $u_d$ and $u_h$ arises at some critical  $\beta_c$ at which $u_d$ and $u_h$ switch role between global and local minimizers: around $\beta_c$ the least favored state remains a local minimizer of Eq.~\eqref{eq:MFenergy}, indicative of the transition being first order. The phase diagram shown in Fig.~\ref{fig:pd1} was obtained this way.

We performed the numerical simulations in $d=2$ dimension using $L=1$ and $a=1/(2\pi) \approx 0.16$ in \eqref{eq:shortr}. To obtain the phase diagram shown in Fig.~\ref{fig:pd1} we first  calculated $u_d(x)$ via iteration of Eq.~\eqref{eq:drop}, i.e. starting from  $u_0(x)$ given in Eq.~\eqref{eq:drop2}, and iterating upon
\begin{equation}
    \label{eq:iterud}
    u_k (x) = C_k  \exp\left(-\beta \int_{[0,1]^2} W(x-y) u_{k-1}(y)dy\right), \qquad k=1,\ldots
\end{equation}
with $C_k=\int_{[0,1]^2}\exp\left(-\beta \int_{[0,1]^2} W(x-y) u_{k-1}(y)dy\right)dx$ until convergence. The stability of $u_d(x)$ and $u_h(x) = 1$ was assessed by using perturbations of these state as initial conditions for the McKean-Vlasov Eq.~\eqref{eq:McKV} and solving this equation numerically using a pseudospecral code with an exponential integrator based on the diffusion term. 

\begin{figure}
    \centering
    \includegraphics[width=\linewidth]{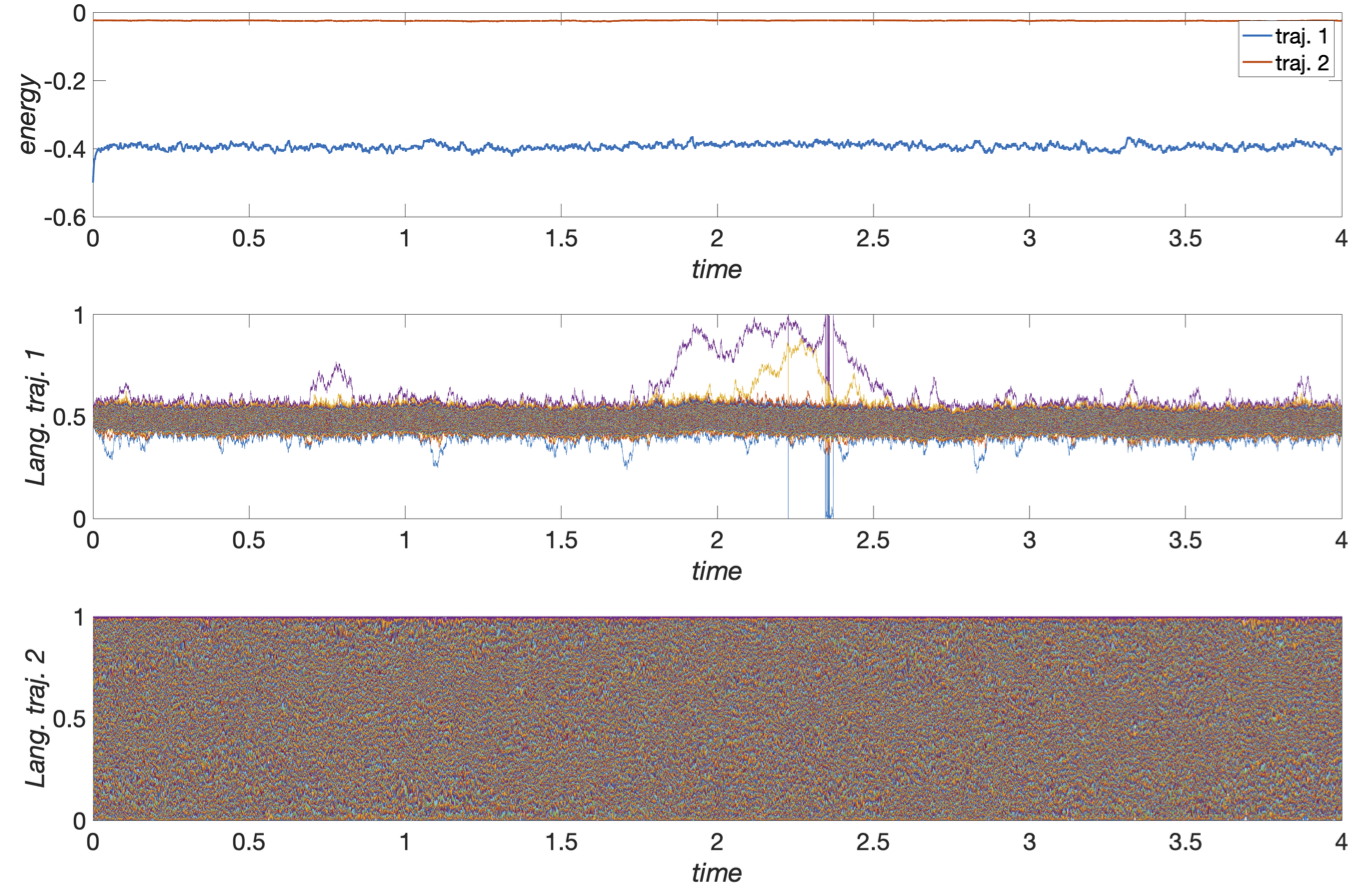}
    \caption{Upper panel: the interacting particle system's energy along two trajectories obtained by brute-force simulation of \eqref{eq:Lpart} with $1/\beta = 0.08$, starting with initial positions  starting with initial positions at the center of the box  (traj. 1) or  drawn uniformly in it (traj. 2). Lower two panels: the $x$-coordinates of the 200 particle's trajectories, showing that the particles started in the clustered state remain clustered, and those started in the homogeneous state remain homogeneous. }
    \label{fig:Lanbf}
\end{figure}

The molecular dynamics simulations were conducted with $n=200$ particles. Fig.~\ref{fig:Lanbf} shows the result of the brute-force simulations of \eqref{eq:Lpart} with time step $h=1e-4$ and $1/\beta = 0.08$, starting with initial positions at the center of the box or  drawn uniformly in it. As can be seen, the system remains trapped in the homogeneous or droplet state it was started in. 

\begin{figure}
    \centering
    \includegraphics[width=.5\linewidth]{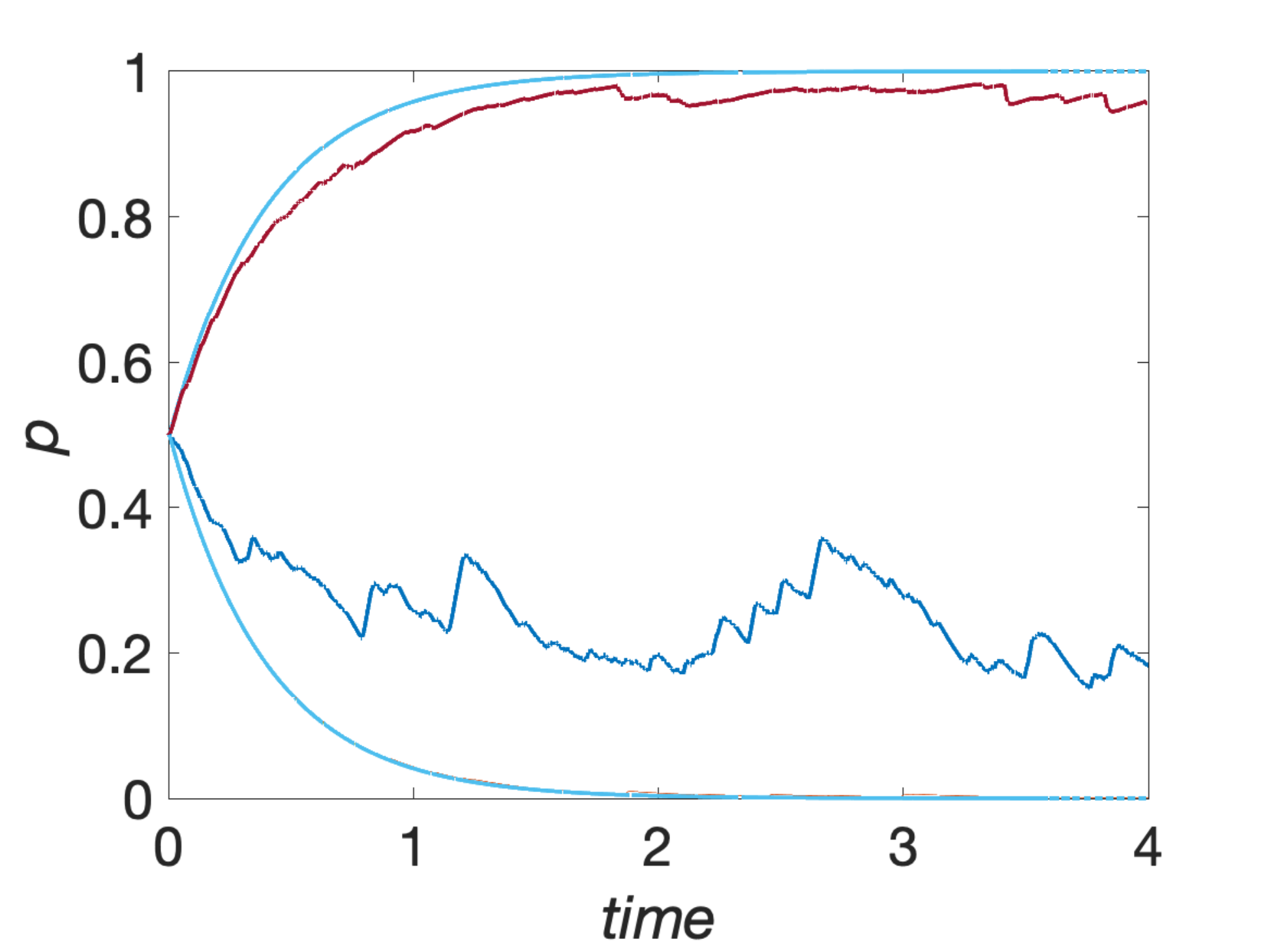}~~\includegraphics[width=.5\linewidth]{phased.pdf}
    \caption{Left panel: The evolution of $p$ by gradient descent on $S(X) \approx\log\rhohat(X)$ for the 20 temperatures shown in Fig.~\ref{fig:pd1} (repeated here as right panel). Almost all $p$'s evolve either quickly to 1 or 0 (and their timeseries  superpose). The number of iterations is the same here as in Fig.~\ref{fig:Lanbf}.). Right panel: repeat of Fig.~\ref{fig:pd1}.}
    \label{fig:pevol}
\end{figure}

In the adaptive simulations, we performed 4 steps of simulations of  \eqref{eq:Lpart} with time step $h=10^{-4}$ followed by one attempt of resampling. The gradient descent of $p$  was conducted simultaneously using a time step $dt = 10^{-3}$. The evolution of $p$ for the twenty temperatures shown in Fig.~\ref{fig:pd1} is shown in Fig.~\ref{fig:pevol}: the average acceptance rate of the resampling steps was between  35\% and 55\% in all these runs.

\paragraph{Code availability}

All code and model parameters are available under an MIT license in a github repository for this project: \url{https://github.com/marylou-gabrie/flonaco}, archived at \url{https://doi.org/10.5281/zenodo.4783701}. 

\section{Computing free energy differences}
\label{app:deltaF}
Once the normalizing flow has been learned to assist sampling, there are two possible methods for estimating expectations. 
Ref.~\cite{nicoli_estimation_2021} trains the map with a different procedure, but compares empirical averages from Metropolized unbiased samples with an importance sampling reweighing procedure with direct samples from the push-forward $\rhohat$. 

Here we focus on the special case of computing free energy differences between metastable states. Denoting by $A$ and $B$ two sets of configurations in $\Omega$, the difference between their free energies is given by
\begin{equation}
\label{eq:fe}
\begin{aligned}
        -\Delta F_{AB} &= \log \frac{\int_{\RR^d} \mathbbm{1}_A(x) e^{-U_*(x)} dx}{\int_{\RR^d} \mathbbm{1}_B(x) e^{-U_*(x)} dx} \\
        &= \log \EE_* (\mathbbm{1}_A)  - \log\EE_* (\mathbbm{1}_B).
\end{aligned}
\end{equation}
In Fig.~\ref{fig:phi4} we used the set $A  = \{\phi: \int_0^1 \phi(s)ds>0\}$ and $B  = \{\phi: \int_0^1 \phi(s)ds<0\}$ and denote $F_+ - F_- = \Delta F_{AB}$

We can obtain unbiased samples from $\rhostar$ by running a Metropolis-Hasting MCMC with the fast mixing kernel $\hat \pi_T$ \eqref{eq:nfker}. From $n$ samples $\{x_i\}_{i=1}^n$, an estimator of the partial partition function and its variance are
\begin{equation}
    \begin{aligned}
    \EE_* (\mathbbm{1}_A) & \approx \hat Z_A^{\rm  MC} = \frac1n \sum_{i=1}^n \mathbbm{1}_A (x_i),\\
    {\rm Var}(\hat Z_A^{\rm MC}) & \approx \frac{\widehat {\rm Var}_n( \mathbbm{1}_A (x_i))}{n_{\rm eff}} = \frac{\widehat {\rm Var}_n( \mathbbm{1}_A (x_i))}{n / \tau_{\rm eff}}
    \end{aligned}
\end{equation}
where $n_{\rm eff}$ is the effective sample size deduced from the autocorrelation time of the chains $\tau_{\rm eff}$ (see e.g. \cite{levy_generalizing_2018b} for details). While the map allows fast mixing between modes, it need not be perfect as the local MCMC kernel can compensate when some mass is not well represented by the map. 

Alternatively, we can use an importance sampling estimator drawing $\{x_i\}_{i=1}^n$ from $\rhohat$,
\begin{equation}
\EE_* (\mathbbm{1}_A) \approx  \hat Z_A^{\rm IS} / Z_* = \frac 1 Z_* \sum_{i=1}^n \mathbbm{1}_A (x_i) \hat w (x_i),
\end{equation}
where we use unnormalized weights  $\hat w_i = e^{-U_*(x_i)}/ \rhohat(x_i)$, and the unknown $Z_*$ cancels out in the free energy difference estimator:
\begin{equation}
    -\Delta F_{AB} \approx \log\left( \sum_{i=1}^n \mathbbm{1}_A (x_i) \hat w _i \right) - \log\left(\sum_{i=1}^n \mathbbm{1}_B (x_i) \hat w _i\right).
\end{equation}
Here again the quality of the estimator can be monitored using an estimate of the effective sample size
\begin{equation}
    n_{\rm eff} = \frac{\left(\sum_{i=1}^n \hat w_i\right)^2}{\sum_{i=1}^n \hat w_i^2}.
\end{equation}

For both types of estimators, the map learned in certain thermodynamics conditions can be leveraged to sample in distinct conditions without relearning. This is possible thanks to the Metropolis-Hasting accept-reject in the MCMC and the importance sampling weights that correct mismatches between the target and push-forward distributions. As conditions are moved away from the training setting, quality of the Monte Carlo estimators can be assessed via the variance estimators above.

\end{document}